\documentclass[preprint,11pt]{aastex62}
\usepackage{multirow}
\usepackage{longtable}
\usepackage{booktabs}

\usepackage[caption=false]{subfig}
\usepackage{subfig}

\usepackage{etoolbox}

\graphicspath{{./}{figures/}}

\shorttitle{VLA observations of Quasars}
\shortauthors{Wardle et al.} 

\begin{document}

\title{VLA Observations at 6 and 19 GHz of a Complete Sample of Radio Loud Quasars with redshifts between 2.5 and 5.28: II. Sample definition, radio images, and analysis .}

\correspondingauthor{John Wardle}
\email{wardle@brandeis.edu}

\author{John F.C. Wardle}
\affil{Department of Physics, MS-057 Brandeis University,\\
Waltham, MA, USA}

\author{Mark Birkinshaw}\thanks{deceased}
\affiliation{H.H. Wills Physics Laboratory, University of Bristol, Bristol, UK}

\author{Diana M. Worrall}
\affiliation{H.H. Wills Physics Laboratory, University of Bristol, Bristol, UK}

\author{C. C. Teddy Cheung}
\affiliation{Space Science Division, Naval Research Laboratory, Washington DC, USA}

\author{Doug Gobeille}
\affiliation{Physics Department, University of Rhode Island, Kingston, RI, USA}

\author{Daisy Matthews}
\affil{Department of Physics, MS-057 Brandeis University, Waltham, MA, USA}

\author{Jesse Reeves}
\affil{Department of Physics, MS-057 Brandeis University, Waltham, MA, USA}



\begin{abstract}

We present high resolution (subarcsecond) observations at 6.2 and 19.6 GHz made with the Karl G. Jansky Very Large Array of 113 radio-loud quasars that form a complete flux limited sample ($\geq 70$ mJy at 1.4 GHz), with spectroscopic redshifts between 2.5 and 5.28. These redshifts correspond to ages since the big bang of 1.1 to 2.6 Gyr, or more colloquially, from Cosmic Dawn to Cosmic High Noon. This is when large scale structure formation and galaxy formation were proceeding  at an ever increasing pace, and this sample appears to be unique (for now) for spanning this era. We show images of the significantly resolved sources, and list structural properties of all of them. Spectroscopic redshifts (mostly from the SDSS) are available for every source, as are higher resolution VLBI images for most of them. We also compare the physical properties of our sample with those of two other complete samples at lower redshifts to broaden the study of redshift dependent effects.  With only one exception, every source in the high-redshift sample displays a compact radio source coincident with the optical position. About half of the sources are well resolved, with 13 of them displaying a prominent jet. We suggest that these jets (and their associated lobes), which reach far outside their embryonic host galaxies, can serve as useful probes of the circumgalactic environment.




%



\end{abstract}


\keywords{galaxies: high-redshift --- galaxies: jets ---  quasars: general --- radio continuum: galaxies}


\section{Introduction} \label{sec:intro}

We have previously posted on astro-ph a paper reporting heterogeneous Very Large Array (VLA) observations of a complete sample of high redshift quasars based on a mixture of archival and new observations \citep[Paper I;][]{GWC14}; see also \citet{GO11}. Here we present uniform short VLA observations (snapshots) of the entire sample at 6.2 GHz (C-band) in the A configuration, and somewhat longer observations of 44 of them at 19.6 GHz (K-band) in the B configuration. The resolution in both sets of observations was about 0.4 arcsec.  This work represents a significant increase in our knowledge of the structural properties of radio-loud quasars in the high redshift Universe. 

The high-redshift Universe is essential for understanding the formation and evolution of galaxies and their active nuclei (quasars). At redshifts of 4 to 5 we can observe the formation of the first quasars and their central engines, as well as study how their properties change over cosmic time. We also know that the high-redshift environment is very different than that of low-redshift sources. The mean density of the intergalactic medium (IGM) scales with redshift as $(1+z)^3$ and the energy density of the Cosmic Microwave Background (CMB) radiation scales with redshift as $(1+z)^4$. 

Since we have spectroscopic redshifts for all our sources, the sample is well defined in terms of cosmological age. Currently, the highest redshift quasar known is J0313-1806 at $z = 7.642$ \citep{Wang2021}. Our sample ranges in redshift from $z = 2.50$ to $z = 5.28$. These correspond to ages since the Big Bang of 2.6 Gyr to 1.1 Gyr, respectively \citep[using Ned Wright’s Cosmic Calculator;][]{Wri06} with $\Omega_{\rm m} =0.3$, $\Omega_{\rm vac} = 0.7$, and $H_{\rm 0} = 70$ km s$^{-1}$ Mpc$^{-1}$.  
This makes the point, as made by many authors, that galaxy formation and black hole/AGN formation took place extremely rapidly in the young universe \citep[e.g.,][]{KH13,Hec14,INA20}.

We wish to point out a second aspect of the extreme youthfulness  of our sample. The epoch 1.08 Gyr after the Big Bang is close in time to the end of the "dark ages" when the very first massive stars started to ionize the universe \citep{Fan2023}. This is sometimes referred to as cosmic dawn \citep[e.g.,][]{Bow2018,Bos2022}, and is when star formation proceeded at an ever increasing pace. The peak in the formation rate of galaxies and stars is around $z = 2$ to 3, and is sometimes referred to as cosmic high noon \citep[e.g.,][]{Mad2014,Som2015}. Our \textit {lowest} redshift is $z = 2.5$. We therefore see our entire sample at cosmic epochs when galaxies and clusters are still in the process of formation. It is possible that the radio structures we observe can help shed new light on this era. As far as we know, there is no other complete, flux limited sample of quasars, above $z = 2.5$, that also has sub-arcsecond scale radio imaging and spectroscopic redshifts. This makes our present sample (at least for the time being) unique.


Finally, it is well established that X-ray emission is a common feature of kiloparsec-scale radio jets \citep{Har2006, Worrall2009}. The two main mechanisms that have been proposed for powerful jets are the scattering of photons of the CMB by the low-energy end of the population of electrons in the jet that generates the radio emission, or alternatively, direct synchrotron radiation from an additional population of high energy electrons in the jet. Deciding between these models is not straightforward \citep{Bre2023}, but the predicted redshift dependencies of the two models are very different \citep{Sch20} because of the strong redshift dependence of the energy density of the CMB. The present sample could help distinguish between the two models for the jet X-ray emission.

\section{A New High Redshift Sample} 

The sample used in Paper I suffered from extremely inhomogeneous observations. It contained a mixture of archival VLA observations at a variety of wavelengths, array configurations and times on source. New observations in the A array at 5 GHz were made only for sources that had never previously been observed at the VLA.

 Here we have constructed a new complete sample of high redshift radio-loud quasars with a radio flux density limit of 70 mJy at 1.4 GHz, with uniform optical data and spectroscopic redshifts. 
 We chose an area of sky in common between the VLA-FIRST \citep{BWH95} and SDSS \citep{A03} fields, ensuring excellent optical data on every source, as well as wide-field radio images from FIRST. The overlapping areas of these surveys resulted in a sample field from 7 to 17.5 hours in right ascension and 0 to 65 degrees in declination.

A query of the NASA/IPAC Extragalactic Database (NED) for sources in our field with redshifts $z \geq 2.5$ produced 137 objects. Twenty four sources were rejected from the sample for various reasons. They are listed in Table 3 as a convenience for other observers.
The edges of the area covered by SDSS are somewhat ragged. There are five sources in the field defined above that do not have SDSS spectra, but all of them do have spectroscopic redshifts and fit our selection criteria in every other respect.
We ended up with a sample of 113 quasars with redshifts ranging 2.5 and 5.28, with 46 of them at $z > 3.0$. 

The sample discussed in the present work is selected at 1.4 GHz, which is a high enough frequency that emission from the flat-spectrum cores cannot be ignored. Since core radio emission (widely believed to be beamed) becomes increasingly dominant at high frequencies, and even more so at large redshifts due to the K-correction, we will want to construct complete sub-samples  that are free of this orientation bias. This can be done by constructing sub-samples of sources whose extended radio emission (widely believed to be unbeamed) is sufficient by itself to exceed the flux limit of the sample. This is analogous to what was done at lower redshift for 3CR quasars by \citet{HR89}.

 \subsection{Sample Completeness}

The radio flux limit of 70 mJy at 1.4 GHz corresponds to a strong source in the FIRST survey and is far above both the noise limit and the confusion limit (their limiting flux density is about 1 mJy). The completeness of our sample therefore hinges mainly on the algorithms used by SDSS to select quasar candidates for spectroscopic observation. 

The quasar selection algorithm (which determines whether or not to take a spectrum) is described in detail by \citet{R02}. They do not try to see redshifts above $z = 5.8$ (which is where the minimum transmission between the {\em i-} and {\em z-} band filter curves at 8280 \AA \,\, is straddled by Ly $\alpha$ emission) and they require flux in at least two optical bands. After editing bad data, they preferentially target point-source matches to FIRST radio sources (within 2 arcsec) without reference to the optical colors. The optical flux limit is set by the number of optical fibers per plate (typically 80) that are assigned to quasar candidates. That limit is {\em i} = 19.1 for redshifts up to $z = 3.0$, and {\em i} = 20.2 for redshifts $z \geq 3.0$. Our sample therefore has different optical flux limits for the 67 quasars with $2.5 \leq z < 3.0$ and the 46 quasars with $z \geq 3.0$.

This algorithm should be extremely efficient at finding radio-loud quasars, because of the positional match with a FIRST radio source, and the absence of any color bias. The completeness of our sample is therefore determined by how often more than 80 fibers were required for a plate. We believe that this is a rare occurrence and that the completeness is close to 100 \%.


Completeness refers to sources that are not in the sample but should be, according to the stated selection criteria. Reliability refers to sources that appear to satisfy the selection criteria but should not be included in the sample. This can be for a variety of reasons and all 24 rejects are listed in Table 3. We are not able to make a numerical estimate of the sample reliability, but we have made every effort to find all the sources that should not be in the sample.

\section{Observations and Data Reduction}
\label{s:obs}

\subsection{The 19.6 GHz (K-band) observations} 

Observations of 42 members of the sample were made in May 2012 in the B configuration (Project code 12A-149, PI Worrall). The bandwidth was 2048 MHz and the effective observing frequency was 19.6 GHz. The sources observed were those for which, at the time, it was unclear if resolved structures were present at the resolution of the Chandra Observatory, and we wished to identify new candidates for observing in the X-ray.  The data were edited, calibrated and imaged in CASA by Mark Birkinshaw with a restoring beam of 0.32 arcsec.

\subsection{The 6.2 GHz (C-band) observations}

The entire sample was observed with the JVLA in the A configuration, equipped with new receivers of greater sensitivity and wider bandwidth than our previous observations. The observations were made over 4 days in November 2012 (project code 12B-230, PI Wardle). The observing bands were centered on 5.0 GHz and 7.4 GHz, each with a bandwidth of 1.024 GHz, giving an effective observing frequency of 6.2 GHz. The resolution was about 0.27 arcsec.

The data were edited, calibrated and imaged with two independent software packages CASA \citep{BBC2022} and AIPS/DIFMAP \citep{GR90,G00,S97}. The reason for this was that CASA was relatively new at the time and known to contain bugs.  We wanted to gain experience with CASA and to compare its results with those from AIPS/DIFMAP.

The data reduction in CASA was carried out by David Matthews and Jesse Reeves, as their senior research projects at Brandeis University. Somewhat later, the same observations were analyzed using calibration provided by Erik Carlson in CASA using the current VLA calibration pipeline and imaged by Doug Gobeille at the University of Rhode Island using DIFMAP.

\section{Results and Images}
\label{s:results}

\subsection{The whole sample}

In this section we present the whole sample of 113 high redshift ($z \geq 2.5$) sources. They are listed in Table 1.\\

--Table 1 here--\\

The columns list the following quantities.\\

\textit{Column (1)} - J2000 Source name in IAU format.

\textit{Column (2)} - Redshift from NED. Nearly all of them come from the SDSS, and we use what is currently listed as the preferred value.

\textit{Columns (3) and (4)} - Right Ascension and Declination (epoch J2000) of the {\em pointing position} of the observation. Where we show an image, it is also the coordinates of the center of the field of view. Invariably, this is coincident with the compact radio core. We do not show the "preferred" position listed in NED, because in general the radio positions are more accurate than the SDSS optical positions. (And sometimes the "preferred position" is taken from the FIRST catalog, which has about ten times worse resolution than our observations.)

\textit{Column (5)} - {\em i}-band magnitude taken from the SDSS. For a few sources noted with asterisks*, they are $R2$-band magnitudes from the USNO-B1.0 \citep{MLC03}.

\textit{Column (6)} - Morphological type. Briefly, these are unresolved Cores (C); Cores with a small extension (E); Doubles (D), consisting of a compact radio core coincident with the optical ID, together with extended emission on only one side of the core; Triples (T), consisting of a compact radio core coincident with the optical ID, and with extended emission on both sides of the core.
Note that every source except one (J1528+5310) exhibited a compact radio component coincident with the optical quasar. Entries with an asterisk indicates a change from the morphology listed in paper I.

\textit{Column (7)} - Total observed source flux density at 1.4 GHz: the total source fluxes were obtained from the final images. 

\textit{Column (8)} - Core flux density at 1.4 GHz. These were also obtained from the final images. In columns (7) and (8), a conservative estimate of the error is $\pm 5\%$, which should be applied to all the listed flux densities.

\textit{Columns (9) and (10)} - The corresponding information at 6.2 GHz.

\textit{Columns (11) and (12)} - The corresponding information at 19.6 GHz.

\textit{Columns (13)} - References to VLBI and X-ray observations. "V" denotes that there are images at VLBI resolution. These can be found in the remarkable Astrogeo website (http://astrogeo.org/) maintained by Leonid Petrov \citep[see][]{PKP19,PK25}.  "X" denotes that the source is in the \textit{Chandra} Source Catalog \citep{EPG10}, where we use Release 2.1 from 2024 (https://cxc.cfa.harvard.edu/csc/). Chandra is the only X-ray observatory with spatial resolution sufficiently high to be matched with the resolution of the VLA radio data.\\

\subsection{The Images}

6.2-GHz images for 48 members of the sample are displayed in Figures 1-6, and their tabular data are listed in Table 2. We do not show images of compact (C) or slightly extended (E) sources.

For each image the half-power restoring beam is shown as a filled ellipse in the lower left hand corner. 
The contours are logarithmic and increase by factors of $\sqrt(2)$ for the lowest four contours, and by factors of $2$ thereafter.\\

--Table 2 here --
\\

\textit{Columns (1), (2) and (3)} are taken directly from Table 1. 

\textit{Column (4)} gives the bottom contour level in the image as a percentage of the peak flux density listed in \textit{Column (5)}

\textit{Columns (6) and (7)} - $\theta_B$ and $\theta_F$: the angular distances from the core to the brighter hot spot and from the core to the fainter hot spot, respectively.

\textit{Column (8)} - Bending Angle. This is defined as the complement of the angle between the lines from the peaks in the radio source extremities to the radio core.

\textit{Column (9)} - Core Spectral Index, $\alpha$ (defined as $S_{\nu} \propto \nu^{-\alpha}$) between 1.4 GHz and 6.2 GHz.

\textit{Column (10)} - Core Spectral Index between 6.2 GHz and 19.6 GHz.

\subsection{Rejected Sources}

The NED inevitably contains a number of misidentifications, positional inaccuracies, erroneous redshifts and poorly measured flux densities. These usually come from older observations and are superseded by more recent data from FIRST and SDSS. It was important to check each source to verify its inclusion in the sample.

This led to 24 sources in the original sample being rejected for various reasons. Checking can be time consuming, so as a convenience to other observers we list them in Table 3, together with a brief reason for why they should not be included. Where a source has been identified as both a galaxy and a quasar in NED we defer to the classification by SDSS.\\

-- Table 3 here  --\\

\section{Discussion: Properties of the High Redshift Sample}
\label{s:Discussion}

An examination of Tables 1 and 2 and Figures 1-6 shows that the sample contains a wide variety of radio structures. In this section we will discuss the radio structures and compare them to those found in samples of lower redshift quasars. The purpose is to see if the high redshift sources are different, and what they can tell us about their environment and the evolution of AGN at early times in the universe.

\subsection{The redshift distribution for the high redshift sample and its dependence on morphology}

We have classified our images into four morphological types: Compact (C, angular size less than 0.3 arcsec, unresolved by the VLA at 6.2 GHz), Extended (E, image slightly resolved, with angular size less than 1.0 arcsec), Double (D, extended emission on only one side of the core; in Miley’s classification (BM88 and prior) this would be called a d2 double), and Triple (T, with extended radio emission on both sides of the core). We note that nearly every source in our sample shows radio emission from a compact radio component that is coincident with the optical quasar. In what follows, we will count the Compact and Extended sources together, because we have the least to say about them.

\begin{figure}
\figurenum{7}
\plotone{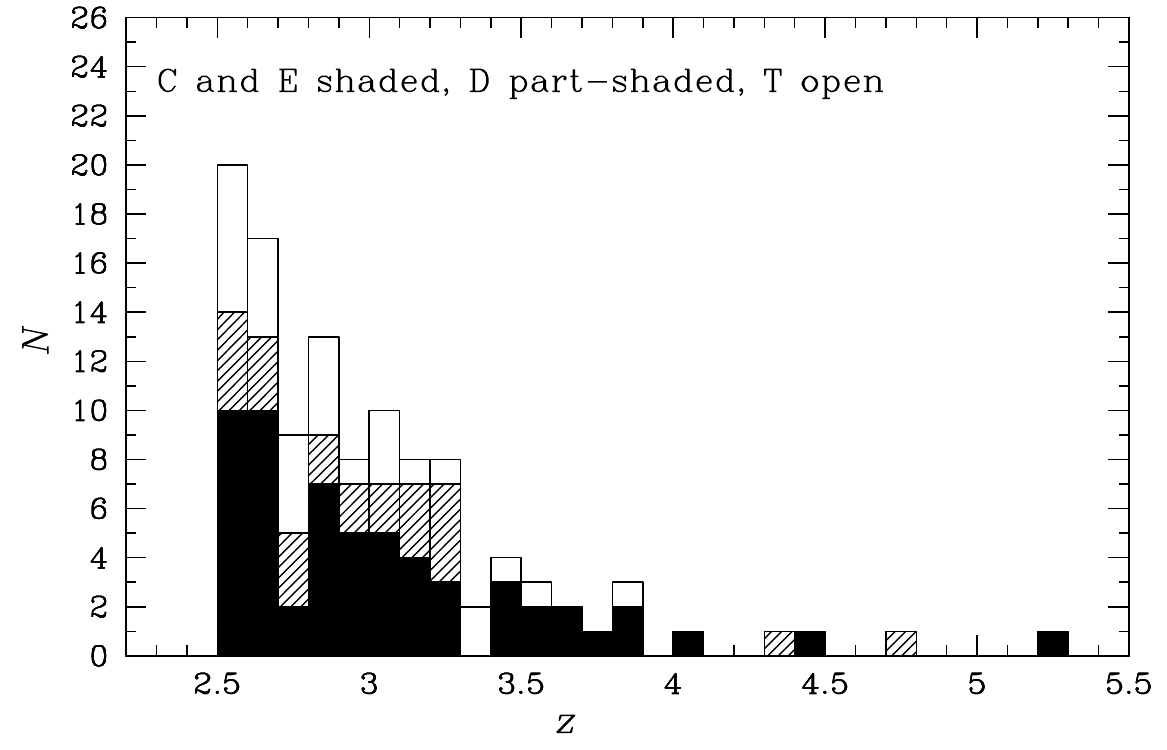}
\caption{Redshift distribution of the 113 source $z \geq 2.5$ sample in bins of $\Delta z=0.1$.}
\end{figure}

Figure 7 shows the redshift distribution for the 113 quasars in our sample, distinguished by morphology.
There does not seem to be much dependence of morphology on redshift. We note that the second highest redshift source in the sample (J1430+4204 at $z = 4.715$) exhibits a double (D) structure. The highest redshift source that exhibits a triple (T) structure has a redshift of 3.82 (J1242+3720). There are only 5 sources with higher redshifts, and the lack of another exhibiting triple structure is probably not significant.\\

\subsection{The three point spectral index distribution for the whole sample and its dependence on redshift and morphology}

Figure 8 shows the core spectral indices between 1.4 and 6.2 GHz plotted against those between 6.2 and 19.6 GHz.  Spectral index is defined by $S_{\nu} \propto \nu^{-\alpha}$. Two sources are dropped from this plot.  One is J1528+5310 because its anomalously large value of $\alpha^{CK}$ causes bunching of the other sources (it lies well to the right of line of equal spectral index), and the other is 1204+5228 because its small-scale resolved emission renders the 1.4-GHz core flux density from FIRST unreliable for this purpose (it lies somewhat to the left of the line of equal spectral index).

We find no significant dependence on redshift for either spectral index. But the median spectral index between 6.2 and 19.6 GHz is 0.6, while the median spectral index between 1.4 and 6.2 GHz is 0.3. This steepening of the spectra at higher frequencies is very evident in Figure 8, where 80\% of the points lie below the line that corresponds to straight spectra. Variability is a factor in this Figure, since the L-band data date from about 8 years earlier than the C-band and K-band data.  However, this should be random and not affect the overall trend for most data points to lie to the left of the lie of equality. 

The steepening of the spectra of core components at higher frequencies is likely due to radiative (synchrotron/IC) losses, and may be exacerbated in high redshift sources where the emitted frequencies are higher. We see that almost no cores have the often assumed flat spectral index, $\alpha = 0$. Care should therefore be exercised when applying K corrections.

\begin{figure}
\figurenum{8}
\plotone{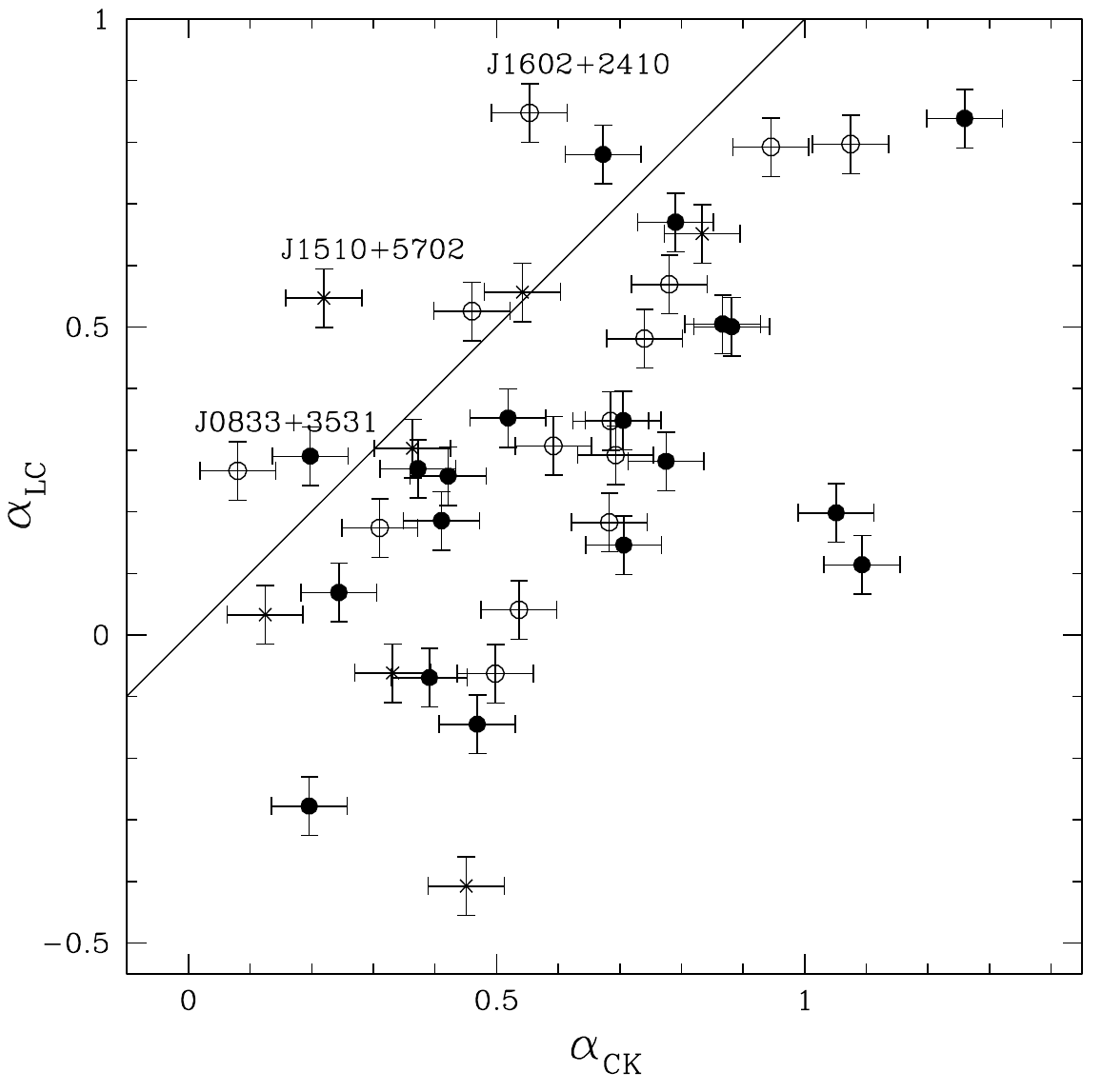}
\caption{1.4-6.2 GHz core spectral index versus 6.2-19.6 GHz core spectral index for 40 sources. A 5\% error has been assumed for all the flux densities. The symbols are similar to Figure 1: C and E filled circles, D crosses, T open circles. Three sources with anomalous 'concave' spectral curvature are identified in the figure.}
\end{figure}

\subsection{Comparing our sample to lower redshift samples}

 Although our sample spans $z = 2.50$ to $z = 5.28$, the \textit{median} redshift is $z = 2.89$ and 82\% of them lie between $z = 2.5$ and $z = 3.3$. This rather narrow range limits the possibilities for analyzing for redshift dependent effects within the sample. We therefore also consider two other samples which are equally well observed but are at lower redshift. These are 3CRR \citep{LRL83} and the sample of intermediate redshift quasars in \citet{BM88}.
 
3CRR contains 42 quasars between redshifts 0.311 (3C249.1) and 2.019 (3C9). The median redshift is 0.915.
The flux limit is 10.9 Jy at 178 MHz. The entire catalog, together with a huge amount of ancillary data and latest updates is available online at an outstanding website maintained by Martin Hardcastle at: https://3crr.extragalactic.info/.

The Barthel-Miley sample of 80 intermediate redshift quasars (hereafter designated as BM88) is listed in their Nature paper. They select on spectral index ($\alpha > 0.6$ between 1.4 and 5.0 GHz), and their redshifts range from 1.50 to 2.91, with a median of 1.88. There is a small amount of overlap between the samples. The six highest redshift quasars in 3CRR are also in BM88. At the high redshift end of BM88, there are six quasars with $z > 2.5$ four of which are also in our sample. 
 
 We are not trying to combine the three samples into a single larger sample. Rather, we use 3CRR and BM88 to illustrate the properties of other complete samples in different redshift ranges.  To remove the overlapping quasars from any of the three samples, would spoil the careful (and different) selection criteria of each. Rather, one must simply take care not to count them twice in any analysis.

\subsection{The dependence of morphology on redshift}

--Table 4--here\\

Table 4 shows how the occurrence of the three main morphological types varies between the three samples. Figure 1 showed the fine grained distribution of morphologies within the present high redshift sample.  Table 4 is coarse grained where the data are aggregated over each redshift range. The numbers list the median values and the inter-quartile range for each quantity. 

There are obvious differences between the three samples. First, the fraction of compact + slightly extended sources increases by a factor of two between 3CRR and this paper. This is mainly due to the K correction,  which selects flatter spectrum sources at higher redshift.

Second, 49\% of the sources in our sample are well resolved by the VLA, with structures that are either one-sided (D) or two-sided (T). Thirteen of them exhibit well defined jets (see Table 2, and the images in Figures 1-6). In all cases, the jets are one sided, regardless of whether the overall structure is one-sided (4 cases) or two-sided (9 cases). There are no visible ‘counterjets’ which is always the case for high power FR II radio sources at any redshift \citep[e.g.,][]{Bridle84}. This is  attributed to relativistic boosting of the forward pointing jet, and corresponding dimming of the backward pointing jet. \citet{WA97} analyzed the jet-to-counterjet flux ratios measured from the deep VLA images of 13 3CRR quasars made by \citet{Bridle1994}. They found that rather modest jet speeds of about 0.6c – 0.7c were adequate to account for the observed jet to counterjet flux ratios and upper limits. Observations of superluminal motion on pc scales \citep[e.g. the MOJAVE program;][]{Lis13} require much higher speeds in the nuclei of most quasars and radio galaxies. This implies that there is significant deceleration between pc and kpc scales.

An unexpected feature in Table 4 is how the incidence of \textit{one-sided} (D) structures changes dramatically with redshift. This changes from almost equal numbers of D’s and T’s in the present high redshift sample (each about 25\% of the total) to only 11\% of D's at intermediate redshift (BM88) to none in the low redshift 3CRR sample. This is  interesting and may indicate that there are some truly one-sided sources at high redshift. Alternatively, our snapshots may have insufficient dynamic range to reveal the counter-hotspots. (Note that these cannot be easily hidden by Doppler dimming, since the typical advance speeds of hotspots were shown by \citet{Sch1995} to be $< 0.1 c$. So, one would expect a counter hot spot to be visible, even though the counter jet is not.)

\subsection{The dependence of linear size and bending angle on redshift}

BM88 claimed that the largest linear size (hotspot to hotspot) of quasars decreases with redshift, and the incidence of distorted structures (indicated by bending angle) increases with redshift. They attributed both effects to a denser intergalactic medium $(\propto (1+z)^3$, and to a more turbulent intergalactic medium at early epochs when galaxy formation is still taking place.

They plot the largest angular size against redshift in their figure 2, and then add lines corresponding to a 500 kpc "meter rule" in 3 different cosmologies. We have the luxury of being able to choose a single concordance model with a reasonable degree of confidence. Using Ned Wright’s Cosmic Calculator \citep{Wri06}, we choose the WMAP cosmology \citep{Sper03} with $\Omega_{\rm m} =0.3$, $\Omega_{\rm vac} = 0.7$, and $H_{\rm 0} = 70$ km s$^{-1}$ Mpc$^{-1}$. (Note that in this context, $H_{\rm 0}$ is simply a scaling factor and does not affect the \textit{shape} of the largest angular size redshift relation.)

The concordance cosmology gives a shape for the Largest Angular Size - Redshift relation that is very similar to Barthel \& Miley's curve (b), which corresponds to a deceleration parameter $q_{\rm 0} = 0.05$.  Scaling by $70/75$ for the different choices for $H_{\rm 0}$, the curve becomes a plausible upper envelope for most of the points on their plot. It can then be seen that it is only at redshifts greater than two that there appears to be a dearth of large sources. At lower redshifts, there is far less evidence for a dependence of linear size on redshift.\\

Many workers have examined different samples of radio sources for the dependence of linear size on redshift and also on intrinsic luminosity, with a variety of results. \citet{BRW99}, in a very useful section 7 of that paper, summarize the different and contradictory results that have been obtained.\\

A more recent paper \citep{deV06} makes use of a huge dataset from SDSS cross matched with FIRST . Starting with 44,984 quasars from SDSS, they end up with 547 quasars with radio cores and an FR II morphology. Dividing the sample in half at the median redshift of 1.4, they show (their figure 6) that the histograms of the linear size distributions are essentially indistinguishable between the low and high redshift halves. In other words, there is little or no evidence for a redshift dependence of linear size, up to more than $z = 2.1$. (This is a lower limit. The redshift $z = 2.1$ is the mean redshift of the high redshift half of their sample. They do not give enough information to determine the upper range.) They also state that they find no significant difference  between the mean bending angles of the high and low redshift halves of their sample.\\

-- Table 5 here --\\

In Table 5, we compare the physical properties of our high redshift sample to those of 3CRR and BM88. It is striking that the median linear sizes and the median bending angles are so similar for 3CRR (lower redshift) and BM88 (intermediate redshift) samples. At high redshifts $z > 2.5$, they are quite different. The linear sizes are smaller by a factor of about two and the bending angles are larger by a similar factor. \\

These two findings are not independent of each other because they both depend on orientation to the line of sight. The observed linear size is seen in projection against the sky (image) plane and is always seen foreshortened. Similarly, if the two arms (core to hotspot) are not perfectly aligned, then the misalignment is also exaggerated by projection. Indeed, a Spearman rank correlation test shows that in our sample the linear sizes and bending angles are anti-correlated at the 99\% level, but there is no significant correlation between linear size and redshift or bending angle and redshift. 

It may be difficult to unravel the roles played by orientation and cosmological density without a much larger sample. In particular, we are not able to explain why these effects become more prominent at redshifts greater than 2.5, but we note that we are comparing samples on either side of cosmic high-noon, when the environments are expected to be very different. \\

\section{Summary}
\label{s:conclude}

We have presented VLA snapshot observations at 6.2 GHz and 19.6 GHz of a complete sample of 113 high redshift ($ z \ge 2.5 $) quasars. This redshift range spans the era from one to two billion years after the Big Bang (or colloquially, from Cosmic Dawn to Cosmic High Noon) We show images of 48 sources that are significantly resolved, and list structural properties for all of them. We also compare our sample to two lower redshift complete samples: 3CRR \citep{LRL83} and Barthel and Miley \citep{BM88}. Among the sources that are well resolved in our sample (we show images of them in Figures 2-7), we see almost equal numbers of one-sided and two-sided sources, both of which can display prominent jets. It is also the case that many resolved sources have distorted or very asymmetric morphologies, indicating a turbulent or chaotic environment.\\

The distribution of morphological types for our sample is very different from what is seen in the lower redshift samples. We see a steady increase with redshift of the fraction of sources that are unresolved or nearly so (Table 4). But this is readily understood in terms of the increasing emission frequency at increasing redshift. This selects a higher proportion of compact flat spectrum sources due to the K-correction.  An unexpected result is the increase in the fraction of one-sided sources with redshift. In 3CRR there are none. In BM88 they constitute 14\% of the sample and in our high redshift sample they constitute 24\%.\\

A plot of the core spectral indices between 1.4 and 6.2 GHz plotted against those between 6.2 and 19.6 GHz shows that most core spectra steepen at high frequencies, probably due to radiative losses. Almost no cores exhibit the often assumed flat spectral index $\alpha = 0$. Care should therefore be taken when K correcting flux densities.

BM88 claimed that the largest linear size of quasars decreases with redshift and that the bending angle or distortion increases with redshift. Table 5 shows that both these effects are visible in our sample, at redshifts greater than 2.5, but essentially absent at lower redshifts (i.e. after Cosmic High Noon). \\

Finally, we stress that high redshift quasars are small enough that sub-arcsecond imaging is essential to discern their radio morphology correctly. The median angular size of the triple sources in our sample is 5.7\arcsec. The angular resolution of the venerable FIRST survey is about 5\arcsec. Nearly half of the well resolved sources in our high redshift sample would be classified as compact or unresolved by workers who used FIRST to determine radio structures.\\

\acknowledgments
\section{Acknowledgments}
\label{s:Acknowledgments}


This work has been supported by NSF grants 0607453 and 1009261. C.C.C. was supported at NRL by NASA DPR S-15633-Y.

The National Radio Astronomy Observatory is a facility of the National Science Foundation operated under cooperative agreement by Associated Universities, Inc.

This research has made use of the NASA/IPAC Extragalactic Database (NED) which is operated by the Jet Propulsion Laboratory, California Institute of Technology, under contract with the National Aeronautics and Space Administration. 

Funding for SDSS-III has been provided by the Alfred P. Sloan Foundation, the Participating Institutions, the National Science Foundation, and the U.S. Department of Energy Office of Science. The SDSS-III web site is http://www.sdss3.org/.

SDSS-III is managed by the Astrophysical Research Consortium for the Participating Institutions of the SDSS-III Collaboration including the University of Arizona, the Brazilian Participation Group, Brookhaven National Laboratory, Carnegie Mellon University, University of Florida, the French Participation Group, the German Participation Group, Harvard University, the Instituto de Astrofisica de Canarias, the Michigan State/Notre Dame/JINA Participation Group, Johns Hopkins University, Lawrence Berkeley National Laboratory, Max Planck Institute for Astrophysics, Max Planck Institute for Extraterrestrial Physics, New Mexico State University, New York University, Ohio State University, Pennsylvania State University, University of Portsmouth, Princeton University, the Spanish Participation Group, University of Tokyo, University of Utah, Vanderbilt University, University of Virginia, University of Washington, and Yale University. 

\appendix
\label{s:comments}
\section{Comments on Individual Sources}

This section is longer than the corresponding section in Paper 1 for several reasons. First, we have added VLA observations at 19.5 GHz (K-band). These were made in the B array so that the resolution would be similar to that of the 6.2 GHz (C-band) A array observations.

Also, there are now much more extensive X-ray and VLBI data available on most of these sources.  We are  particularly interested in \textit{radio jets} that exhibit detectable X-ray emission, as well as those that don't. For VLBI information, we quote the 5 GHz information where possible (usually the VIPS survey as this also followed the SDSS footprint \citep{HT07}). Otherwise we use information from the Astrogeo website (http://astrogeo.org/). 

There are sources shown in Paper 1 that exhibit extended structure at 1.4 GHz, but show only an unresolved core at 6.2 GHz. In these cases, the extended features are resolved or have steep spectra. In the present paper, for consistency, we always show the 6.2 GHz image, and use the corresponding morphological classification. Where it differs from the classification in Paper 1, this is indicated by an asterisk in the tables.\\

\indent \textit{J0730+4049:} (D) This is a very interesting source. In a \textit{Chandra} observation that targeted a nearby galaxy cluster, a 12 \arcsec-long, bright X-ray jet  was serendipitously detected in this $z=2.5$ quasar\citep{Sim16} . There is no detectable radio counterpart to the X-ray jet, other than the compact feature at a core distance of 1.4 arcsec. This places the jet in  the X-ray bright - radio faint category predicted by \citet{Sch02} for high redshift objects.

The projected length of the 12 arcsec X-ray jet is 100 kpc. There is also a "hot spot" in the \textit{Chandra} image at a core distance of 35 arcsec (280 kpc projected) that has no detectable radio or optical emission. The excellent alignment of this hot spot with the 12 arcsec X-ray jet strongly suggests that they are all part of the same physical jet.

The maximum orientation of the jet to the line of sight is constrained by VLBI monitoring observations made by \citet{BR08}. One of the mas-scale components exhibits superluminal motion of $\beta_{app} = 6.6 c$. The maximum angle that the jet makes to the line of sight is therefore $\theta_{max} < 2 \arctan (1/\beta_{app}) = 17.2^\circ$ \citep{CU84}. The minimum \textit{deprojected} lengths of the 12 arcsec jet and the 35 arcsec hot spot are therefore at least 348 kpc and 946 kpc respectively. This puts \textit{J0730+4049} firmly in the category of Giant Radio Quasars \citep{KJ21}  \\
\newline
\indent \textit{J0733+2536:} (T, jet)
The 19.5 GHz image of this source is very similar to the 6.2 GHz image shown here, indicating a rather uniform spectral index distribution over the source, including the three compact components near the map center. The component at (0,0) is the most compact and it is coincident  with the optical position. But it is interesting that it has a relatively steep and straight, optically thin spectral indices  of 0.79 between 1.4 GHz and 6.2 GHz, and 0.94 between 6.2 GHz and 19.6 GHz.

 The radio jet was one of two objects in our sample targeted in an early \textit{Chandra} survey of intermediate redshift ($z = 2-3$) quasars \citep[]{Mck16} and X-ray emission from the radio jet was clearly detected.\\
\newline 
\indent \textit{J0733+2721:} (C*) This source is an unresolved core source. Ten arcsec to the SE is a prominent but probably unrelated foreground source identified in SDSS as J073321.12+272054.4. NED lists a {\em g} = 22.7 magnitude for it but gives no other information. It has the appearance of a foreground radio galaxy (see the 1.4 GHz image in Paper I). Both sources are visible in the FIRST image cutout. At 6.2 GHz, only the core of the radio galaxy is visible, so we do not show it here. \\
\newline
\indent \textit{J0807+0432:} (T, jet) This source has a beautiful, curving, 4 arcsecond-long jet. All parts of the 6.2 GHz image are visible in the 19.5 GHz image.\\

\indent \textit{J0833+3531:} (T) In Paper 1, only a single low resolution 1.4 GHz image was available to us. At 6.2 GHz, in the A- array, it is clearly a triple source. The southern lobe bends through about $90^{\circ}$, giving the source a whimsical appearance. The bend is also visible in the 19.5 GHz image. There is no corresponding bend on the northern side, so it is unlikely to be caused by a changing jet orientation. 
\\

\indent \textit{J0905+4850:} (D) (The Paper 1 image at 4.8 GHz and the 6.2 GHz image shown here are, of course, very similar. What is remarkable is the absence of any detectable counter-feature on the northern side of the quasar. The ratio of the flux density of the southern component to the corresponding region on the northern side is $>17:1$ in the 6.2 GHz image shown here. In an archival 1.5 GHz image (not shown) it is $>250:1$.
VLBI observations (astrogeo.org) at 2.3 GHz show a 70 mas jet in PA $170^{\circ}$ that points nearly directly at the southern component, indicating that it is on the near side of the quasar. 
\\

\indent \textit{J0909+0354:} (D) This is another compact double source; the secondary component 2.3 arcsec to the NW in position angle $ -15^{\circ}$ is also detectable at 1.5 and 19.5 GHz. 
 An extended X-ray jet found in the \textit{Chandra} X-ray survey \citep{Sni21} is coincident with the 2.3 arcsec distant radio knot, and was confirmed in a deeper X-ray observation \citep{Per21}. They also detected X-ray emission at a distance of 6.5 arcsec to the north-east. \\ 


\indent \textit{J0918+5332:} (T, jet) This triple source exhibits a prominent curving jet with compact hot spots. It is not listed in the Astrogeo.org database, nor in the \textit{Chandra} X-ray Catalog.
\\
\newline
\indent \textit{J0934+4908:} (D) This very one sided double source has a stubby jet that points close to the direction of the hot spot. That jet is also visible at 19.5 GHz. 
VLBI monitoring by \citet{BR08} finds that one component exhibits a proper motion of $\beta_{app} = 13.9 \pm 3.1$. \\
\newline
\indent \textit{J0934+3050:} (T) This unusual looking source has emission emanating from both sides of the core. The quasar core is at (0,0) in the image shown, and a strong jet propagates to the north-west. But there is also jet-like emission on the opposite side of the core that terminates in a bright hot spot at a distance of 2 arcsec. All of these features are visible in the 19.6 GHz image.\\


\indent \textit{J0944+2554:} (D, jet) This source consists of a bright core and a bright bent jet (the most bent jet in the whole sample). There is no detectable emission on the counter-jet side, hence the D classification. The conventional interpretation of such a bent structure is that the jet is oriented at a small angle to the line of sight, so that any intrinsic bending is greatly exaggerated by projection effects. 
\\
\newline
\indent \textit{J0947+6328:} (T, jet) The only image available to us for Paper I was from 1.4 GHz B-array observations. The radio source appeared to be a triple, but grossly under resolved. The A-array 6.2 GHz image presented here reveals a beautiful triple source with a slender jet, traced out by the brighter knots, pointing to the southwest hot-spot. The northeast hot-spot appears to be double. The bright, compact component at (0,0) is the core and is coincident with the optical ID. 
\\


\indent \textit{J1007+1356:} (T, jet) This source has an unusual Z-shaped morphology with a prominent jet which does not point directly towards either lobe.  There are extensive VLBI observations of this source to be found on Astrogeo.org. \citet{Sok11} made images at nine frequencies from 1.4 GHz to 19.5 GHz, in order to measure the core shift \citep{Lob98}.
\\
\newline
\indent \textit{J1026+2542:} (E*) At $z = 5.267$ this is the highest redshift quasar in our sample. It was classified as compact in Paper I. At 6.2 GHz, it is clearly extended towards the west, hence the E* classification. The VIPS survey found a 21 mas jet pointing in the same direction (PA $-78^{\circ}$).\\

\indent \textit{J1049+1332:} (T) This is a small, asymmetric triple source. In Paper I, the 1.4 GHz image showed another radio source about 20 arcsec due south of the core. Our 6.2 GHz observations show this southern component itself to be a small double source. It is most likely a foreground radio galaxy and unrelated to the quasar. 
\\
\newline 
\indent \textit{J1050+3430:} (T*) This very small, core dominated and unusual looking source was classified as compact (C) in Paper I. At 6.5 GHz, there is clearly emission on both sides of the core, which is also visible at 19.5 GHz, changing its classification from C to T*. Astrogeo has many images at both S- and X-band. 
\\
\newline 
\indent \textit{J1057+0324:} (T) A collinear triple source. Its central component is clearly resolved. 
\\ 

\indent \textit{J1204+5228:} (T) This is an interesting object. Some authors have classified it as a high redshift radio galaxy, but the SDSS spectrum shows broad lines of C III, C IV and Ly $\alpha$. \citet{P00} suggest that the brightest component, seen at the center of our image, is a hot spot. However, the SDSS position is coincident with this component, as is also the VLBI core observed by \citet{HT07} in the VIPS survey. They see a 15 mas jet pointing north in position angle $-10^{\circ}$. We classify this source as a very asymmetric triple source with the northern (and fainter) hot spot close to the core. The diffuse flux to the SW in the 5 GHz image has no easy explanation. It is plainly visible at 1.4, 4.7 (Paper I) and 6.2 GHz but not at 8, 15 or 19.5 GHz, and probably has a steep spectrum. The SE hot spot is clearly visible at 19.5 GHz. 
\\
\newline 
\indent \textit{J1213+3247:} (T) This is a very core-dominated asymmetric triple source. The map center (0,0) is slightly displaced from the peak due to a scheduling error. The peak coincides with the SDSS and WISEA positions, and the position listed in Table 1 is correct. 
\\ 
\newline
\indent \textit{J1217+5835:} (T) We have classified this unusual source as a highly bent triple, even though all the extended structure is seen on only one side of the core. The source appears to consist of a core with a south pointing jet. There is a third component to the southeast which may be a counter-hotspot. There is no sign of emission connecting it to either the jet or the core. This is one of the most core-dominated sources in the sample, indicating  a small angle to the line of sight and consistent with exaggerated projection effects. \\ 
\newline
\indent \textit{J1217+3435:} (T, jet) For Paper I, the only available image (at 1.4 GHz) was of insufficient resolution to identify which emission peak was the core. The 6.2 GHz image shown here places the optical ID at (0,0) which also coincides with the brightest emission peak. The 19.5 GHz image shows exactly the same morphology with the core more prominent due to a flatter spectral index. \\

\indent \textit{J1356+2918:} (T) This unusual triple source shows a short jet pointing to the northeast at 6.2 and 19.5 GHz. In both images, the jet then turns to the west without terminating in a hot-spot. (The 1.4 GHz image shown in Paper I is badly under resolved, and should be disregarded.) 
    \\
\newline 
\indent \textit{J1400+0425:} (T, jet) Our image at 6.2 GHz shows a long, sinuous jet stretching northwest of the core, and terminating in a hot-spot. The contrast in arm lengths on the two sides of the core is striking and may suggest a very asymmetric environment. \\ 

\indent \textit{J1405+0415:} (D) This is a core-dominated double source. The core at (0,0) is coincident with the optical quasar. The emission to the southwest of the core is presumably a jet that leads to a hot-spot at a distance of 3.3 arcsec.
This source has been extensively observed with VLBI. The time evolution of the 15 GHz structure can be seen on the MOJAVE site (https://www.cv.nrao.edu/MOJAVE/). Based on five moving features, \citet{Lis19} find superluminal motion up to $9.2 \pm 1.8 c$. \\ 
The initial evidence for a faint X-ray jet in the distance range of 1''-3'' from the quasar core \citep[]{Sch20} was confirmed in deeper \textit{Chandra} observations \citep{Sni22}. The X-ray extension is in the direction of the inner $\sim$1 arcsec radio jet.\\
\newline 
\indent \textit{J1429+2607:} (T) This is a very asymmetric, but collinear triple source. All the features of the 6.2 GHz image are also visible at 19.5 GHz. There is a small extension to the core pointing towards the SE component (which is in the lower left corner of the image, and should not to be confused with the restoring beam, which is the solid black ellipse). That this is a jet is confirmed by VLBI observations \citep{HT07} that show a fine 50 mas long jet pointing in the same direction (position angle $120^{\circ}$). 
\\


\indent \textit{J1430+4204:} (D) At 6.2 GHz, this is a very core-dominated double source with a separation of 3.6 arcsec in PA $-70^{\circ}$. An X-ray jet in this $z=4.72$ quasar was discovered \citep{Che12} as part of a snapshot imaging program of high-redshift flat-spectrum radio quasars with radio jets \citep[see also][]{Mck16}. The weaker component of this double is coincident with the extended X-ray emission and is most likely a jet knot. \\ 

\indent \textit{J1450+0910:} (D*) At 6.2 GHz, we see a bright core with a 1.5 arcsec jet pointing south. The jet then appears to bend sharply to the west, terminating in a hotspot.  The 1.4 GHz PaperI image indicates there is also diffuse structure to the north-east, that is completely resolved at 6.2 GHz. The core is very self absorbed at 1.4 GHz ($\alpha_{1.4}^{4.9} < -0.9$). 
\\

\indent \textit{J1459+3253:} (T) This source has an intriguing morphology. It is clearly a triple (T) source with extended emission on both sides of the core. On the north side there is  an extension from the core which is likely the base of a jet that extends to a hot-spot at a distance of 4 arcsec. 
On the south side of the core, there is a hot spot at about one arcsec south of the core, and diffuse emission extending to three arcsec further south. 
Note the additional diffuse emission in the south east and in the north west. These may be relics of an earlier phase in the life of the quasar, and that the axis of the central quasar has turned through some $20^{\circ}$, possibly the result of a galaxy merger.\\

\indent \textit{J1502+5521:} (T) The true nature of this compact triple source is unclear. In Paper I, the very elongated beam made it difficult to see detail. At 6.2 GHz, there is a not very prominent core component, and a probable hot spot to the north west. On the south east side, there is a smooth tongue of emission but no discernible hot spot. At 19.5 GHz we see the same extended features, but with the core being more prominent.\\ 
\newline
\indent \textit{J1528+5310:} (T) This is an unusual source because the optical positions found in the literature (SDSS and WISEA) agree with each other but not with the location of either emission peak in the 6.2 GHz radio image. The 19.6 GHz radio image is in complete agreement with that at the lower frequency. Further, both radio components have steep spectral indices ($\alpha \sim 0.85$), and there is no compact radio component between them  at the optical position. For clarity, the image we show includes the optical position.
J1528+5310 is a broad absorption line (BAL) quasar \citep{Trump06}. The faintness of the radio core could be due to absorption, or perhaps a larger angle to the line of sight.\\
\newline 
\indent \textit{J1540+4738:} (T,jet) The paper I image made at 1.4 GHz showed a fat jet, unresolved in the transverse direction. Our 6.2 GHz image shows an elegant, curving 6 arcsec jet that terminates in a bright hot spot. The "counter hot spot" is visible but much fainter. The same structure is visible at 19.5 GHz. 
\\
 

\indent \textit{J1602+2410:}(T, jet) This triple source has a prominent 5 arcsec jet with four bright knots. An unusual feature is that the lobe/hot spot on the side with the jet is much fainter than the lobe/hot spot on the opposite side. 
The core appears to exhibit a rather steep spectrum ($\alpha = 0.85$) between 1.4 and 6.2 GHz, but this is probably not reliable due to the lack of a high resolution image at 1.4 GHz.
\\
\newline 
\indent \textit{J1610+1811:} (D) At first sight, this is another one-sided double source, with a separation of 5 arcsec in PA $-45^{\circ}$.
 Initial \textit{Chandra} observations \citep[]{Sch20} found an excess of X-rays (1.6 arcsec to 6.0 arcsec from the core) in the direction of the terminal radio feature seen in VLA maps. This emission was confirmed in a deeper \textit{Chandra} observation by \citet{Sni22}. A deeper VLA observation by \citet{Mai25} shows that this source is in fact two sided. There is radio emission to the southeast of the core that is coincident with the single contour seen in our image.\\
\newline 
\indent \textit{J1612+2758:} (T) At 6.2 GHz, we see a small triple source. Oddly, the NW component is not visible in an 8 GHz image. There is a paucity of radio observations of this source. 
\\


\indent \textit{J1655+3242:} (D, jet) This double source features a prominent 3 arcsecond jet with a strong terminal lobe and hot spot. 
The paper I image at 4.86 GHz shows an absence of any emission on the counter-jet side of the core. We see the same at 6.2 GHz (the single contour of emission is probably not significant). No data at other radio frequencies are available. 
\\


\section{Visualizing 3C~273 at Higher Redshift}

Jets in AGN are customarily defined as an observed feature (or series of knots) well separated from the core, with an overall length at least four times their width \citep[e.g.,][]{Bridle84}. 
However, the extended radio emission imaged in a number of the systems in this study, particularly at the highest redshifts, consist of only a single, isolated arcsecond-scale feature.
This observational limitation at high redshifts is well known, and due to three dominant effects which we demonstrate in the following.

Figure~A1 shows the powerful, nearby ($z = 0.158$) radio jet in 3C~273 \citep{Conway93,Perley17} as it would appear at a higher redshift ($z=3.6$ in this visualization) accounting for the three observational effects. 
The data were taken from an archival VLA 1.5 GHz observation obtained on 1985 January 31 (program AC120; PI: R.G. Conway). 
In the 1.5'' resolution map (panel a), the well-known bright, one-sided jet is seen out to $\sim$23'' in the SW direction, surrounded by some diffuse lobe emission (the feature to the NW is an artifact). 

The successive images then show the effects of changing the angular scale at the higher redshift (panel b), cosmological dimming (panel c), and finally after $K$-correction (panel d). 
These estimates are based on adopting a cosmology with, $H_{\rm 0} = 70\,{\rm~km\,s}^{-1}\,{\rm~Mpc}^{-1}$, $\Omega_{\rm m}=0.3$, and $\Omega_{\rm vac}=0.7$, resulting in the parameters: (physical scale, $D_{\rm A}$, $D_{\rm L}$) = (2.729 kpc/'', 563.0 Mpc, 754.9 Mpc) for $z=0.158$ and (7.245 kpc/'', 1494.4 Mpc, 31620.5 Mpc) for $z=3.6$.
Thus in panel (b), the $z=3.4$ image is smaller by 1/2.655 in both RA and Dec by taking the ratio of the physical scales; this is the same as the ratio of the angular size distances.
Recalling that surface brightness decreases as $(1+z)^{-4}$, the map in panel (c) is decreased by a factor [(1+3.6)/(1+0.158)]$^{4}$ = 249.
In the final panel (d), the $K$-correction assumes $\alpha_{\rm core} = 0$ and $\alpha_{\rm extended} = 1.0$ \citep[see][]{Perley17}, with the net effect being that the core flux is unchanged (flat spectrum) and the jet is diminished by a further factor of $(\nu_{\rm 2}/\nu_{\rm 1})^{-1.0} = [(1+z_{2})/(1+z_{1})]^{-1.0}$ = 0.25.


\newpage
\clearpage

\begin{table*}
\scriptsize
\begin{center}
\caption{Positions and Fluxes for the whole sample}
    \tabcolsep 4.5pt
\begin{tabular}{rrccccrrrrrrr}
\hline\hline
\multicolumn{1}{c}{IAU name} &
\multicolumn{1}{c}{$z$} &
\multicolumn{1}{c}{R.A.} &
\multicolumn{1}{c}{Decl.} &
\multicolumn{1}{c}{mag} &
\multicolumn{1}{c}{morph-} &
\multicolumn{1}{c}{$S_{\rm 1.4}^{\rm total}$} &
\multicolumn{1}{c}{$S_{\rm 1.4}^{\rm core}$} &
\multicolumn{1}{c}{$S_{\rm 6.2}^{\rm total}$} &
\multicolumn{1}{c}{$S_{\rm 6.2}^{\rm core}$} &
\multicolumn{1}{c}{$S_{\rm 19.6}^{\rm total}$} &
\multicolumn{1}{c}{$S_{\rm 19.6}^{\rm core}$} &
\multicolumn{1}{c}{VLBI/}  \\
\multicolumn{1}{c}{} &
\multicolumn{1}{c}{} &
\multicolumn{1}{c}{(J2000)} &
\multicolumn{1}{c}{(J2000)} &
\multicolumn{1}{c}{} & 
\multicolumn{1}{c}{ology} & 
\multicolumn{1}{c}{(mJy)} &
\multicolumn{1}{c}{(mJy)} &
\multicolumn{1}{c}{(mJy)} &
\multicolumn{1}{c}{(mJy)} &
\multicolumn{1}{c}{(mJy)} &
\multicolumn{1}{c}{(mJy)} &
\multicolumn{1}{c}{X-ray} \\
\hline
J0730+4049 & 2.501 	& 07 30 51.346 & +40 49 50.83 & 18.55 & D & 381 & 374 & 176 & 175 &  &   &  X\\ 
J0733+2536 & 2.689 	& 07 33 08.794 & +25 36 25.06 & 19.38 & T & 578 &  26 & 133 & 8.0 & 32.1 & 2.7 &  X  \\
J0733+2721 & 2.941 	& 07 33 20.491 & +27 21 03.53 & 19.50 & C* & 216 & 214 & 101 & 101 & 37.6 & 37.3 &   \\
J0746+2549 & 2.987 	& 07 46 25.874 & +25 49 02.13 & 19.43 & C & 428 & 428 & 322 & 322 &  &  & V\\  
J0752+5808 & 2.940 	& 07 52 09.679 & +58 08 52.26 & 19.62* & C & 152 & 152 & 90 & 90  & 50.0 & 49.6 & V\\ 
J0753+4231 & 3.595 	& 07 53 03.337 & +42 31 30.77 & 17.77 & C & 709 & 709 & 369 & 369 &  &  & V\\  
J0756+3714 & 2.515 	& 07 56 28.249 & +37 14 55.65 & 19.85 & C & 235 & 235 & 175 & 175 & 52.6 & 52.3 &  V\\ 
J0801+1153 & 2.677 	& 08 01 00.484 & +11 53 23.59 & 19.13 & T* & 117 & 53 & 42 & 25.4    &  &  & V\\ 
J0801+4725 & 3.218 	& 08 01 37.682 & +47 25 28.24 & 19.17 & D & 71 & 68 & 26.3 & 25.8 & 10.1 & 9.9 & X \\  
J0805+6144 & 3.033 	& 08 05 18.180 & +61 44 23.70 & 19.64* & D & 813 & 798 & 883 & 875 & - & 598 & VX\\  
J0807+0432 & 2.877 	& 08 07 57.538 & +04 32 34.53 & 17.78* & T & 530 & 326 & 396 & 358 & 212 & 202 & VX  \\
J0821+3107 & 2.613 	& 08 21 07.616 & +31 07 51.17 & 16.87 & C & 91 & 91 & 62 & 62 & 38.7 & 38.2 & V \\
J0824+2341 & 2.611 	& 08 24 28.022 & +23 41 07.95 & 18.75 & C* & 146 & 145 & 87 & 87 &  &  &  V\\  
J0833+3531 & 2.710 	& 08 33 01.658 & +35 31 33.68 & 19.96 & T & 75 & 22 & 28.3 & 14.8 & 17.6 & 13.5 & \\ 
J0833+1123 & 2.981 	& 08 33 14.367 & +11 23 36.23 & 18.30 & D & 392 & 361 & 260 & 249 &  &  & V\\ 
J0833+0959 & 3.715 	& 08 33 22.514 & +09 59 41.14 & 18.70 & C* & 122 & 121 & 81 & 81 & 53.2 & 52.8 & VX\\  
J0847+3831 & 3.186 	& 08 47 15.169 & +38 31 09.98 & 18.31 & C & 142 & 142 & 124 & 124 &  &  & V\\  
J0905+4850 & 2.686 	& 09 05 27.464 & +48 50 49.97 & 17.39 & D & 646 & 505 & 378 & 340 &  &  & V\\ 
J0905+3555 & 2.824 	& 09 05 36.065 & +35 55 51.68 & 18.27 & D* & 87 & 87 & 48.3 & 45.7 &  &  & V\\  
J0909+0354 & 3.288 	& 09 09 15.915 & +03 54 42.98 & 19.73 & D & 132 & 126 & 123 & 120 & 106 & 104 & VX\\ 
J0910+2539 & 2.752 	& 09 10 55.237 & +25 39 21.50 & 18.38 & D* & 179 & 179 & 81 & 80 &  &  &  V\\ 
J0915+0007 & 3.056 	& 09 15 51.696 & +00 07 13.30 & 20.54 & E* & 399 & 386 & 248 & 246 & 166 & 162 &  V\\  
J0918+5332 & 3.011 	& 09 18 57.676 & +53 32 20.02 & 19.68 & T & 90 & 20 & 29.7 & 13.2 &  &  & \\ 
J0933+2845 & 3.431 	& 09 33 37.298 & +28 45 32.24 & 17.83 & E & 121 & 121 & 56 & 55 &  &  & VX\\  
J0934+4908 & 2.586 	& 09 34 15.762 & +49 08 21.74 & 19.13 & D & 802 & 785 & 345 & 343 & 186 & 184 & V\\ 
J0934+3050 & 2.895 	& 09 34 47.245 & +30 50 56.04 & 20.35 & T & 291 & 88 & 114 & 57 & 52.4 & 25.7 &   \\
J0935+3633 & 2.859 	& 09 35 31.842 & +36 33 17.55 & 18.39 & C & 291 & 291 & 240 & 240 &  &  & VX \\  
J0941+1145 & 3.194 	& 09 41 13.558 & +11 45 32.33 & 19.34 & T & 218 & 210 & 124 & 123 &  &  & V\\  
J0944+2554 & 2.904 	& 09 44 42.320 & +25 54 43.31 & 18.38 & D & 827 & 586 & 264 & 205 &  &  & V\\  
J0947+6328 & 2.606 	& 09 47 59.416 & +63 28 03.11 & 19.04 & T & 78 & 33 & 22.7 & 15.1 & 11.2 & 8.9 &  \\  
J0958+3922 & 3.065 	& 09 58 56.780 & +39 22 32.40 & 21.25 & T & 160 & 46 & 30.5 & 15.0 &  &  & \\  
J1007+1356 & 2.715 	& 10 07 41.498 & +13 56 29.60 & 18.47 & T & 914 & 818 & 693 & 675 &  &  & V\\  
J1013+3526 & 2.638 	& 10 13 02.299 & +35 26 05.70 & 17.81 & C & 200 & 200 & 161 & 161 & 72.0 & 71.5 &  V\\  
J1016+2037 & 3.114 	& 10 16 44.322 & +20 37 47.31 & 19.11 & D & 730 & 695 & 469 & 467 &  &  & VX\\  
J1017+6116 & 2.801 	& 10 17 25.887 & +61 16 27.50 & 18.10 & C & 474 & 474 & 658 & 658 &  &  & V\\  
J1026+2542 & 5.266 	& 10 26 23.621 & +25 42 59.40 & 19.96 & E* & 239 & 239 & 103 & 103 &  &  &  VX\\  
J1026+3658 & 3.252 	& 10 26 45.340 & +36 58 25.60 & 19.43 & C & 195 & 195 & 163 & 163 &  &  & \\  
J1036+1326 & 3.097 	& 10 36 26.888 & +13 26 51.77 & 17.78 & T & 100 & 56 & 42.4 & 33.4 & 16.3 & 15.2 & \\  
J1042+0748 & 2.661 	& 10 42 57.589 & +07 48 50.55 & 17.31 & C & 457 & 457 & 163 & 163 &  &  & \\  
J1045+3142 & 3.236 	& 10 45 23.481 & +31 42 31.65 & 18.73 & C & 132 & 132 & 67 & 67 &  &  & V\\  
J1049+1332 & 2.764 	& 10 49 41.098 & +13 32 55.68 & 18.74 & T & 110 & 87 & 52 &  46 &  &  & \\
J1050+3430 & 2.595 	& 10 50 58.123 & +34 30 10.94 & 20.60 & T* & 583 & 583 & 252 & 250 & 103 & 102 & V\\  
J1057+0324 & 2.832 	& 10 57 26.622 & +03 24 48.09 & 19.32 & T & 152 & 88 & 45.1 & 36.1 &  &  & V\\ 
J1101+0010 & 3.694 	& 11 01 47.890 & +00 10 39.44 & 20.14 & E* & 188 & 188 & 58 & 54 & 18.2 & 12.7 & \\  
J1103+0232 & 2.517 	& 11 03 44.540 & +02 32 9.92 & 18.33 & C & 160 & 160 & 76 & 76 & 27.6 & 27.6 & V\\  
\hline\hline
\end{tabular}
\end{center}
\end{table*}

\addtocounter{table}{-1}
\begin{table*}
\scriptsize
\begin{center}
\caption{Positions and Fluxes ({\it continued})}
    \tabcolsep 4.5pt
\begin{tabular}{rrccccrrrrrrr}
\hline\hline
\multicolumn{1}{c}{IAU name} &
\multicolumn{1}{c}{$z$} &
\multicolumn{1}{c}{R.A.} &
\multicolumn{1}{c}{Decl.} &
\multicolumn{1}{c}{mag} &
\multicolumn{1}{c}{morph-} &
\multicolumn{1}{c}{$S_{\rm 1.4}^{\rm total}$} &
\multicolumn{1}{c}{$S_{\rm 1.4}^{\rm core}$} &
\multicolumn{1}{c}{$S_{\rm 6.2}^{\rm total}$} &
\multicolumn{1}{c}{$S_{\rm 6.2}^{\rm core}$} &
\multicolumn{1}{c}{$S_{\rm 19.6}^{\rm total}$} &
\multicolumn{1}{c}{$S_{\rm 19.6}^{\rm core}$} &
\multicolumn{1}{c}{VLBI/}  \\
\multicolumn{1}{c}{} &
\multicolumn{1}{c}{} &
\multicolumn{1}{c}{(J2000)} &
\multicolumn{1}{c}{(J2000)} &
\multicolumn{1}{c}{} & 
\multicolumn{1}{c}{ology} & 
\multicolumn{1}{c}{(mJy)} &
\multicolumn{1}{c}{(mJy)} &
\multicolumn{1}{c}{(mJy)} &
\multicolumn{1}{c}{(mJy)} &
\multicolumn{1}{c}{(mJy)} &
\multicolumn{1}{c}{(mJy)} &
\multicolumn{1}{c}{X-ray} \\
\hline

J1119+6004 & 2.641  & 11 19 14.344 & +60 04 57.19 & 17.06 & C & 109 & 109 & 92 & 92 & 26.2 & 26.2 & V\\ 
J1127+5650 & 2.893  & 11 27 40.135 & +56 50 14.79 & 19.53 & E* & 529 & 510 & 243 & 243 &  &  &  VX \\ 
J1128+2326 & 3.041 	& 11 28 51.697 & +23 26 17.35 & 18.36 & E & 141 & 130 & 95 & 93 &  &  & VX\\  
J1137+2935 & 2.644 	& 11 37 20.880 & +29 35 39.10 & 19.54 & C & 76 & 76 & 69 & 69 &  &  & \\  
J1150+4332 & 3.037 	& 11 50 16.603 & +43 32 05.91 & 19.69 & C & 160 & 160 & 97 & 97 &  &  &  V\\
J1151+4008 & 2.749 	& 11 51 16.927 & +40 08 22.21 & 19.36 & C & 83 & 83 & 92 & 92 & 59.6 & 58.7 & V\\  
J1204+5228 & 2.729 	& 12 04 36.799 & +52 28 41.78 & 18.36 & T & 940 & 632 & 151 & 127 & 100 & 91 &   V\\  
J1213+3247 & 2.516 	& 12 13 03.804 & +32 47 36.74 & 18.90 & T & 129 & 116 & 62 & 58 &  &  &    \\  
J1217+5835 & 2.553 	& 12 17 11.019 & +58 35 26.25 & 19.09 & D & 412 & 412 & 623 & 620 &  &  & VX  \\ 
J1217+3435 & 2.651 	& 12 17 15.160 & +34 35 38.20 & 18.44 & T & 181 & 101 & 90 & 78 & 58.3 & 54.6 &   V\\  
J1217+3305 & 2.609 	& 12 17 32.540 & +33 05 38.12 & 18.37 & D & 208 & 200 & 64 & 62 &  &  &   \\ 
J1223+5037 & 3.491 	& 12 23 43.169 & +50 37 53.40 & 17.41 & E & 178 & 171 & 201 & 199 &  &  &  VX\\  
J1242+3720 & 3.827 	& 12 42 09.812 & +37 20 05.69 & 19.39 & D & 816 & 799 & C* & 752 & 407 & 406 & VX\\  
J1246+0104 & 2.511 	& 12 46 23.732 & +01 04 02.43 & 19.54 & D & 117 & 3  & 33 & 28 &  &  & \\ 
J1301+1905 & 3.069 	& 13 01 21.019 & +19 04 21.39 & 18.18 & C & 146 & 146 & 96 & 96 & 39.6 & 39.4 & \\  
J1310+6229 & 2.614 	& 13 10 46.559 & +62 29 09.05 & 20.02 & C & 292 & 292 & 106 & 106 &  &  & V\\  
J1319+6217 & 3.073 	& 13 19 07.484 & +62 17 21.34 & 19.27 & C & 214 & 214 & 98 & 98 &  &  &   V\\
J1322+3912 & 2.992 	& 13 22 55.664 & +39 12 07.95 & 17.58 & C & 124 & 124 & 134 & 134 &  &  &  V\\  
J1325+1123 & 4.415 	& 13 25 12.494 & +11 23 29.84 & 19.17 & C & 74 & 74 & 49 & 49 &  &  & X\\
J1329+5009 & 2.654 	& 13 29 05.802 & +50 09 26.40 & 20.17 & C & 164 & 164 & 248 & 248 & 199 & 198 &  V\\  
J1337+3152 & 3.182 	& 13 37 24.698 & +31 52 54.59 & 18.47 & C & 84 & 84 & 31 & 31 & 12.7 & 12.5 & \\  
J1339+6328 & 2.562 	& 13 39 23.783 & +63 28 58.42 & 19.25 & C & 480 & 480 & 261 & 261 &  & &  V\\
J1340+3754 & 3.110 	& 13 40 22.952 & +37 54 43.83 & 18.54 & C & 276 & 276 & 195 & 195 &  & &  V\\
J1342+5110 & 2.599 	& 13 42 24.314 & +51 10 12.43 & 19.20 & C & 154 & 154 & 139 & 139 & 107 & 105 & V\\ 
J1346+2900 & 2.721 	& 13 46 37.435 & +29 00 42.41 & 20.92 & D & 96 & 90 & 36 & 36 &  &  &    V\\
J1353+5725 & 3.477 	& 13 53 26.036 & +57 25 52.96 & 19.31 & T & 319 & 207 & 162 & 134 &  &  &   V\\ 
J1356+2918 & 3.243 	& 13 56 52.542 & +29 18 18.75 & 20.36 & T & 125 & 90 & 51 & 44 & 21.3 & 18.8 &   V\\  
J1400+0425 & 2.550 	& 14 00 48.443 & +04 25 30.87 & 20.32 & T & 281 & 170 & 189 & 154 &  &  &   V\\  
J1404+0728 & 2.885 	& 14 04 32.992 & +07 28 46.96 & 18.83 & C & 158 & 158 & 183 & 183 &  &   &  V\\  
J1405+0415 & 3.208 	& 14 05 01.120 & +04 15 35.82 & 19.36 & D & 729 & 580 & 735 & 708 &  &  &   VX  \\  
J1406+3433 & 2.563 	& 14 06 53.847 & +34 33 37.31 & 18.49 & C & 169 & 169 & 271 & 271 &  &  &  V\\  
J1413+4505 & 3.118 	& 14 13 18.864 & +45 05 23.01 & 19.09 & C & 141 & 141 & 139 & 139 &  &  &  V\\  
J1429+5406 & 2.990 	& 14 29 21.879 & +54 06 11.12 & 19.81 & C* & 1036 & 916 & 497 & 497 &  &  &  V\\ 
J1429+2607 & 2.916 	& 14 29 50.916 & +26 07 50.32 & 18.13 & T & 415 & 371 & 246 & 235 & 122 & 119 &  V\\  
J1430+4204 & 4.715 	& 14 30 23.742 & +42 04 36.49 & 19.14 & D & 153 & 151 & 278 & 277 & 165 & 165 &  VX \\  
J1435+5435 & 3.809 	& 14 35 33.779 & +54 35 59.31 & 19.95 & C* & 96 & 83 & 63 & 63 & 40.9 & 39.3  &  VX \\ 
J1445+0958 & 3.552 	& 14 45 16.465 & +09 58 36.07 & 17.88 & C & 2614 & 2614 & 814 & 814 &  &  &   V\\  
J1450+0910 & 2.628 	& 14 50 31.169 & +09 10 27.96 & 19.44 & D* & 157 & 148 & 339 & 337 &  &  &  V\\ 
J1454+5003 & 2.849 	& 14 54 08.322 & +50 03 30.99 & 19.32 & C & 830 & 830 & 260 & 260 & 126 & 120 &  V\\ 
J1455+4431 & 2.689 	& 14 55 54.133 & +44 31 37.65 & 19.49 & C & 288 & 288 & 187 & 187 & 150 & 149 &     V\\ 
J1457+3439 & 2.734 	& 14 57 57.303 & +34 39 50.39 & 19.50 & D & 227 & 184 & 96 & 89 &  &  &   V\\  
J1459+3253 & 3.332 	& 14 59 27.042 & +32 53 57.91 & 19.43 & T & 119 & 53 & 31 & 22 &  &  &    \\  
J1459+4442 & 3.401 	& 14 59 35.458 & +44 42 07.92 & 19.99 & C & 109 & 109 & 182 & 182 &  &  &   V\\  
J1502+5521 & 3.329 	& 15 02 06.524 & +55 21 46.64 & 19.80 & T & 83 & 21 & 28 & 16 & 12.3 & 7.3 &    \\
J1503+0419 & 3.667 	& 15 03 28.888 & +04 19 48.99 & 18.20 & C & 137 & 137 & 205 & 205 &  &   &   V\\
J1510+5702 & 4.313 	& 15 10 02.922 & +57 02 43.38 & 19.89 & C* & 278 & 273 & 121 & 121 & 94 & 94 &  VX  \\  
\hline\hline
\end{tabular}
\end{center}
\end{table*}

\addtocounter{table}{-1}
\begin{table*}
\scriptsize
\begin{center}
\caption{Positions and Fluxes ({\it continued})}
    \tabcolsep 4.5pt
\begin{tabular}{rrccccrrrrrrr}
\hline\hline
\multicolumn{1}{c}{IAU name} &
\multicolumn{1}{c}{$z$} &
\multicolumn{1}{c}{R.A.} &
\multicolumn{1}{c}{Decl.} &
\multicolumn{1}{c}{mag} &
\multicolumn{1}{c}{morph-} &
\multicolumn{1}{c}{$S_{\rm 1.4}^{\rm total}$} &
\multicolumn{1}{c}{$S_{\rm 1.4}^{\rm core}$} &
\multicolumn{1}{c}{$S_{\rm 6.2}^{\rm total}$} &
\multicolumn{1}{c}{$S_{\rm 6.2}^{\rm core}$} &
\multicolumn{1}{c}{$S_{\rm 19.6}^{\rm total}$} &
\multicolumn{1}{c}{$S_{\rm 19.6}^{\rm core}$} &
\multicolumn{1}{c}{VLBI/}  \\
\multicolumn{1}{c}{} &
\multicolumn{1}{c}{} &
\multicolumn{1}{c}{(J2000)} &
\multicolumn{1}{c}{(J2000)} &
\multicolumn{1}{c}{} & 
\multicolumn{1}{c}{ology} & 
\multicolumn{1}{c}{(mJy)} &
\multicolumn{1}{c}{(mJy)} &
\multicolumn{1}{c}{(mJy)} &
\multicolumn{1}{c}{(mJy)} &
\multicolumn{1}{c}{(mJy)} &
\multicolumn{1}{c}{(mJy)} &
\multicolumn{1}{c}{X-ray} \\
\hline

J1520+4732 & 2.815 	& 15 20 43.602 & +47 32 49.14 & 18.56 & C & 94 & 94 & 56 & 56 & 25.2 & 24.9 &   V\\  
J1521+1756 & 3.053 	& 15 21 17.583 & +17 56 01.09 & 19.27 & C & 181 & 181 & - & - &  &  &    V\\  
J1528+5310 & 2.825 	& 15 28 21.659 & +53 10 30.48 & 19.79 & T & 178 & 163 & 51 & 47 & 18.5 & 0.5 &    \\  
J1535+4836 & 2.561 	& 15 35 14.653 & +48 36 59.70 & 17.81 & C & 108 & 108 & 134 & 134 & 78.5 & 78.2 &  V\\  
J1540+4738 & 2.563 	& 15 40 58.710 & +47 38 27.60 & 18.97 & T & 221 & 36 & 58 & 11.0 & 16.6 & 3.2 &   \\  
J1541+5348 & 2.539 	& 15 41 25.461 & +53 48 13.03 & 18.92 & C & 253 & 253 & 212 & 212 &  &  &  V\\  
J1547+4208 & 2.739 	& 15 47 59.043 & +42 08 55.57 & 19.73 & C & 70 & 70 & 92 & 92 &  &  &   V\\  
J1559+0304 & 3.891 	& 15 59 30.973 & +03 04 48.26 & 19.77 & C* & 384 & 377 & 382 & 382 &  &  &    V\\  
J1602+2410 & 2.526 	& 16 02 12.601 & +24 10 11.06 & 18.58 & T & 341 & 18 & 86 & 5.1 & 21.7 & 2.7 &   \\  
J1603+5730 & 2.850 	& 16 03 55.930 & +57 30 54.41 & 17.26 & C & 362 & 362 & 301 & 301 &  &  &    V\\  
J1610+1811 & 3.122 	& 16 10 05.289 & +18 11 43.47 & 18.25 & D & 258 & 223 & 77 & 67 &  &  &  VX \\  
J1612+2758 & 3.544 	& 16 12 53.417 & +27 58 42.57 & 19.60 & T & 70 & 49 & 37 & 34 &  &  &    \\  
J1616+0459 & 3.215 	& 16 16 37.557 & +04 59 32.74 & 19.22 & E* & 416 & 347 & 1091 & 1084 &  &  &  VX \\  
J1625+4134 & 2.550 	& 16 25 57.670 & +41 34 40.63 & 21.58 & E* & 1715 & 1661 & 795 & 781 &  &  &   V\\  
J1632+2643 & 2.683 	& 16 32 21.051 & +26 43 53.46 & 17.64 & C & 267 & 267 & 130 & 130 &  &  &    V\\ 
J1655+3242 & 3.181 	& 16 55 19.225 & +32 42 41.13 & 19.45 & D & 180 & 14 & 55 & 16 &  &  &  X  \\  
J1655+1948 & 3.262   	& 16 55 43.568 & +19 48 47.12 & 19.85 & E & 194 & 188 & 159 & 158 &  &  &   VX\\  
J1656+1826 & 2.546 	& 16 56 34.089 & +18 26 26.35 & 19.99 & C & 221 & 221 & 238 & 238 &  &  &   V\\  
J1704+0134 & 2.842 	& 17 04 07.489 & +01 34 08.47 & 19.62* & C? & 293 & 290 & 216 & 216 &  &  &   V\\  
J1707+0148 & 2.568 	& 17 07 34.415 & +01 48 45.70 & 18.52* & C & 640 & 640 & 894 & 894 &  &  &   V\\  
J1707+1846 & 2.518 	& 17 07 53.748 & +18 46 39.02 & 19.36 & C* & 334 & 332 & 190 & 190 &  &  &   V\\  
J1715+2145 & 4.011 	& 17 15 21.250 & +21 45 31.83 & 21.48 & E* & 387 & 383 & 126? & 113 &  &  & VX     \\  
\hline\hline
\end{tabular}
\end{center}
Magnitudes are SDSS $i$-band unless noted with asterisks*, in which case, they are $R2$-band magnitudes from the USNO-B1.0 \citep{MLC03}.\\

\end{table*}


\begin{table*}
\footnotesize
\begin{center}
\caption{6.2 GHz Image Properties}
\begin{tabular}{cccccccccc}
\hline\hline
\multicolumn{1}{c}{Source} &
\multicolumn{1}{c}{z} &
\multicolumn{1}{c}{Morph-} &
\multicolumn{1}{c}{First Contour} &
\multicolumn{1}{c}{Map Peak} &
\multicolumn{1}{c}{$\theta_B$} &
\multicolumn{1}{c}{$\theta_F$} &
\multicolumn{1}{c}{BA}  &    
\multicolumn{1}{c} {core} & 
\multicolumn{1}{c} {core}   \\
\multicolumn{1}{c}{} &
\multicolumn{1}{c}{} &
\multicolumn{1}{c}{ology} &
\multicolumn{1}{c}{$\%$ of peak} &
\multicolumn{1}{c}{(mJy)} &
\multicolumn{1}{c}{(arcsec)} &
\multicolumn{1}{c}{(arcsec)} &
\multicolumn{1}{c}{(degrees)}    &
\multicolumn{1}{c}  {$\alpha_{\rm 1.4}^{\rm 6.2}$}   &
\multicolumn{1}{c}  {$\alpha_{\rm 6.2}^{\rm 19.6}$}  \\

\hline
J0730+4049 & 2.501 & D & $0.10$ & 175 & 1.3 &  &  & 0.51 &  \\
J0733+2536 & 2.689 & T, jet & $1.3$ & 19 & 4.1 & 3.6 & 41 & 0.79 & 0.94 \\
J0801+1153 & 2.677 & T* & $0.68$ & 25 & 3.1 & 2.2 & 0 & 0.49 &  \\
J0801+4725 & 3.218 & D & $0.30$ & 26 & 2.5 & & & 0.65 & 0.83  \\
J0805+6144 & 3.033 & D & $0.09$ & 875 & 3.8 & & & -0.06 & 0.33\\
J0807+0432 & 2.877 & T, jet & $0.09$ & 356 & 3.9 & 2.7 & 21 & -0.06 & 0.50 \\
J0833+3531 & 2.710 & T & $0.10$ & 15 & 1.4 & 3.2 & 18 & 0.27 & 0.08 \\
J0833+1123 & 2.986 & D & $0.49$ & $249$ & $0.7$ &  & & 0.25 & \\ 
J0905+4850 & 2.686 & D & $0.10$ & 338 & 3.8 & & & 0.27 & \\
J0905+3555 & 2.824 & D* & $0.24$ & 46 & 0.9 & & & 0.43 & \\
J0909+0354 & 3.288 & D & $0.10$ & 120 & 2.3 & & & 0.03 & 0.12 \\
J0910+2539 & 2.752 & D* & $0.10$ & 80 & 0.9 & & & 0.54 & \\
J0918+5332 & 3.011 & T, jet & 0.72 & 13 & 3.4 & 5.8 & 12 &  0.28 & \\
J0934+4908 & 2.586 & D & 0.08 & 342 & 5.0 & & & 0.56 & 0.54 \\
J0934+3050 & 2.895 & T & 0.76 & 44 & 0.8 & 2.8 & 48 & 0.29 & 0.69 \\
J0941+1145 & 3.194 & T & 0.08 & 124  & 2.0 & 5.3 & 20 & 0.36 &   \\
J0944+2554 & 2.904 & D, jet & 0.12 & 197 & 1.5 & & & 0.71\\
J0947+6328 & 2.606 & T, jet & 0.48 & 15 & 8.5 & 7.8 & 24 & 0.53 & 0.46\\
J0958+3922 & 3.065 & T & 0.47 & 16 & 2.8 & 3.7 & 32 & 0.75\\
J1007+1356 & 2.715 & T, jet & 0.11 & 666 & 5.0 & 2.5 & 56 & 0.13 \\
J1016+2037 & 3.114 & D & 0.04 & 463 & 1.2 &  & & 0.27 \\
J1036+1326 & 3.097 & T & 0.25 & 33 & 8.5 & 7.5 & 17 & 0.35 & 0.68 \\
J1049+1332 & 2.764 & T & 0.20 & 45.9 & 5.9 & 1.8 & 35 & 0.43 \\
J1050+3430 & 2.595 & T* & 0.06 & 250 & 0.8 & 1.6 & 51 & 0.57 & 0.78\\
J1057+0325 & 2.832 & T & 0.48 & 31 & 2.2 & 2.3 & 0 & 0.6 \\
J1204+5228 & 2.729 & T & 0.19 & 112 & 0.6 & 4.1 & 40 & (1.08) & 0.29\\
J1213+3247 & 2.516 & T & 0.28 & 54 & 1.2 & 0.7 & 38 & 0.47\\
J1217+5835 & 2.551 & D & 0.04 & 619 & 1.0 & 1.6 & 129 & -0.27\\
J1217+3435 & 2.651 & T, jet & 0.31 & 48.5 & 5.6 & 2.0 & 2 & 0.17 & 0.31\\
J1217+3305 & 2.609 & D & 0.28 & 2.3 & 2.3 & & & 0.79\\
J1246+0104 & 2.510 & D & 0.48 & 21.9 & 5.5 & & & -1.50 \\
J1346+2900 & 2.721 & D, jet & 0.76 & 34.7 & 2.0 & & & 0.62 & \\
\hline\hline
\end{tabular}
\end{center}
\end{table*}

\addtocounter{table}{-1}
\begin{table*}
\footnotesize
\begin{center}
\caption{Image Properties ({\it continued})}
\begin{tabular}{cccccccccc}
\hline\hline
\multicolumn{1}{c}{Source} &
\multicolumn{1}{c}{z} &
\multicolumn{1}{c}{Morph-} &
\multicolumn{1}{c}{First Contour} &
\multicolumn{1}{c}{Map Peak} &
\multicolumn{1}{c}{$\theta_B$} &
\multicolumn{1}{c}{$\theta_F$} &
\multicolumn{1}{c}{BA}  &   
\multicolumn{1}{c}{$\alpha_{\rm 1.4}^{\rm 6.2}$} & 
\multicolumn{1}{c}{$\alpha_{\rm 6.2}^{\rm 19.6}$}  \\
\multicolumn{1}{c}{} &
\multicolumn{1}{c}{} &
\multicolumn{1}{c}{ology} &
\multicolumn{1}{c}{$\%$ of peak} &
\multicolumn{1}{c}{(mJy)} &
\multicolumn{1}{c}{(arcsec)} &
\multicolumn{1}{c}{(arcsec)} &
\multicolumn{1}{c}{(degrees)}    &
\multicolumn{1}{c}{}     &
\multicolumn{1}{c}{} \\
\hline
J1353+5725 & 3.477 & T & 0.43 & 129 & 7.9 & 4.2 & 11 & 0.29\\
J1356+2918 & 3.244 & T & 0.23 & 19.4 & 4.2 & 1.1 & 27 & 0.48 & 0.74\\
J1400+0425 & 2.550 & T, jet & 0.30 & 151 & 2.9 & 1.0 & 29 & 0.07\\
J1405+0415 & 3.208 & D & 0.08 & 694 & 1.4 & & & -0.13\\
J1429+2607 & 2.916 & T & 0.14 & 231 & 9.3 & 2.0 & 0 & 0.31 & 0.59\\
J1430+4204 & 4.715 & D & 0.22 & 276 & 3.9 & & & -0.41 & 0.45 \\
J1450+0910 & 2.628 & D* & 0.07 & 336 & 1.2 & &  & -0.55\\
J1457+3439 & 2.734 & D, jet & 0.19 & 88 & 1.6 & & & 0.49\\
J1459+3253 & 3.332 & T & 0.56 & 18.2 & 1.1 & 3.7 & 19 &  0.59\\
J1502+5521 & 3.329 & T & 0.83 & 13.3 & 0.7 & 0.9 & 35 & 0.18 & 0.68\\
J1528+5310 & 2.825 & T & 0.27 & 40 & 1.8 & & & (0.84) & (3.95)\\
J1540+4738 & 2.563 & T, jet & 0.51 & 26.2 & 8.0 & 4.9 & 17 & 0.80 & 1.07\\
J1602+2410 & 2.536 & T, jet & 0.73 & 28.3 & 1.7 & 4.0 & 11 & 0.85 & 0.55\\
J1610+1811 & 3.122 & D & 0.20 & 66.3 & 4.7 & & & 0.81\\
J1612+2758 & 3.544 & T & 0.23 & 33.2 & 1.8 & 1.3 & 20 & 0.25\\
J1655+3242 & 3.181 & D, jet & 1.70 & 15.0 & 3.2 & & & -0.09\\
\hline\hline
\end{tabular}
\end{center}
\end{table*}

\begin{table*}
\begin{center}
\caption{Rejected Sources}
\begin{tabular}{cc}
\hline\hline
\multicolumn{1}{c}{Source} &
\multicolumn{1}{c}{Reason for Rejection}\\
\hline
J0736+6513 & The northern declination limit of the FIRST survey is +65d 00 m. \\
J0744+2119 & $S_{\rm 1.4 GHz} <$ 70 mJy, source is compact\\
J0747+4618 & radio galaxy\\
J0751+2716 & gravitational lens\\
J0817+2918 & radio galaxy\\
J0823+0611 & $S_{\rm 1.4 GHz} <$ 70 mJy, source is compact\\
J0849+0950 & flux confused by nearby strong source\\
J0850+1529 & flux confused by nearby strong source\\
J0958+2947 & blend of two sources\\
J1020+1039 & radio galaxy\\
J1044+2959 & $S_{\rm 1.4 GHz} <$ 70 mJy; very core dominated structure \\
J1058+0443 & $S_{\rm 1.4 GHz} <$ 70 mJy; source is triple with a jet \\
J1130+0781 & $S_{\rm 1.4 GHz} <$ 70 mJy; source is compact \\
J1215+6422 & $S_{\rm 1.4 GHz} <$ 70 mJy; source is compact \\
J1330+4954 & $S_{\rm 1.4 GHz} <$ 70 mJy; source is compact \\
J1401+1513 & gravitational lens\\
J1424+2256 & gravitational lens\\
J1436+6319 & radio galaxy\\
J1439+1117 & $S_{\rm 1.4 GHz} <$ 70 mJy; source is compact \\
J1506+1029 & radio source and quasar are different objects\\
J1538+0019 & radio galaxy\\
J1600+0412 & original redshifts in NED were in error; SDSS DR10 gives $z = 0.794$\\
J1605+3038 & $S_{\rm 1.4 GHz} <$ 70 mJy; source is compact \\
J1606+3124 & radio galaxy\\
\hline\hline
\end{tabular}
\end{center}
\end{table*}

\newpage
*
\begin{table*}
\begin{center}
\caption{Overview of Morphologies}
\begin{tabular}{lccccc}
\hline\hline
\multicolumn{1}{l}{Sample} &
\multicolumn{1}{c}{Number of} &
\multicolumn{1}{c}{Redshift Range} & 
\multicolumn{3}{c}{Morphological class} \\
\multicolumn{1}{c}{} &
\multicolumn{1}{c}{Quasars} &
\multicolumn{1}{c}{} &
\multicolumn{1}{c}{C+E} &
\multicolumn{1}{c}{\bf{D}} &
\multicolumn{1}{c}{T} \\
\hline 
This paper & 113 & $2.5 \geq z < 5.3$ & 58 (51\%) & \bf {27 (24\%)} & 28 (25\%) \\ 
Barthel \& Miley (1988) & 80 & $1.5 \geq z < 2.9 $ & 23 (29\%) & \bf{11 (14 \%)}  & 46 (58 \%)  \\
3CRR & 42 & $0.37 < z < 2.0 $ & 10 (23 \%) & \bf{0 (0 \%)} & 32 (77 \%)  \\
\hline\hline
\end{tabular}
\end{center}
Tabulated are the number of sources in each class, and the corresponding percentages.
\end{table*}

\begin{table*}
\begin{center}
\caption{Linear Size and Bending Angle for the Triple (two sided) Sources}
\begin{tabular}{lcccc}
\hline\hline
\multicolumn{1}{l}{Sample} &
\multicolumn{1}{c}{Number of} & {Redshift} &
\multicolumn{2}{c}{Physical Properties}\\
\multicolumn{1}{c}{} &
\multicolumn{1}{c}{Triples} & {Range} &
\multicolumn{1}{c}{Linear Size (kpc)} &
\multicolumn{1}{c}{Bending Angle (deg)}\\
\hline 
This paper & 28 & $2.5 \geq z < 5.3$ & 48 (35, 65) & 21.5 (17, 35) \\ 
Barthel \& Miley (1988) & 46 &  $1.5 \geq z < 2.9 $ &  109 (59, 157) & 12 (7, 30) \\
3CRR & 32 & $0.37 < z < 2.0 $ & 110 (50. 260) & 10 (4, 20)\\
\hline\hline
\end{tabular}
\end{center}
Tabulated are the median values of each property, and the inter-quartile range in parentheses. 
\end{table*}


\newpage
\clearpage

\clearpage
\begin{figure}
\figurenum{2}
\centering
\subfloat[Part 1][J0730+4049 z = 2.500 ]
{\hspace{-0.3in}\includegraphics[width=3.0in]{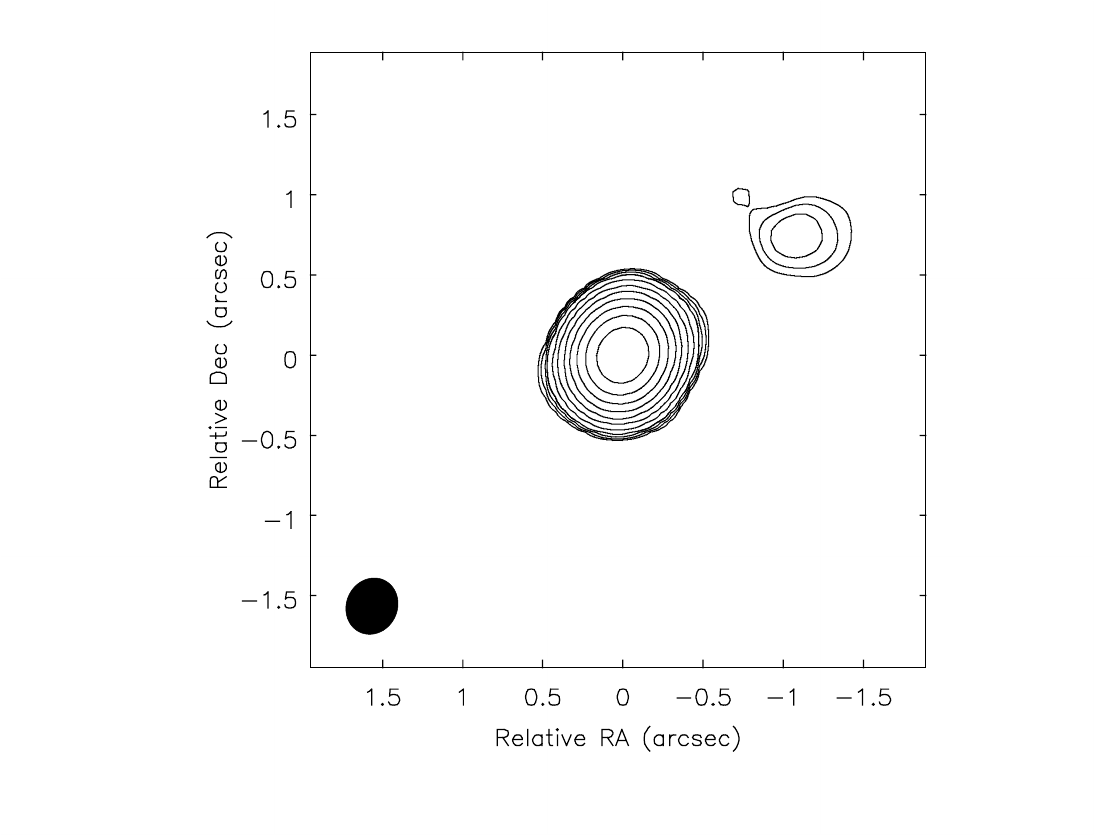}\hspace{-0.3in}
\label{fig:Images1-1}} 
\subfloat[Part 2][J0733+2536 z = 2.686]{\includegraphics[width=2.2in]{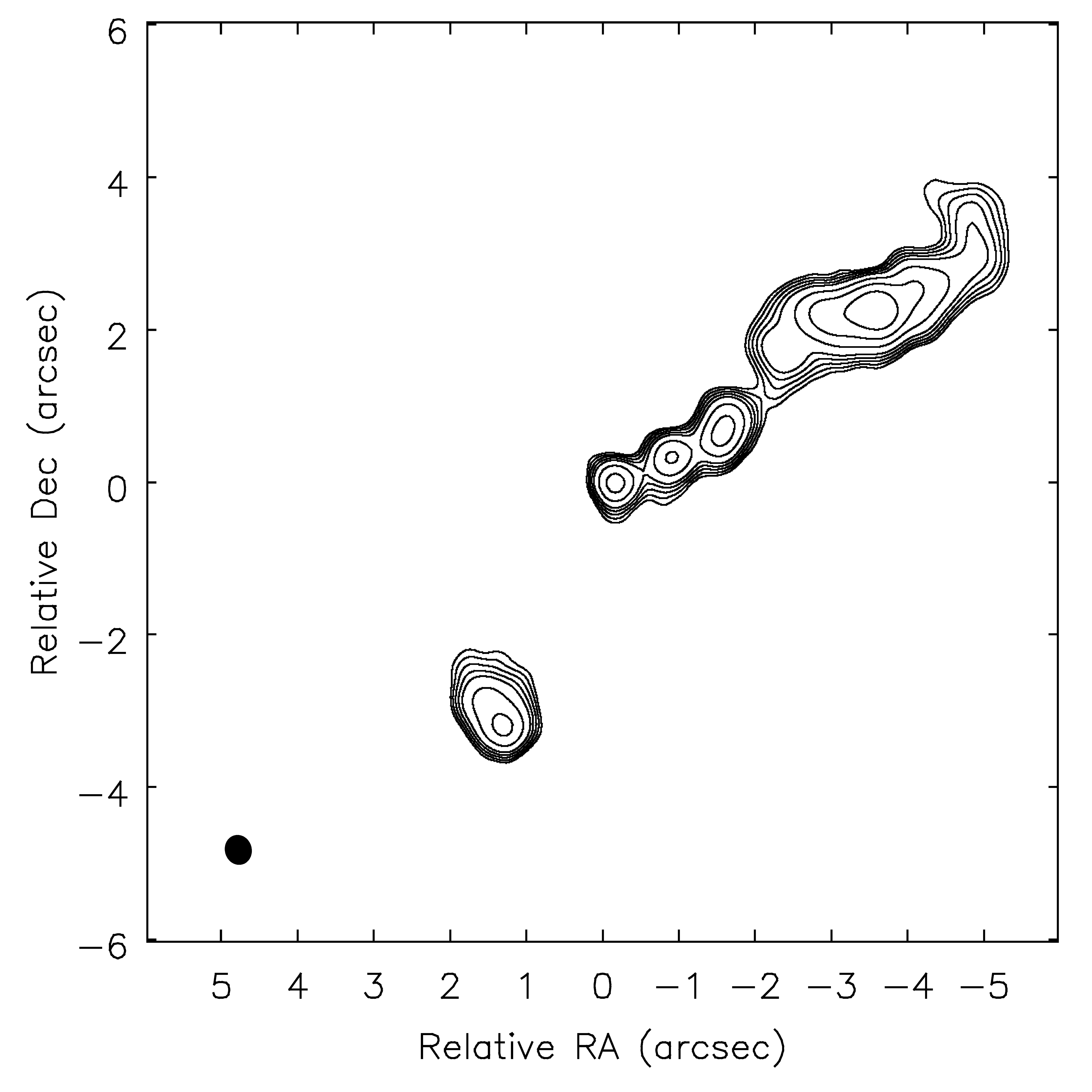} \label{fig:Images1-2}}
\subfloat[Part 3][J0801+1153 z = 2.680]{\includegraphics[width=2.2in]{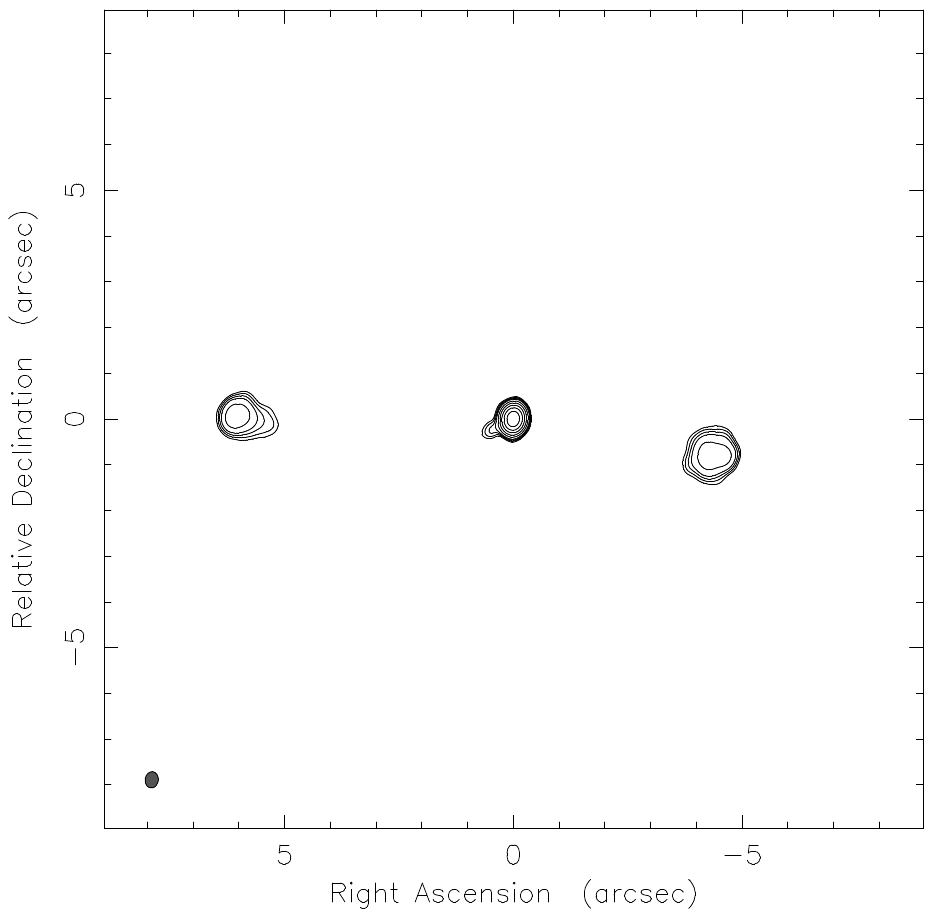} \label{fig:Images1-3}} \\

\subfloat[Part 4][J0801+4725 z = 3.276]{\includegraphics[width=2.2in]{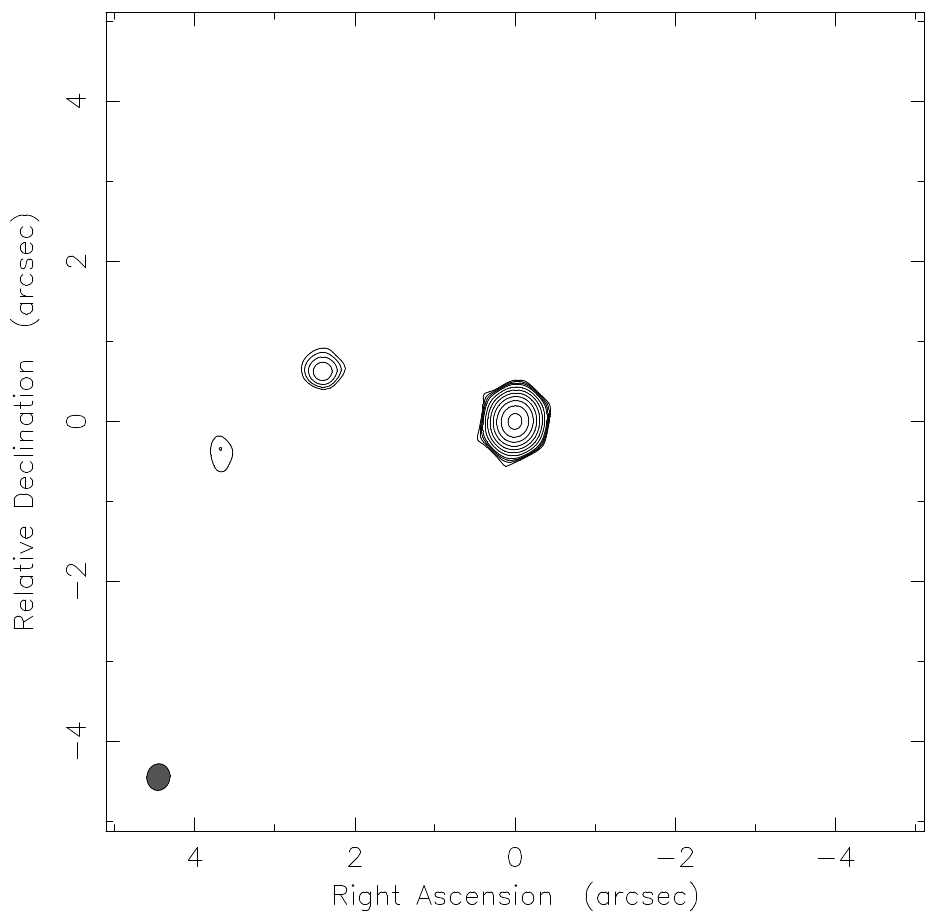} \label{fig:Images1-4}} 
\subfloat[Part 5][J0805+6144 z = 3.033]{\includegraphics[width=2.2in]{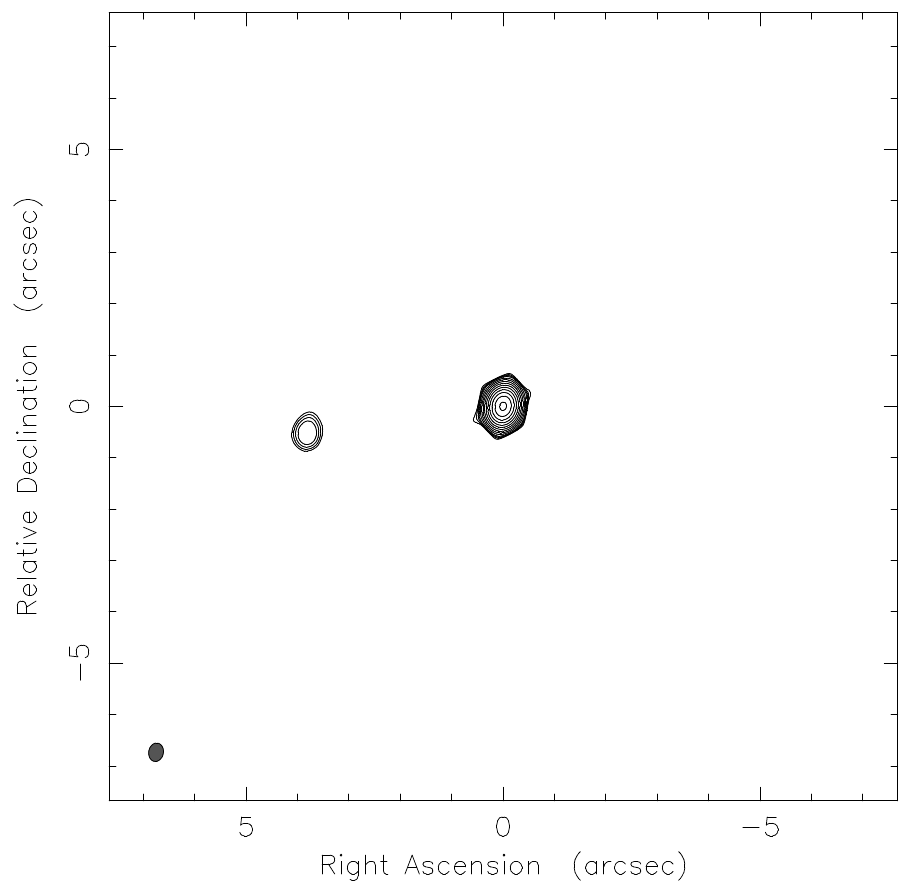} \label{fig:Images1-5}}
\subfloat[Part 6][J0807+0432 z = 2.877]{\includegraphics[width=2.2in]{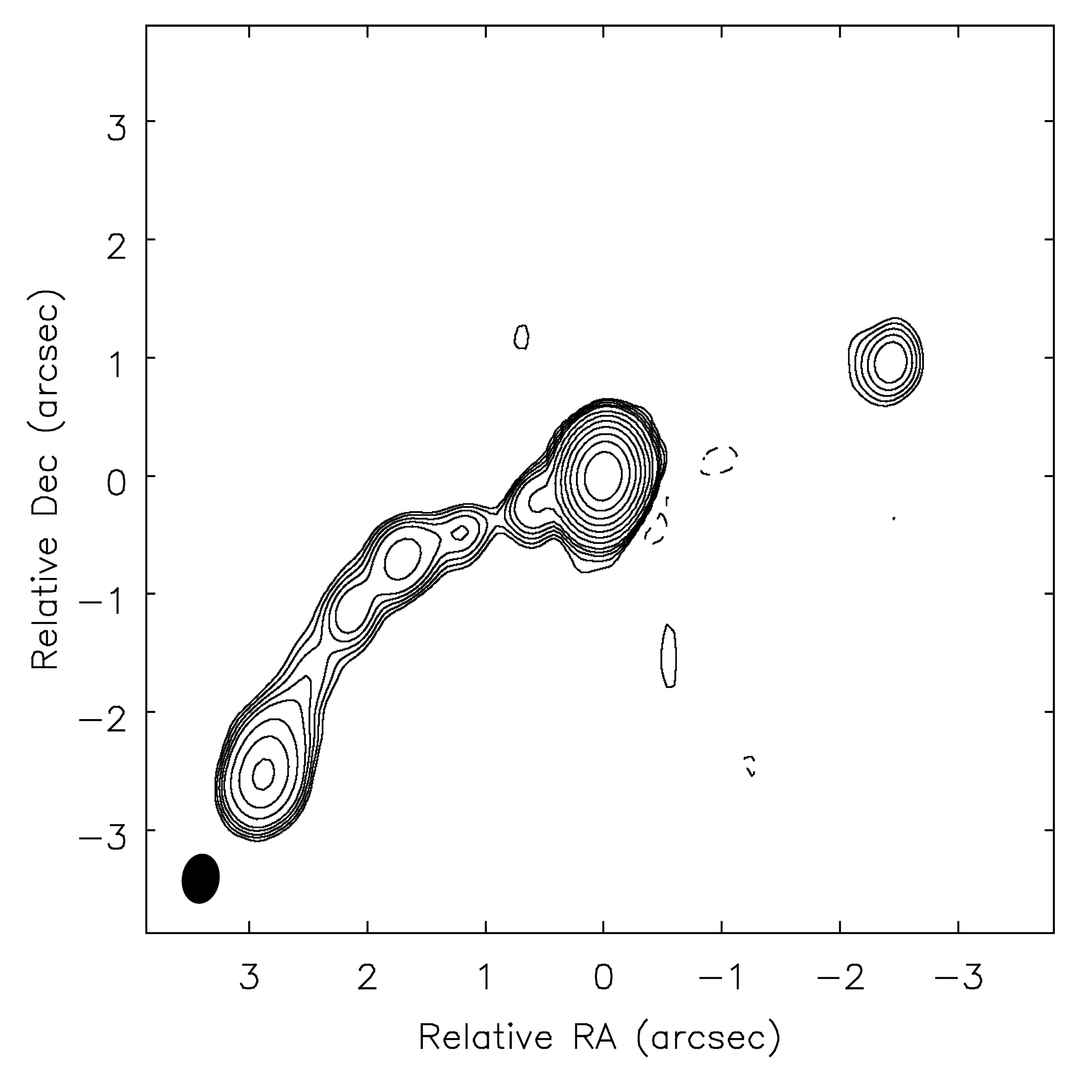} \label{fig:Images1-6}} \\

\subfloat[Part 7][J0833+3531 z = 2.703]{\includegraphics[width=2.2in]{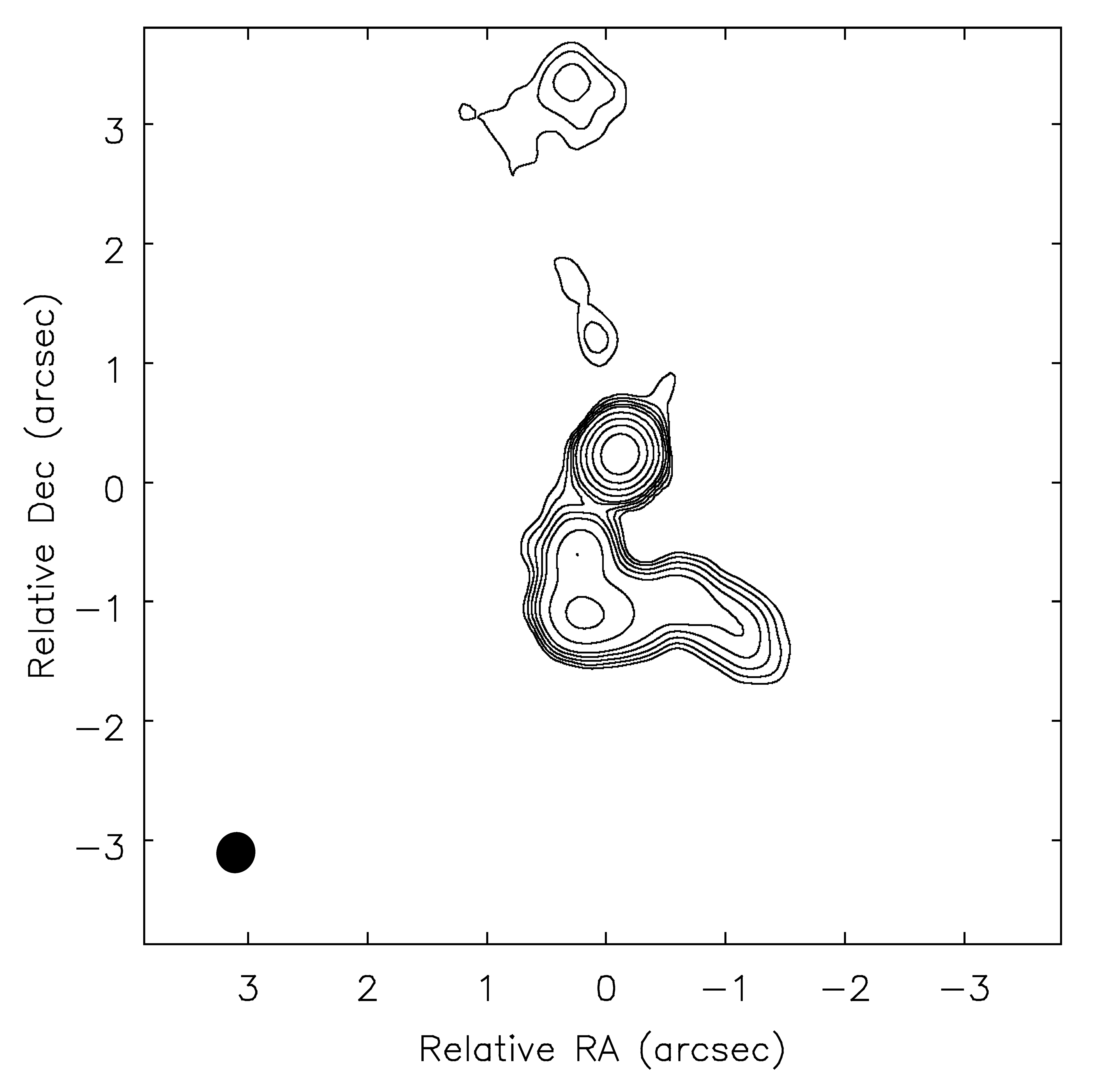}
\label{fig:Images1-7}} 
\subfloat[Part 8][J0833+1123 z = 2.985]{\includegraphics[width=2.2in]{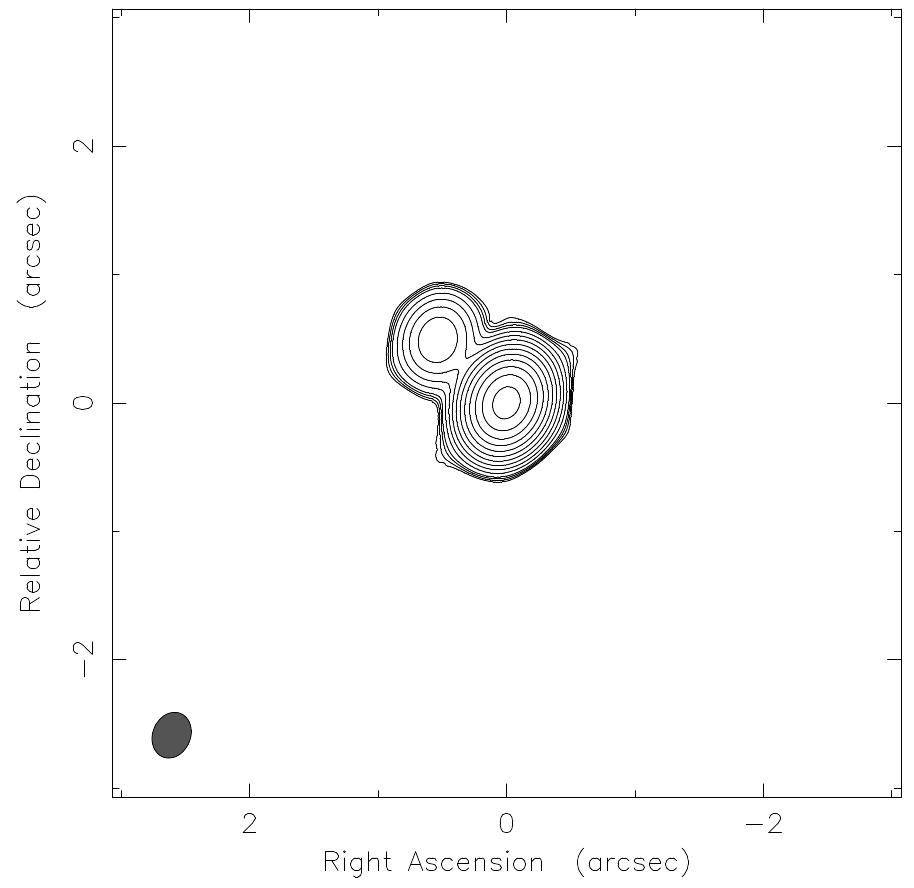} \label{fig:Images1-8}} 
\subfloat[Part 9][J0905+4850 z = 2.697]{\includegraphics[width=2.2in]{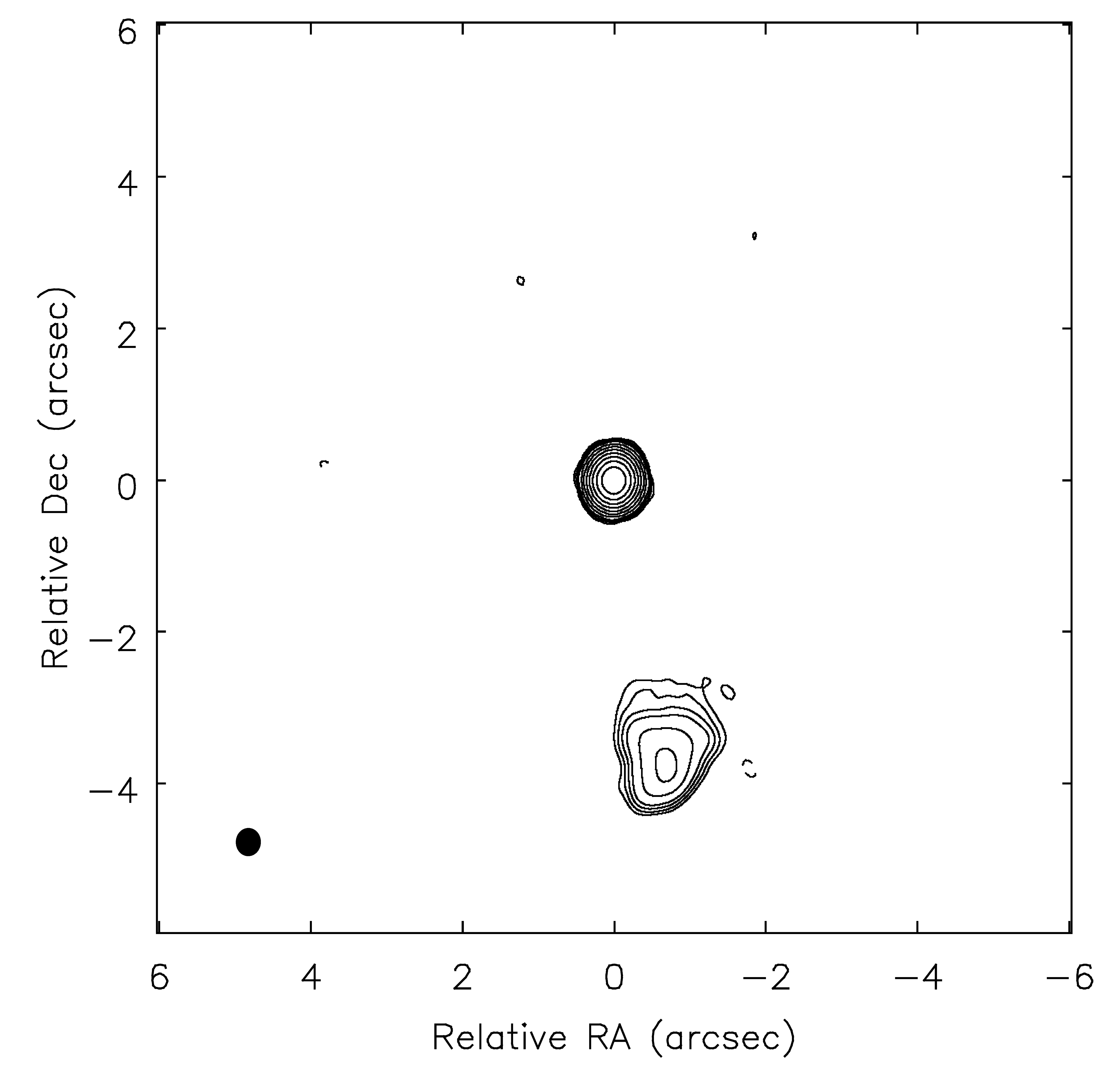} \label{fig:Images1-9}} \\

\end{figure}


\begin{figure}
\figurenum{3}
\centering
\subfloat[Part 1][J0905+3555 z = 2.840]{\includegraphics[width=2.2in]{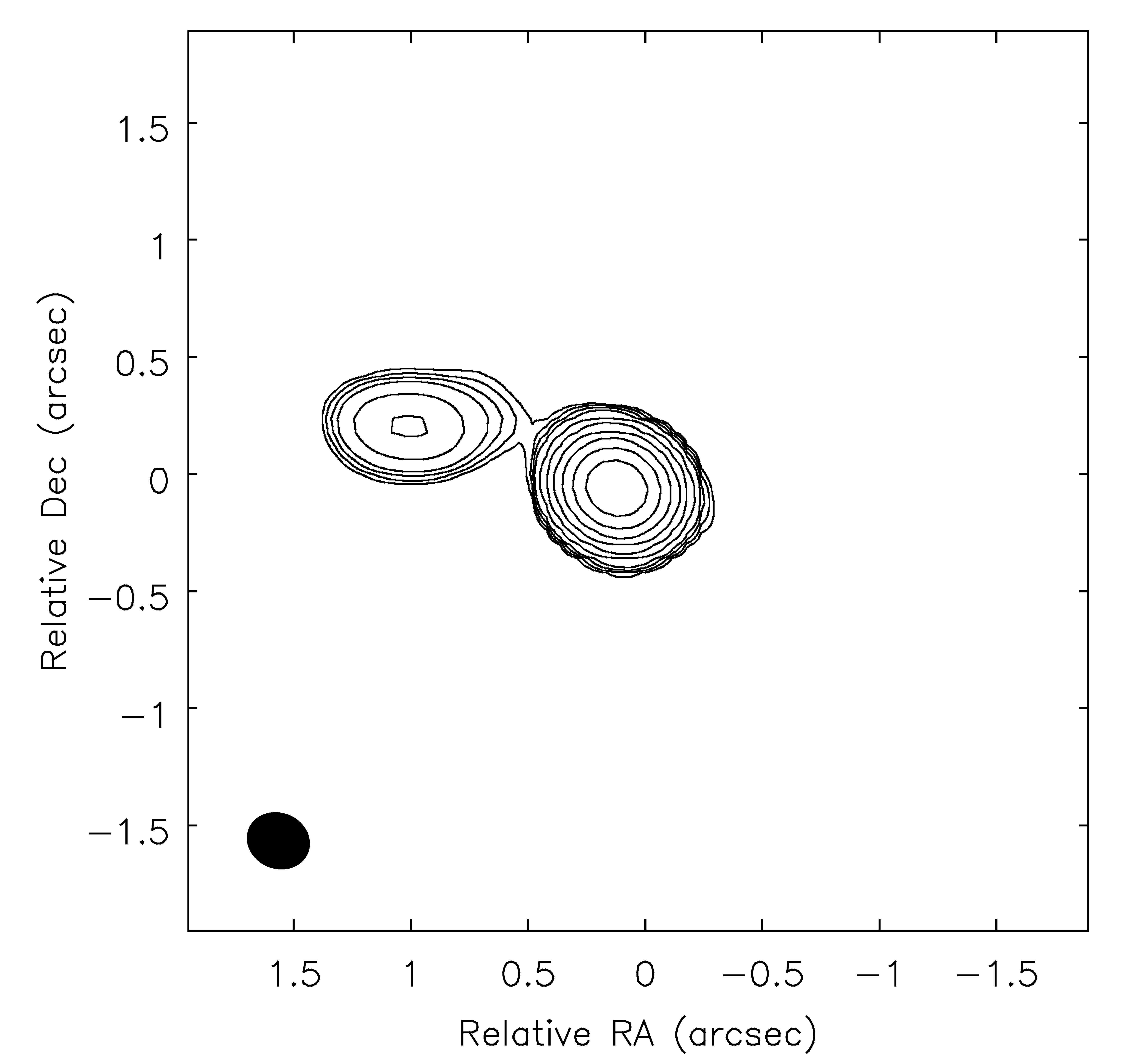} \label{fig:Images2-1}}
\subfloat[Part 2][J0909+0354 z = 3.288]{\includegraphics[width=2.2in]{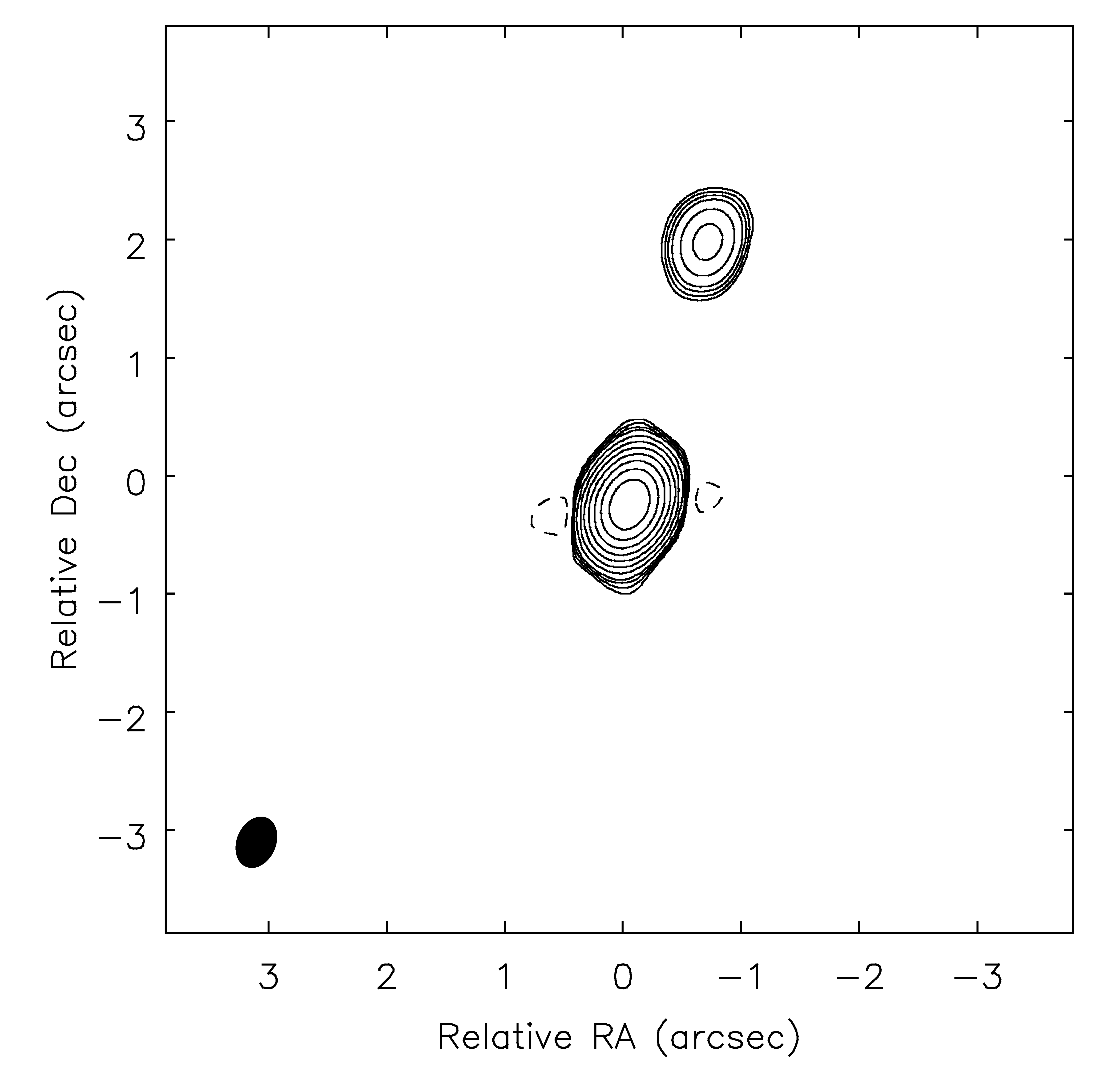} \label{fig:Images2-2}} 
\subfloat[Part 3][J0910+2539 z = 2.753]{\includegraphics[width=2.2in]{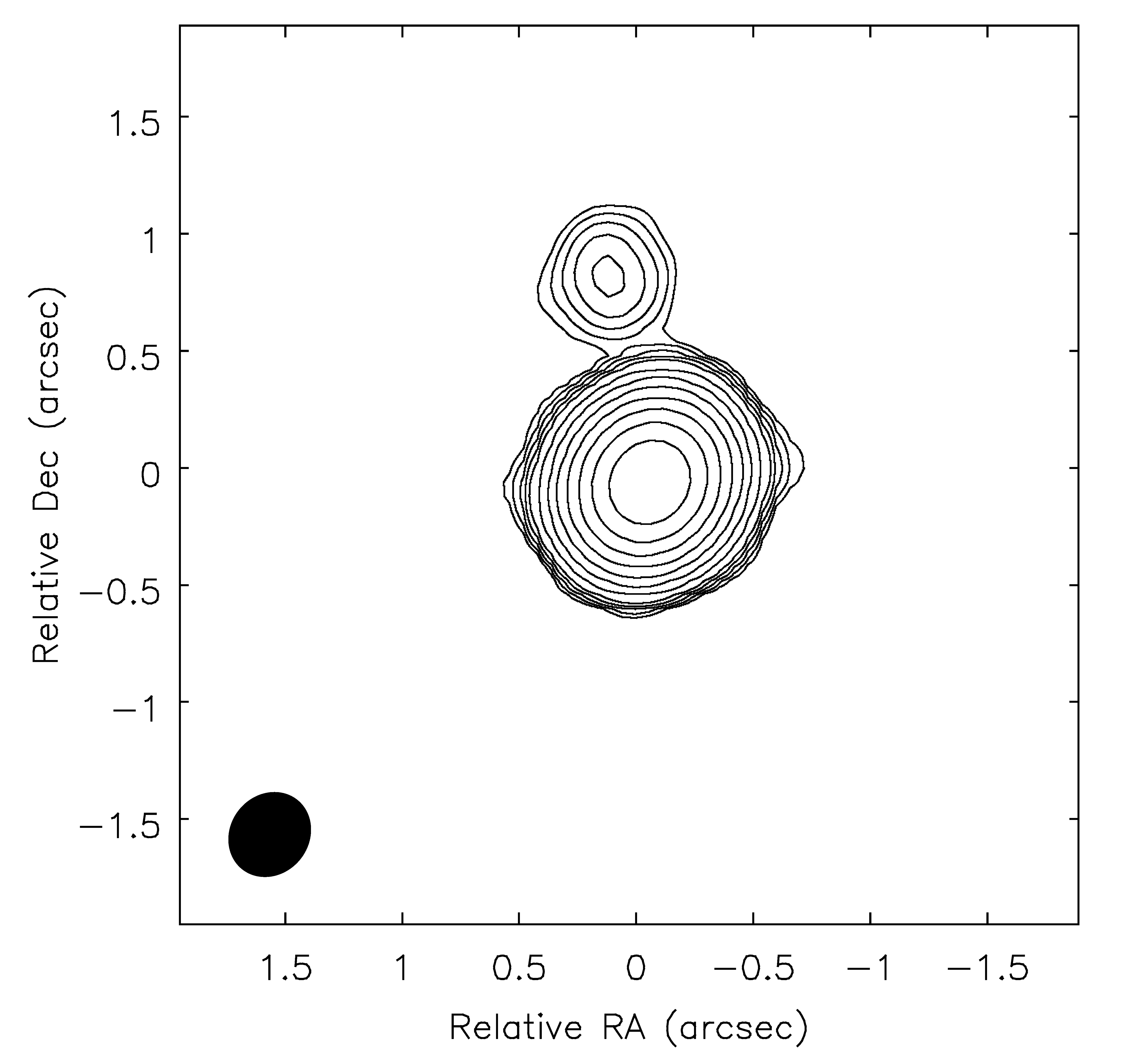} \label{fig:Images2-3}} \\
\subfloat[Part 4][J0918+5332 z = 3.012]
{\includegraphics[width=2.2in]{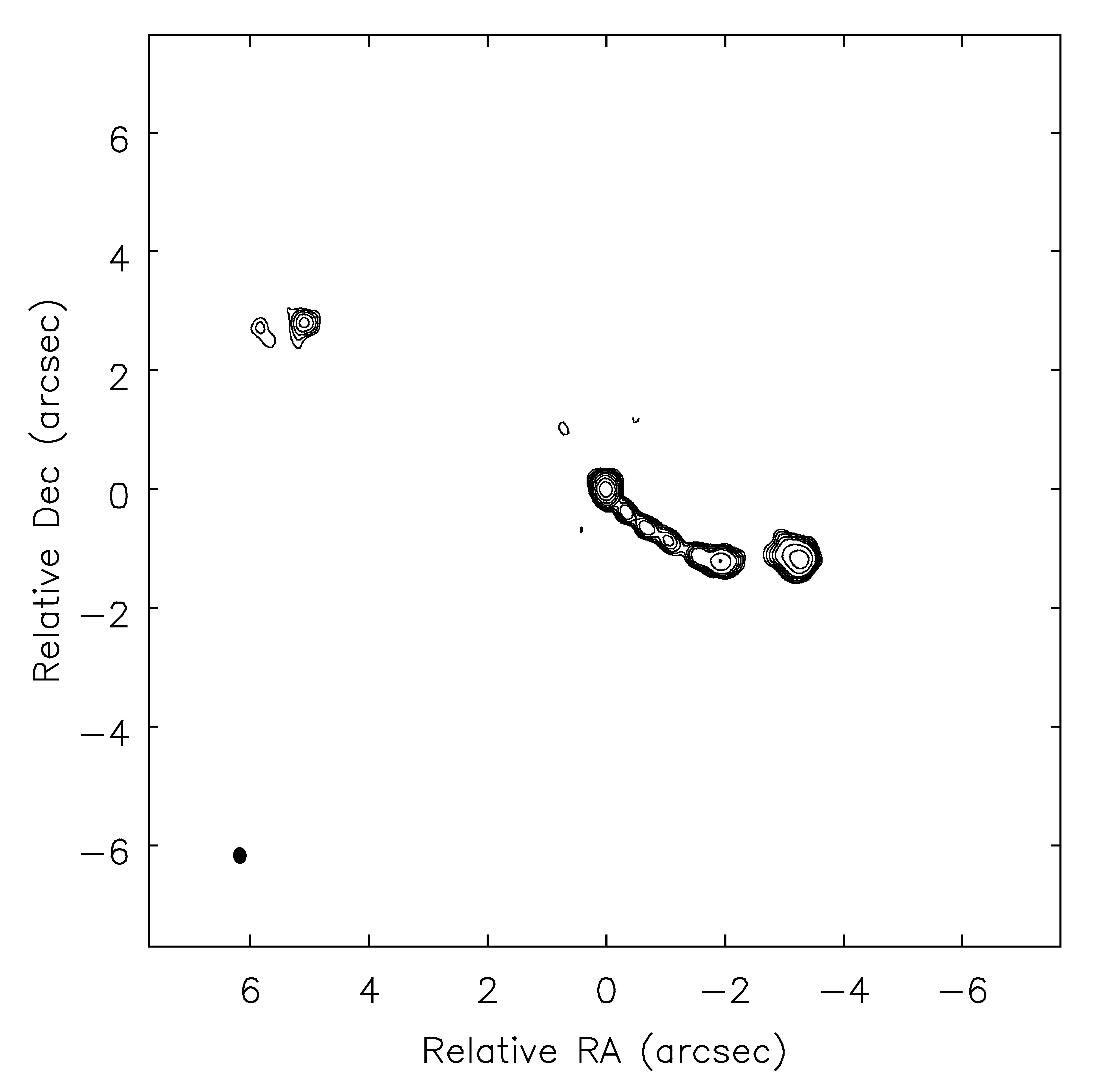} \label{fig:Images2-4}}
\subfloat[Part 5][J0934+4908 z = 2.584]{\hspace{-0.1in}\includegraphics[width=2.9in]{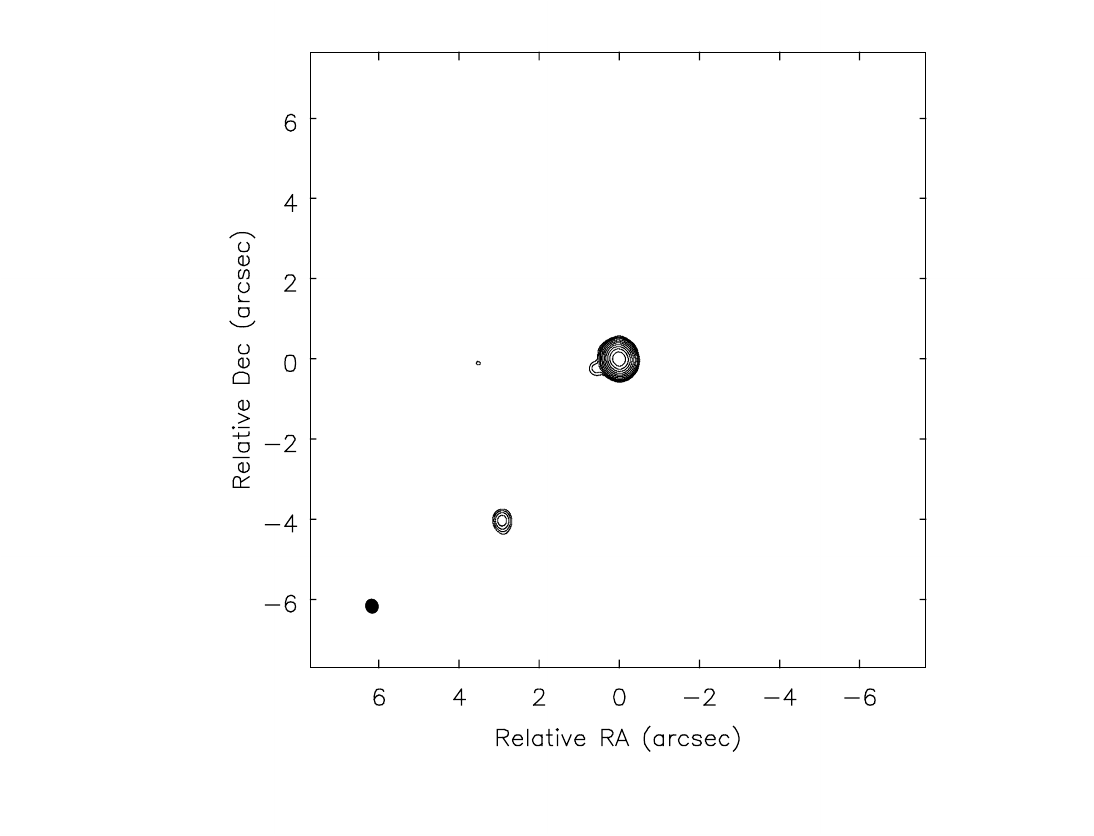}\hspace{-0.3in} \label{fig:Images2-5}} 
\subfloat[Part 6][J0934+3050 z = 2.895]{\includegraphics[width=2.2in]{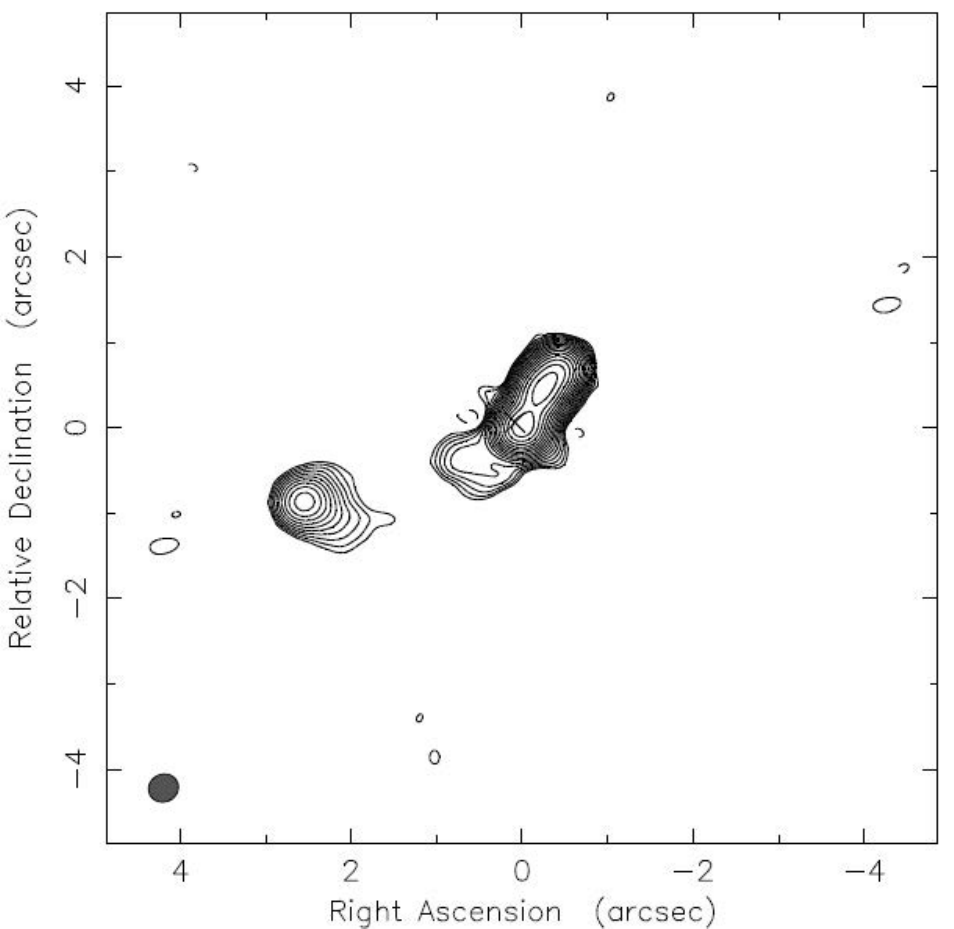} \label{fig:Images2-6}}\\
\subfloat[Part 7][J0941+1145 z = 3.196]{\includegraphics[width=2.2in]{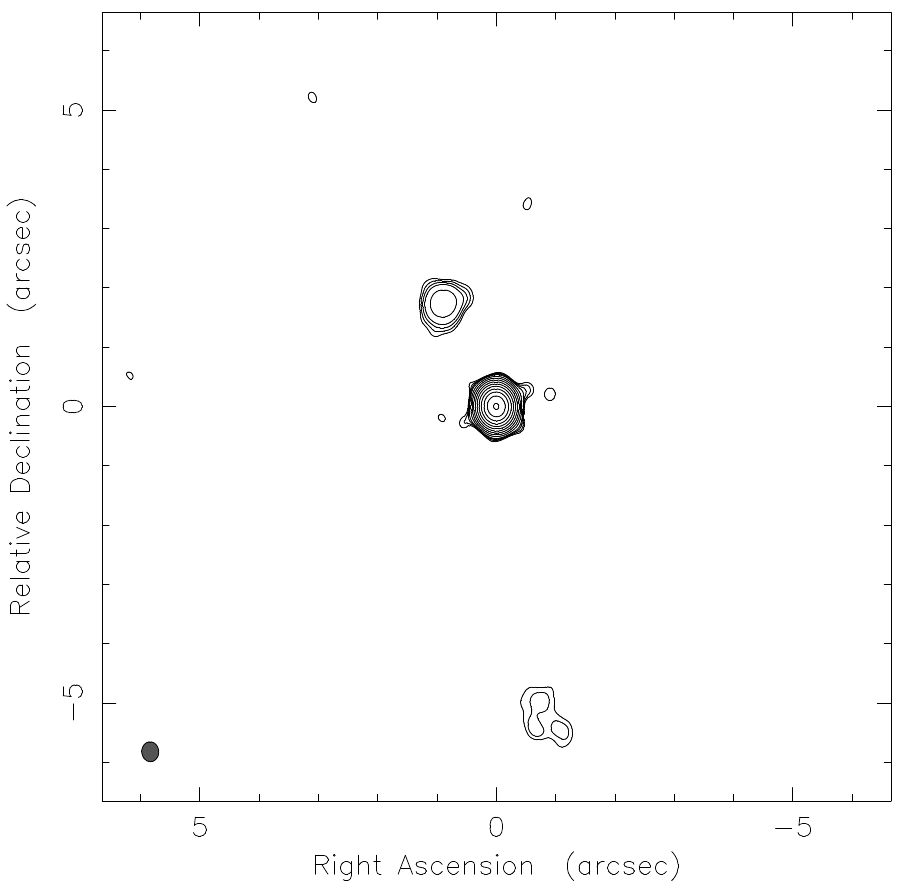} \label{fig:Images2-7}} 
\subfloat[Part 8][J0944+2554 z = 2.916]{\includegraphics[width=2.2in]{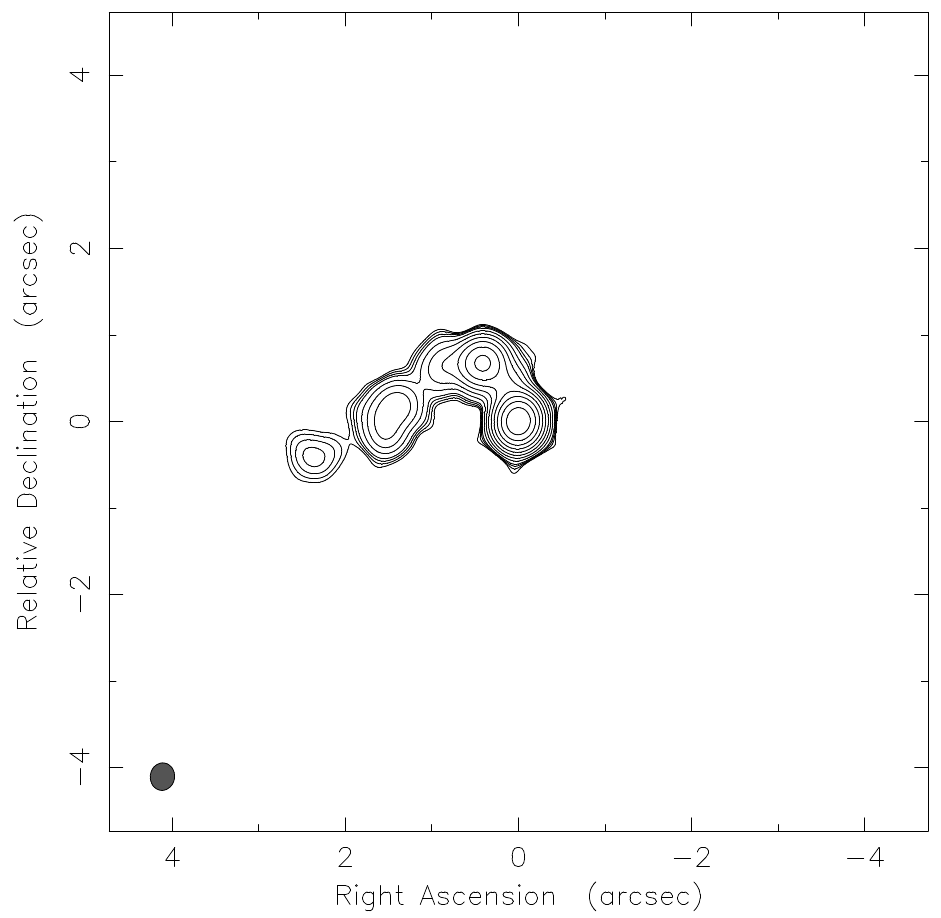} \label{fig:Images2-8}}
\subfloat[Part 9][J0947+6328 z = 2.617]{\includegraphics[width=2.2in]{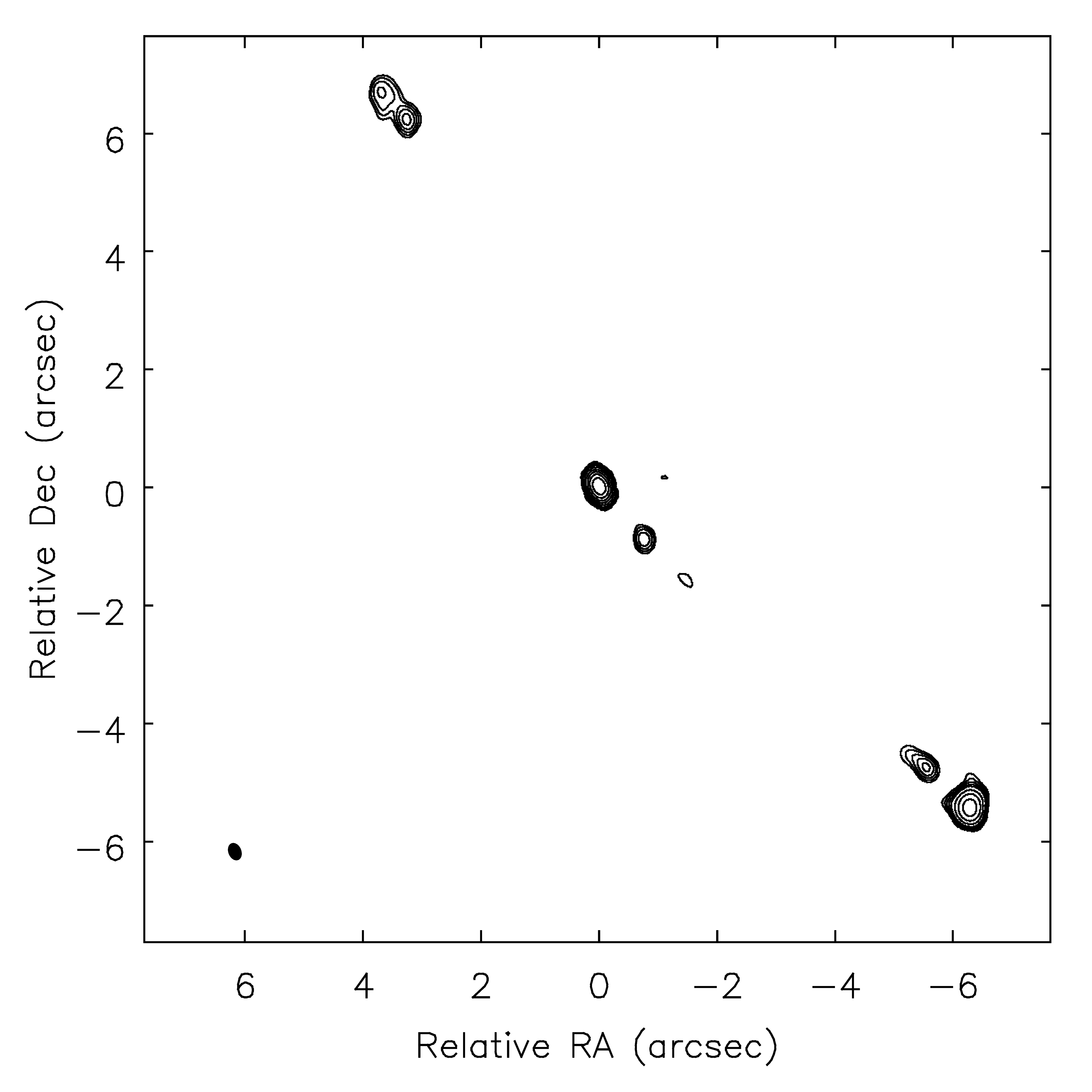} \label{fig:Images2-9}} \\
\caption{}
\label{fig:Images3}
\end{figure}


\begin{figure}
\figurenum{4}
\centering 
\subfloat[Part 1][J0958+3922 z = 2.941]{\includegraphics[width=2.2in]{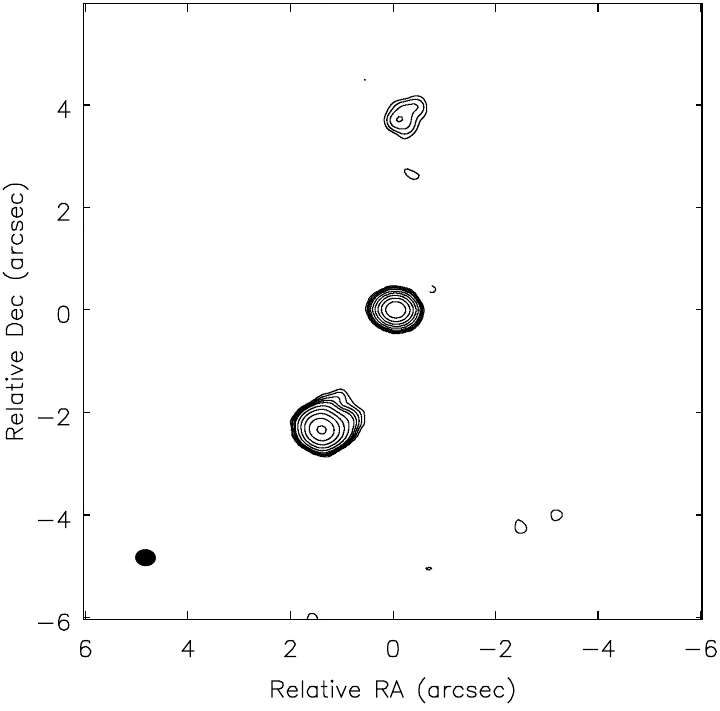} \label{fig:Images3-1}}
\subfloat[Part 2][J1007+1356 z = 2.707]{\includegraphics[width=2.2in]{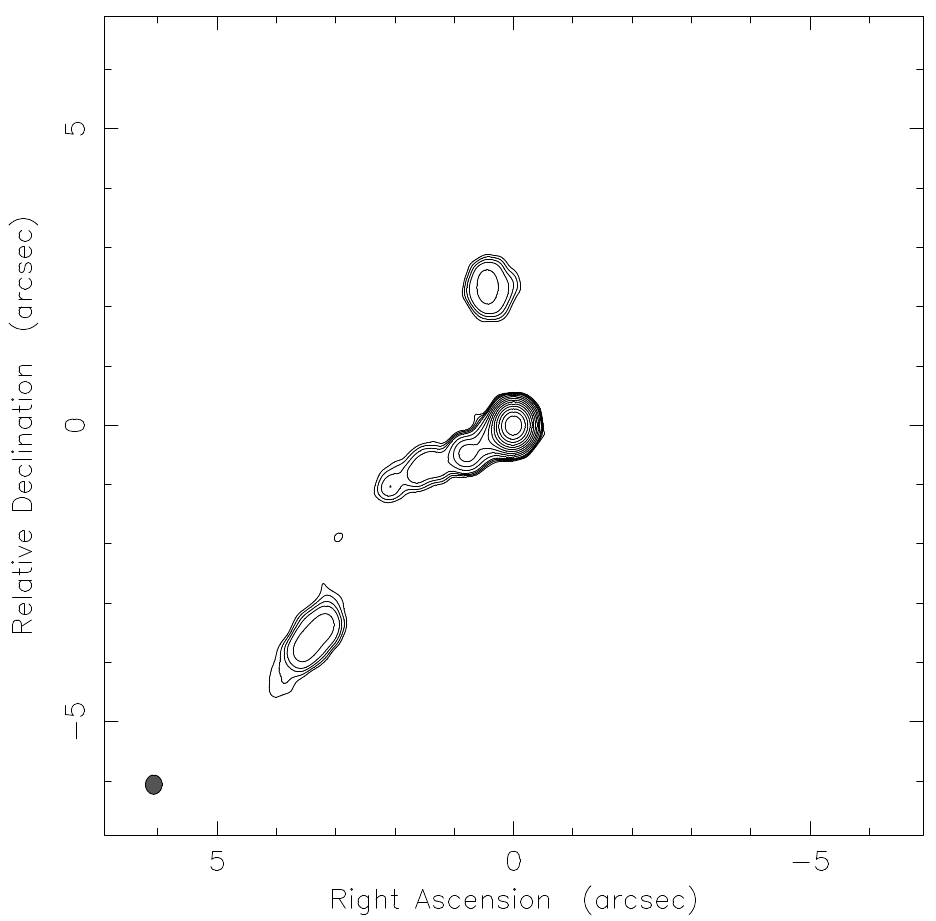}} \label{fig:Images3-2} 
\subfloat[Part 3][J1016+2037 z - 3.119]{\includegraphics[width=2.2in]{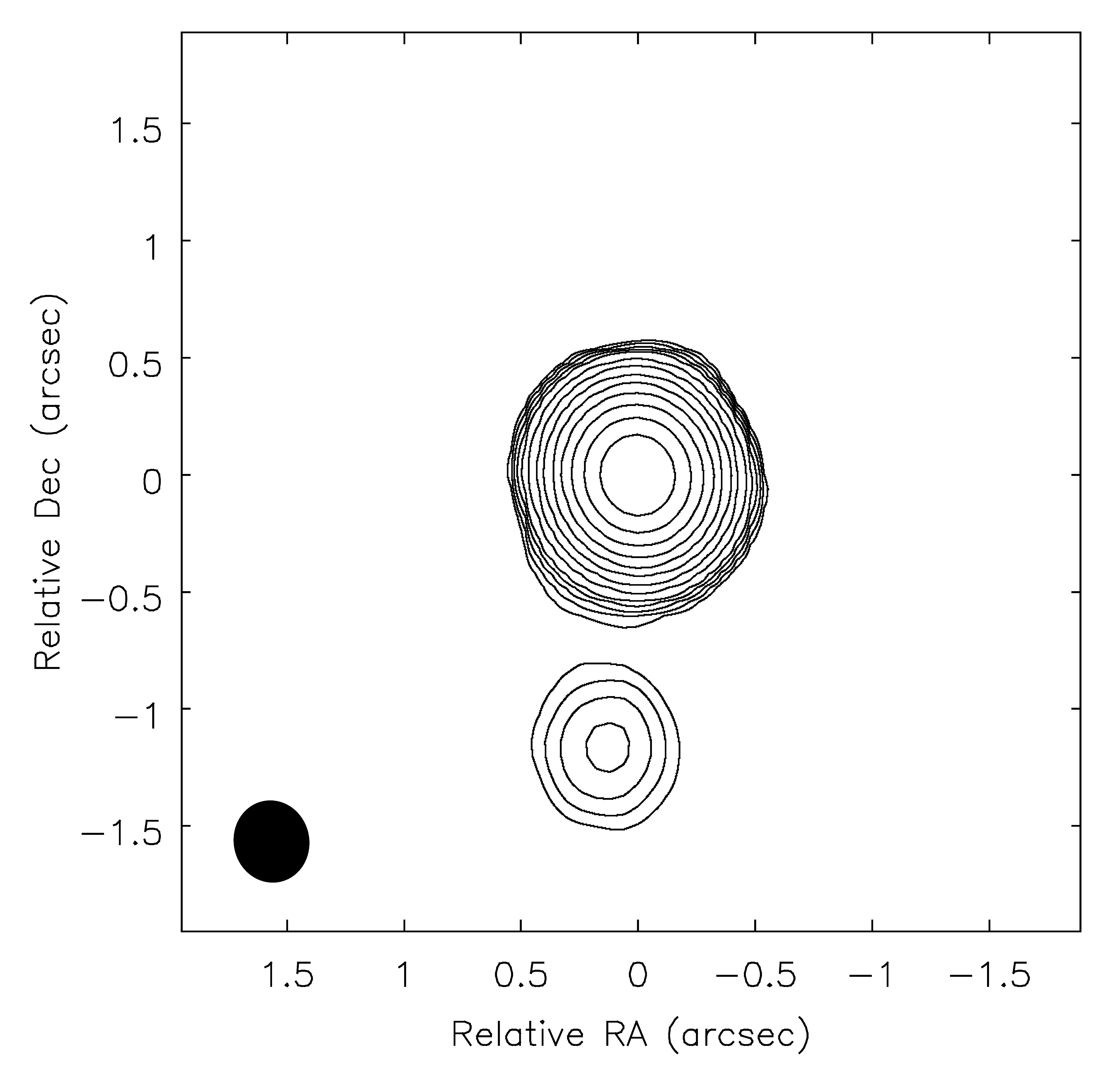} \label{fig:Images3-3}}\\

\subfloat[Part4][J1036+1326 z = 3.076]{\includegraphics[width=2.2in]{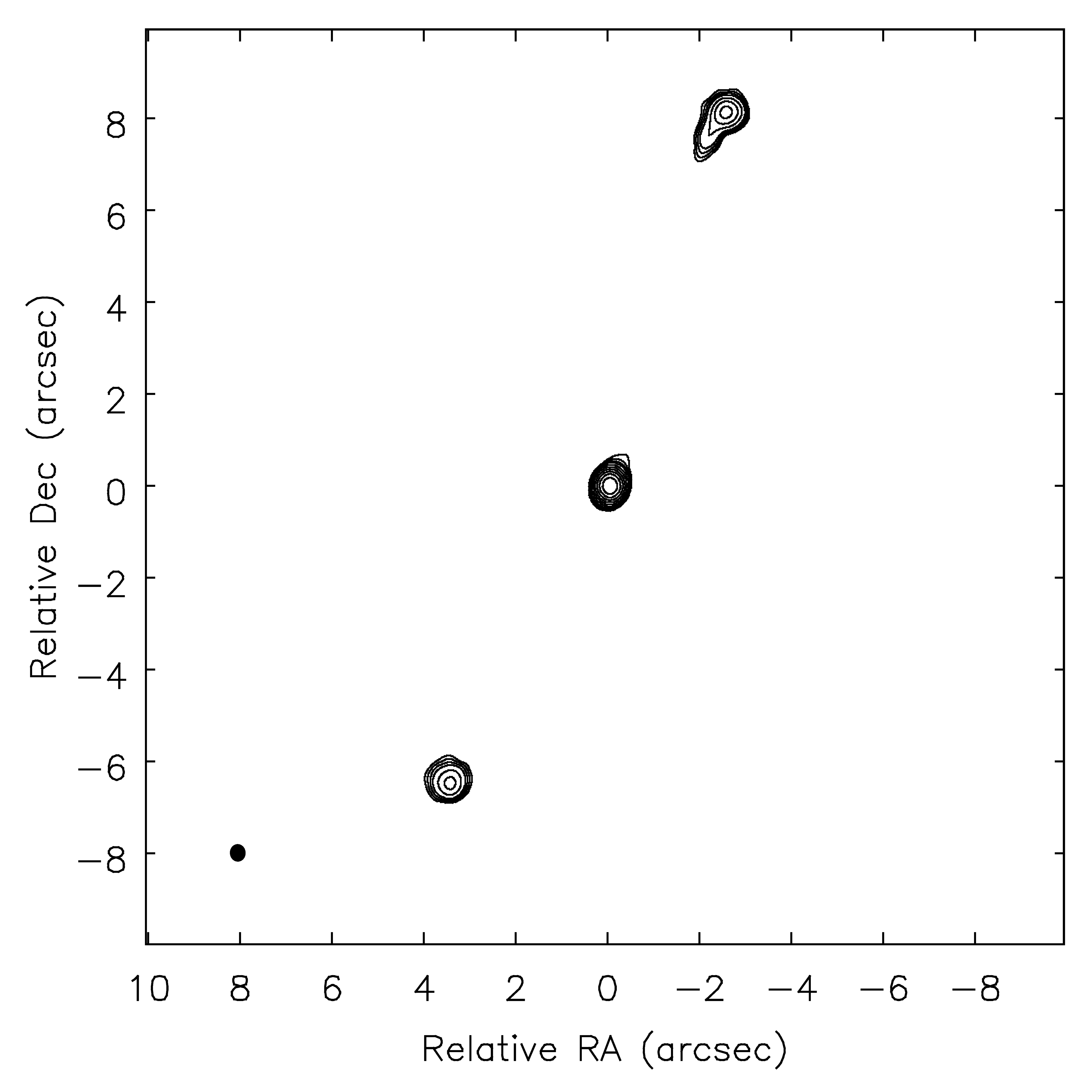} \label{fig:Images3-4}} 
\subfloat[Part 5][J1049+1332 z = 2.764]{\includegraphics[width=2.2in]
{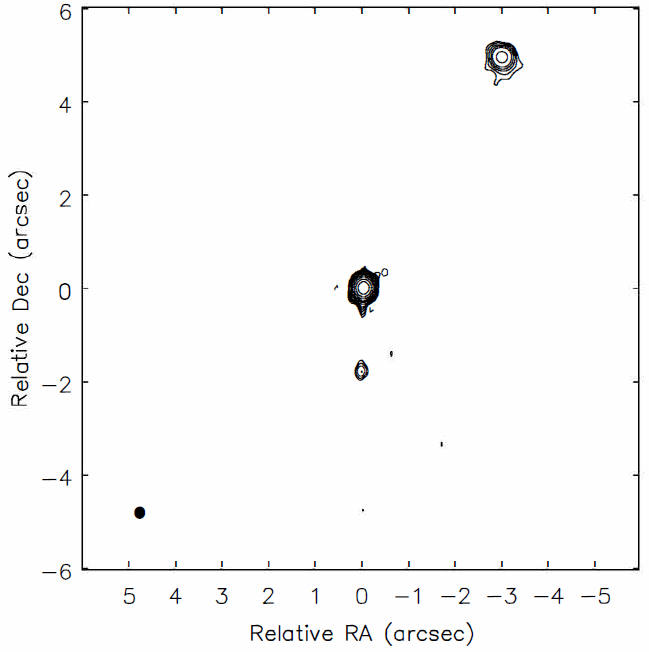} 
\label{fig:Images3-5}}
\subfloat[Part 6][J1050+3430 z = 2.528]
{\includegraphics[width=2.2in]
{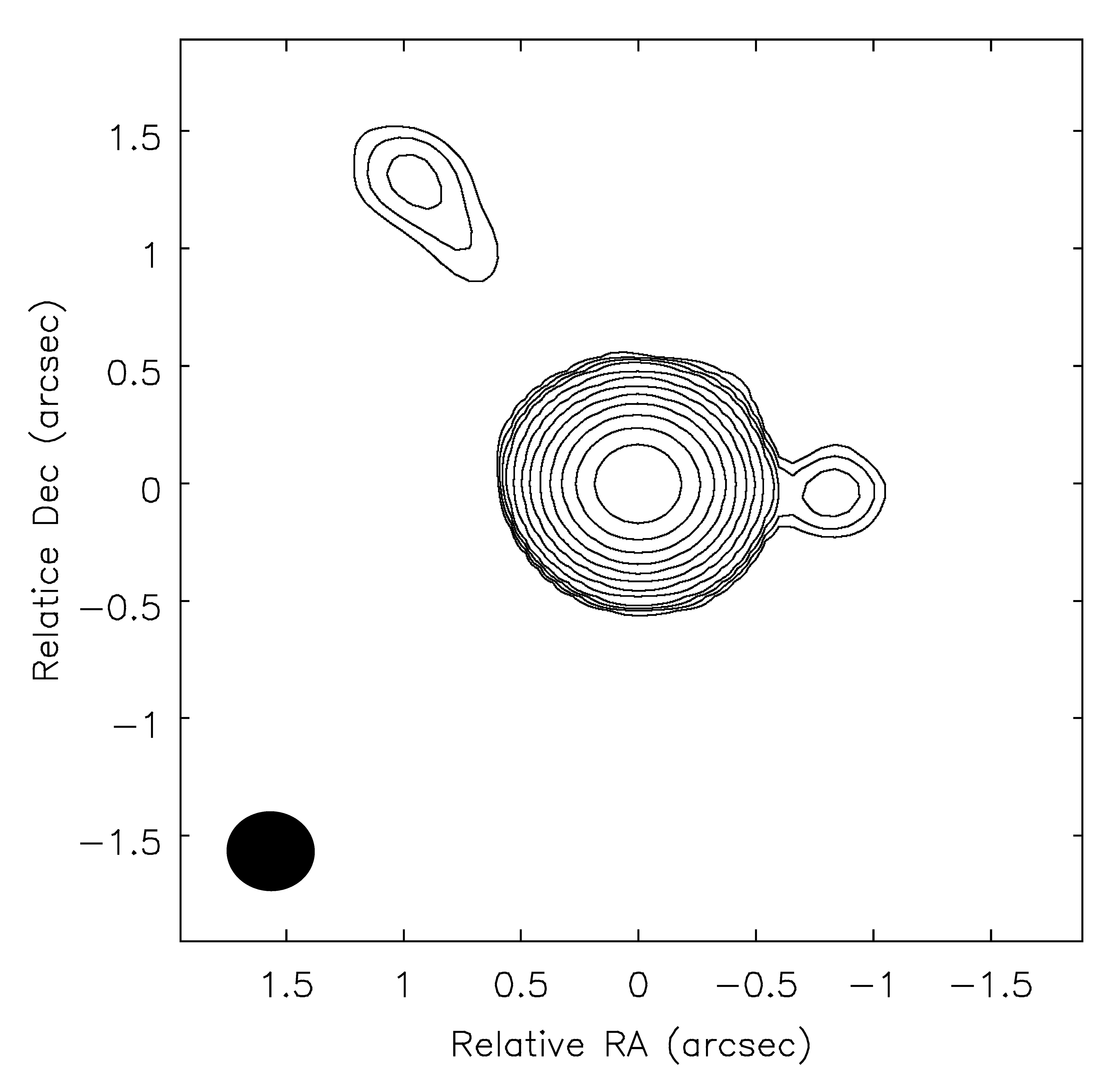} 
\label{fig:Images3-6}} \\

\subfloat[Part 7][J1057+0324 z = 2.832]{\includegraphics[width=2.2in]{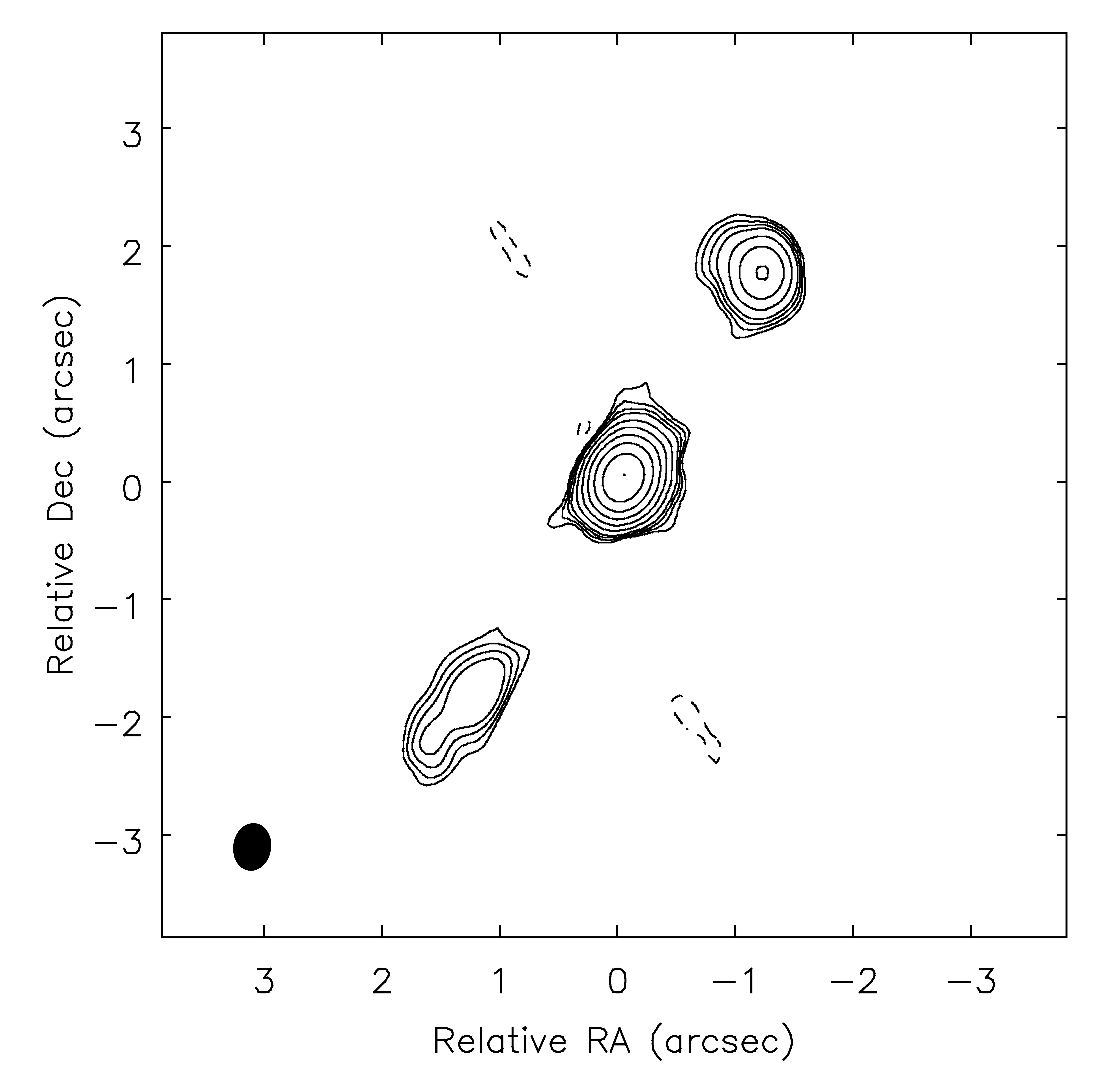} \label{fig:Images3-7}} 
\subfloat[Part 8][J1204+5228 z = 2.73]
{\includegraphics[width=2.2in]{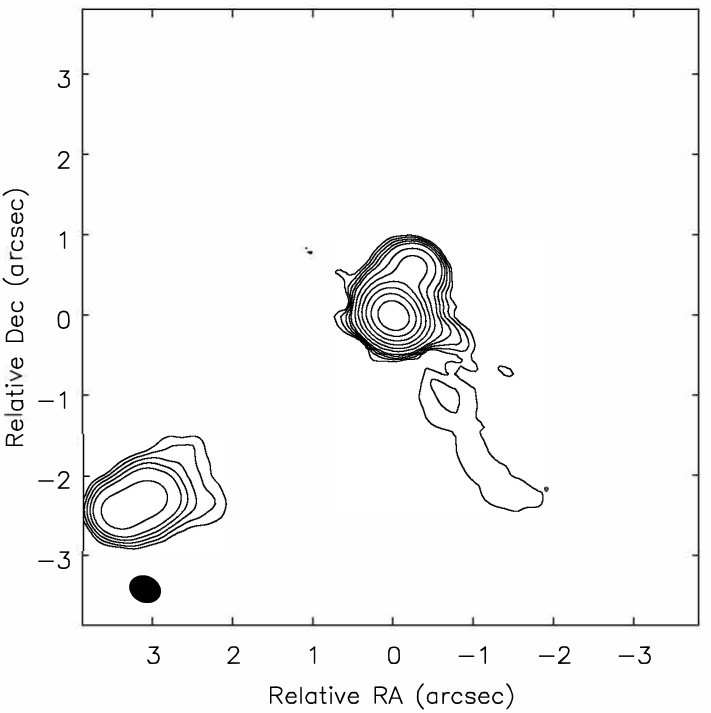} 
\label{fig:Images3-8}}  
\subfloat[Part 9][J1213+3247 z = 2.511]
{\includegraphics[width=2.2in]{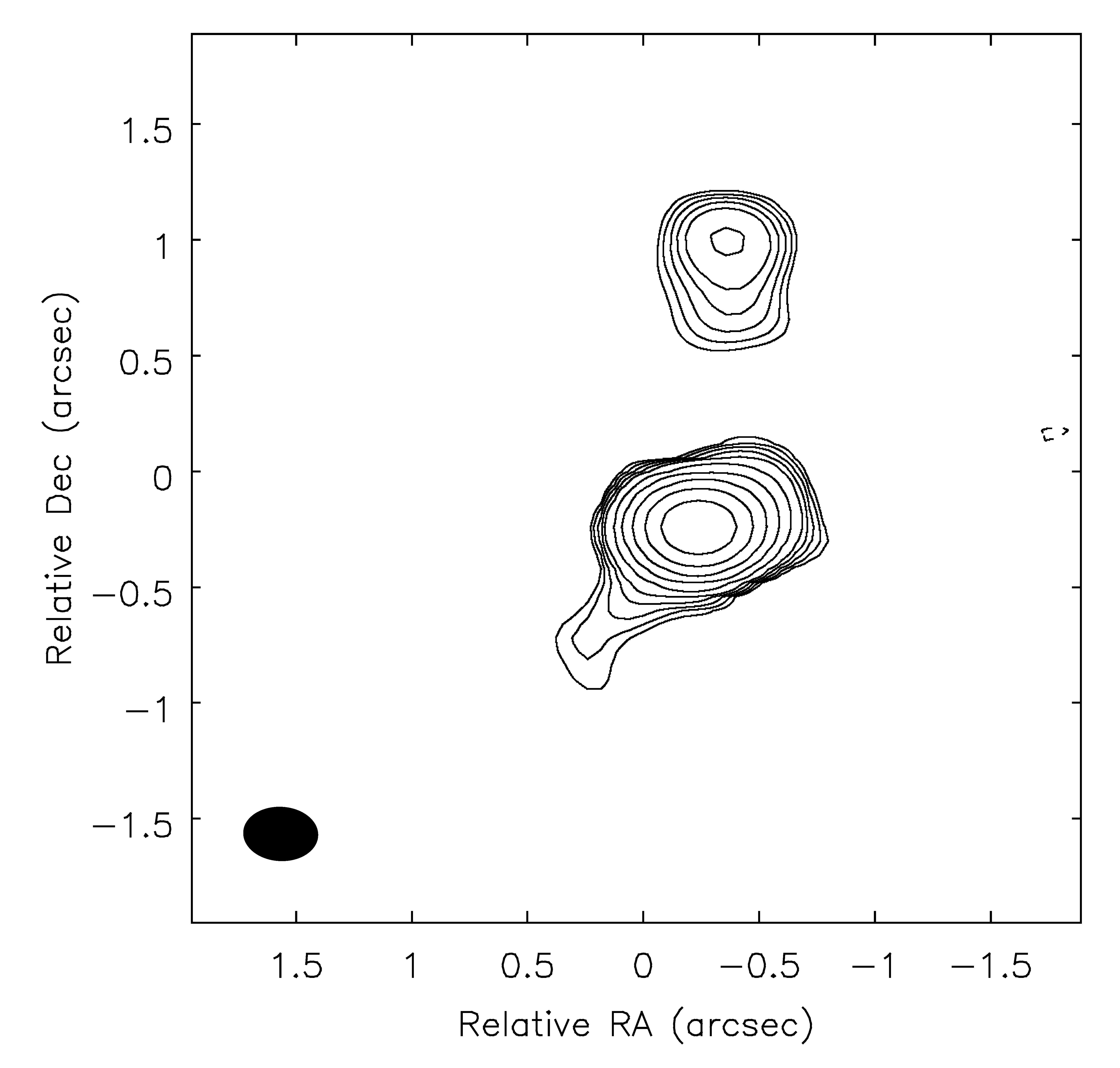} 
\label{fig:Images3-9}}\\

\caption{}
\label{fig:Images4}
\end{figure}


\begin{figure}
\figurenum{5}
\centering

\subfloat[Part 1][J1217+5835 z = 2.547]
{\includegraphics[width=2.2in]{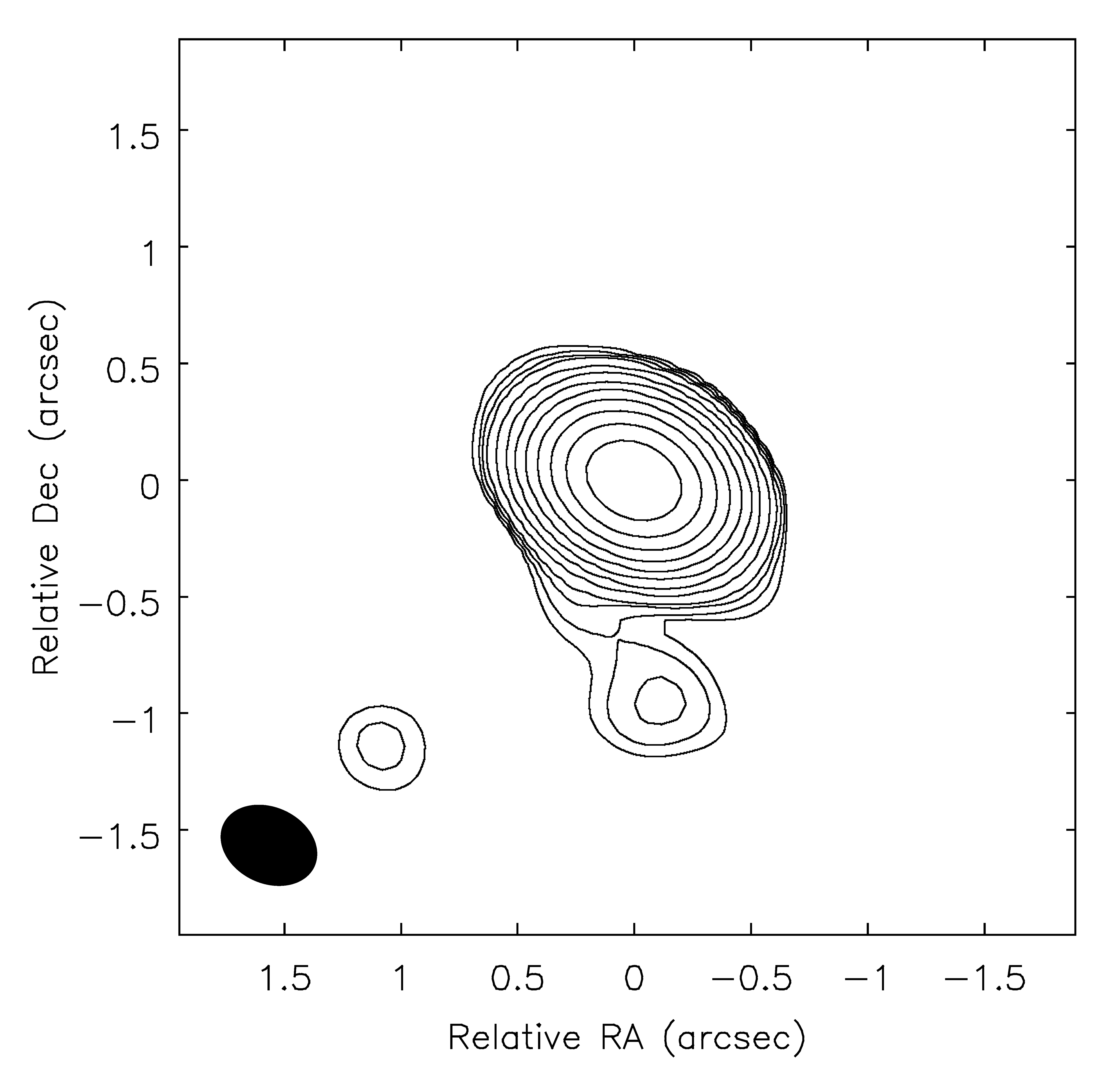} 
\label{fig:Images4-1}} 
\subfloat[Part 2][J1217+3435 z = 2.638]
{\includegraphics[width=2.2in]{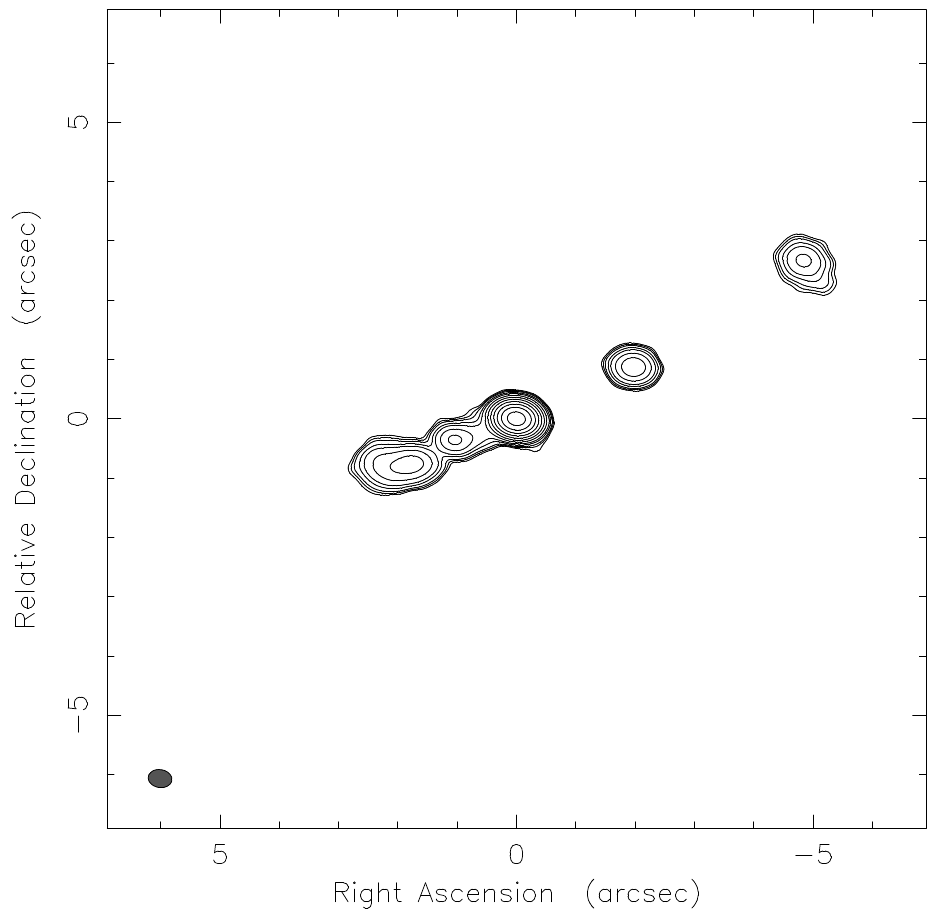} 
\label{fig:Images4-2}} 
\subfloat[Part 3][J1217+3305 z = 2.606]
{\includegraphics[width=2.2in]{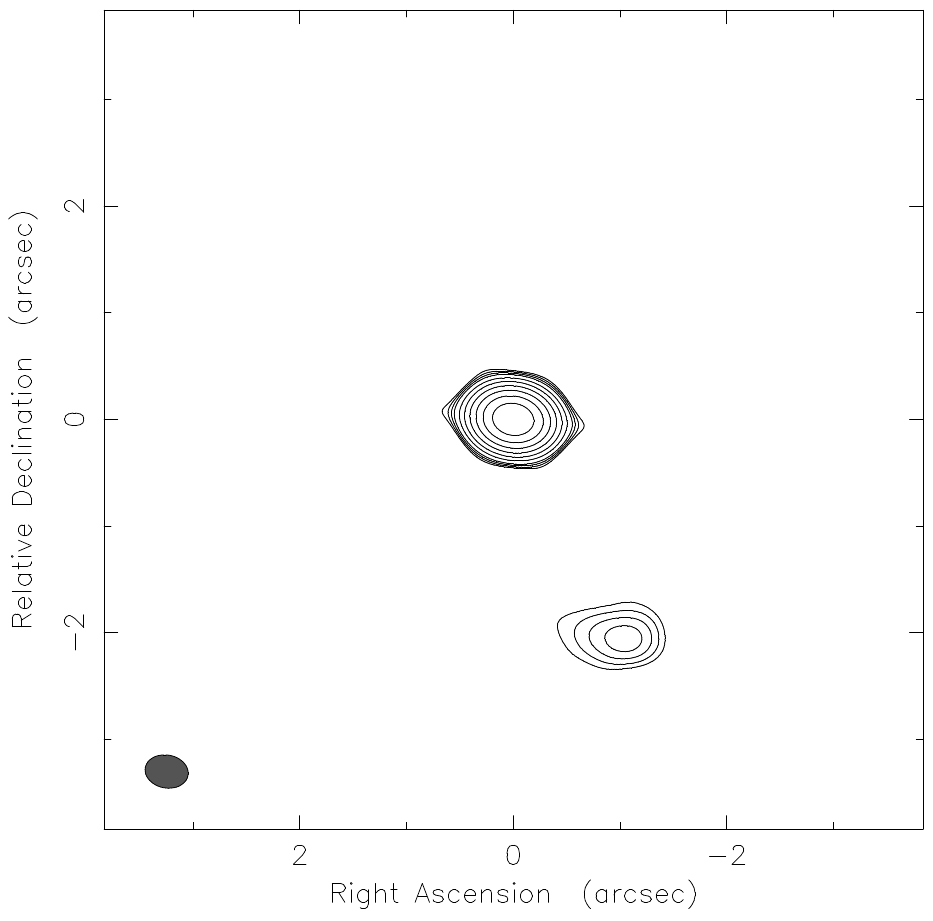} 
\label{fig:Images4-3}}\\
 
\subfloat[Part 4][J1246+0104 z = 2.510]
{\includegraphics[width=2.2in]{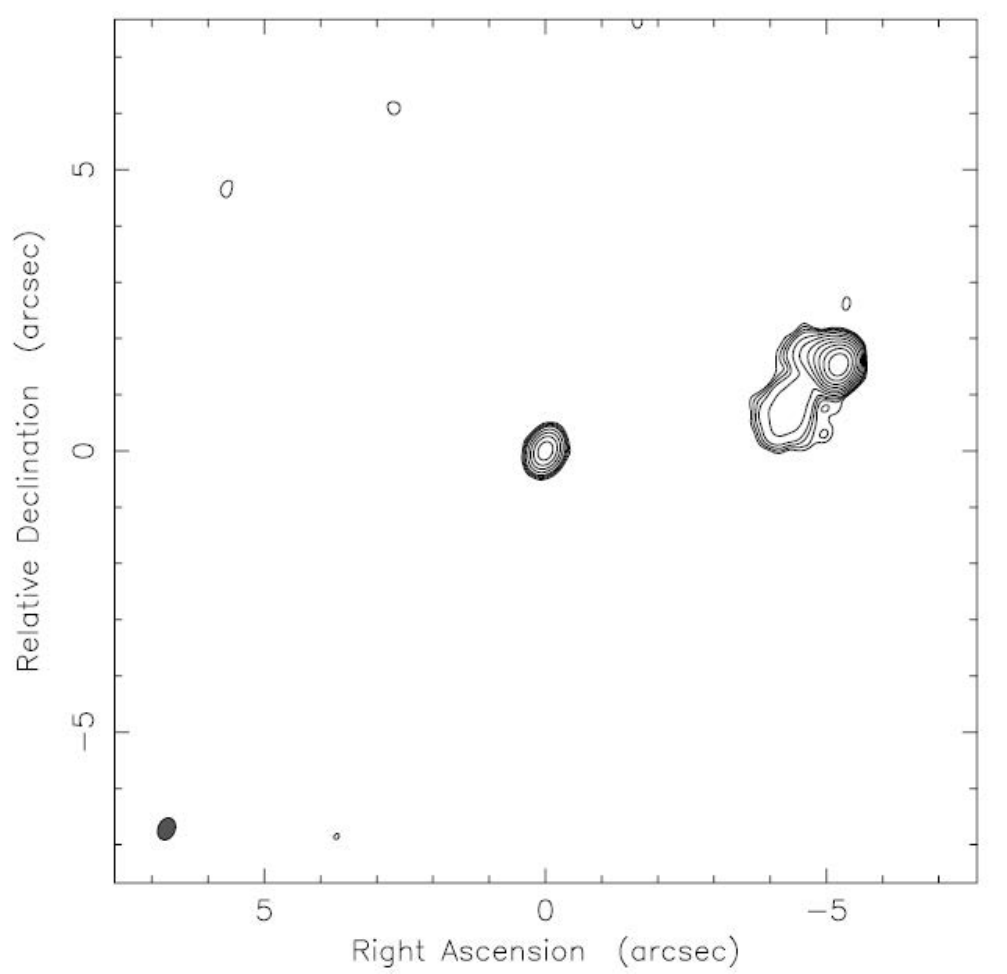} 
\label{fig:Images4-4}} 
\subfloat[Part 5][J1346+2900 z = 2.721]
{\includegraphics[width=2.2in]{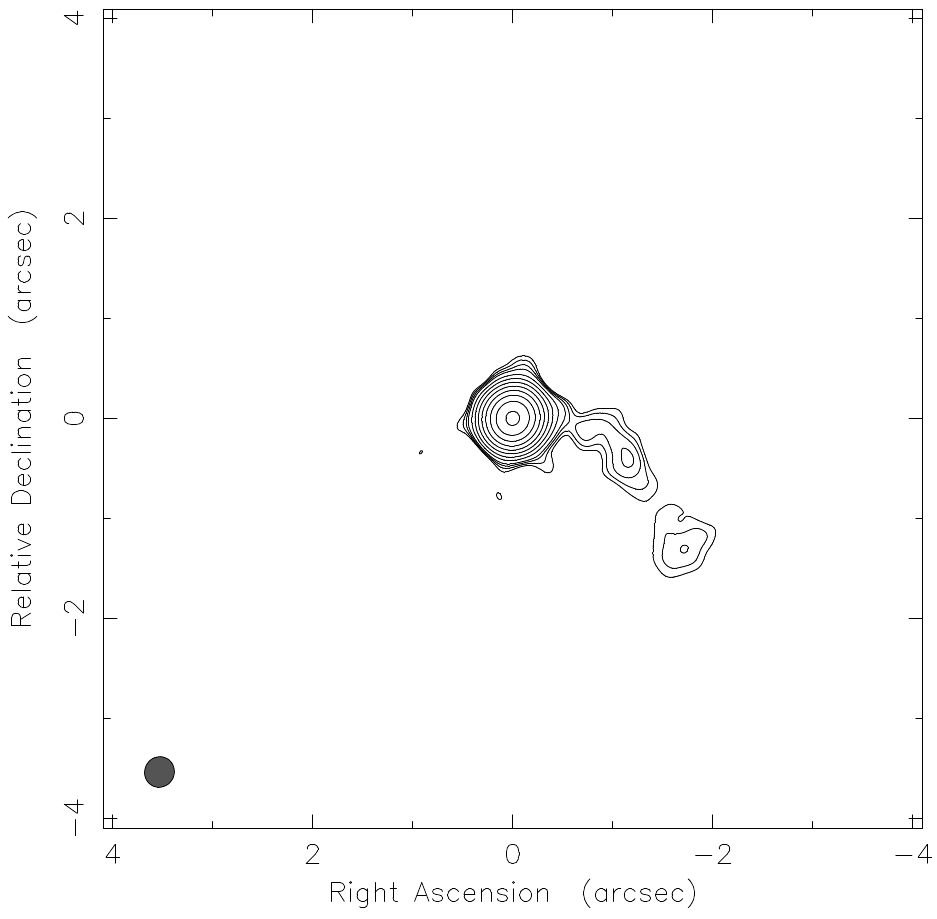} 
\label{fig:Images4-5}} 
\subfloat[Part 6][J1353+5725 z = 3.464]{\includegraphics[width=2.2in]{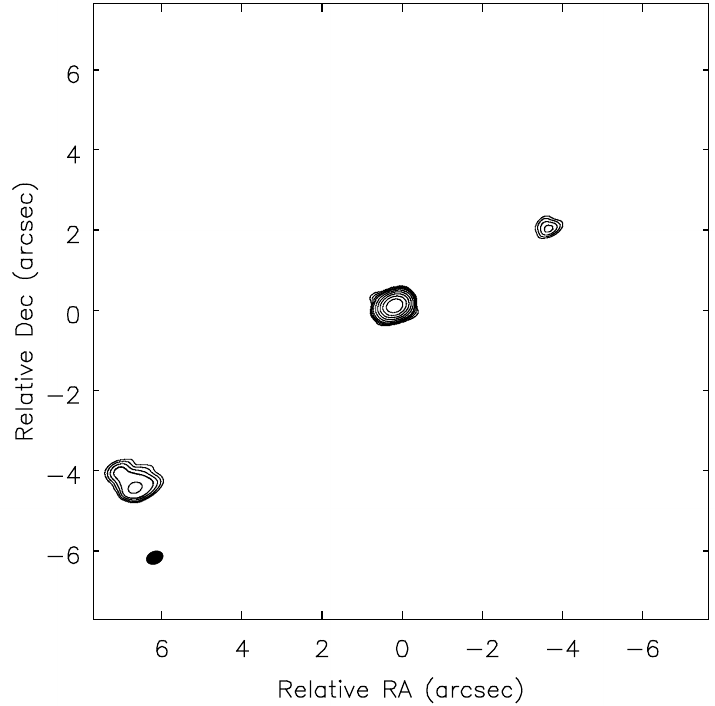} \label{fig:Images4-6}} \\

\subfloat[Part 7][J1356+2918 z = 3.244]{\includegraphics[width=2.2in]{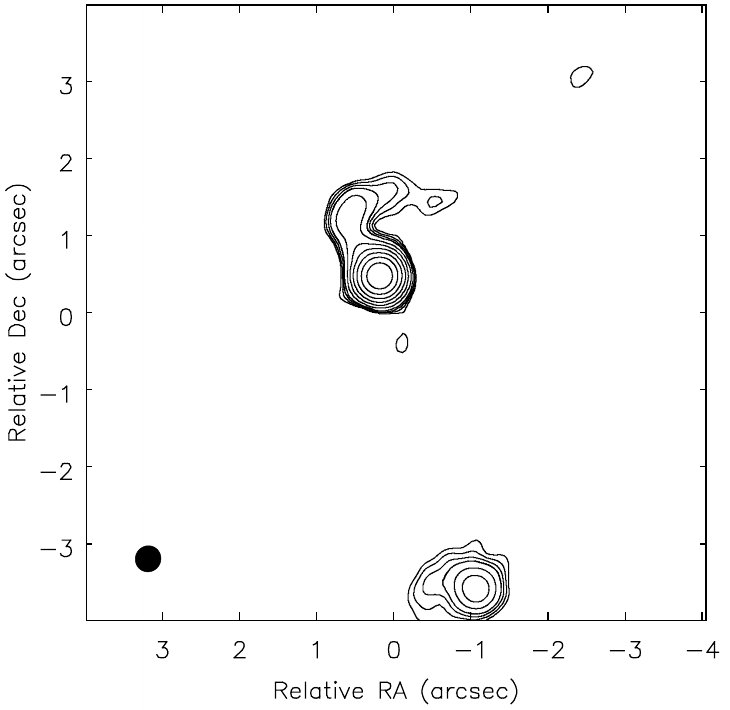} \label{fig:Images4-7}} 
\subfloat[Part 8][J1400+0425 z = 2.550]{\includegraphics[width=2.2in]{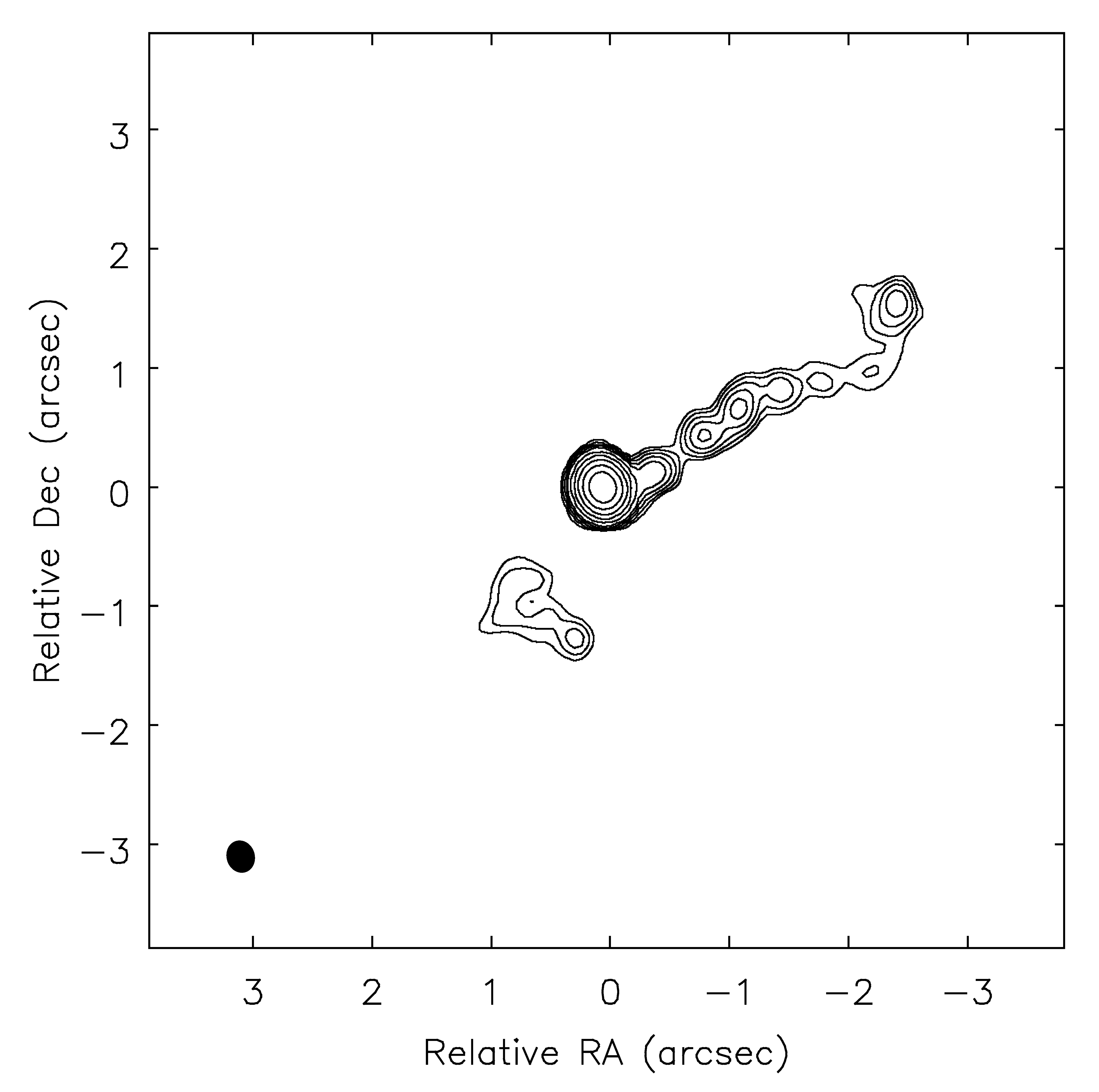} \label{fig:Images4-8}} 
\subfloat[Part 9][J1405+0415 z = 3.215]{\includegraphics[width=2.2in]{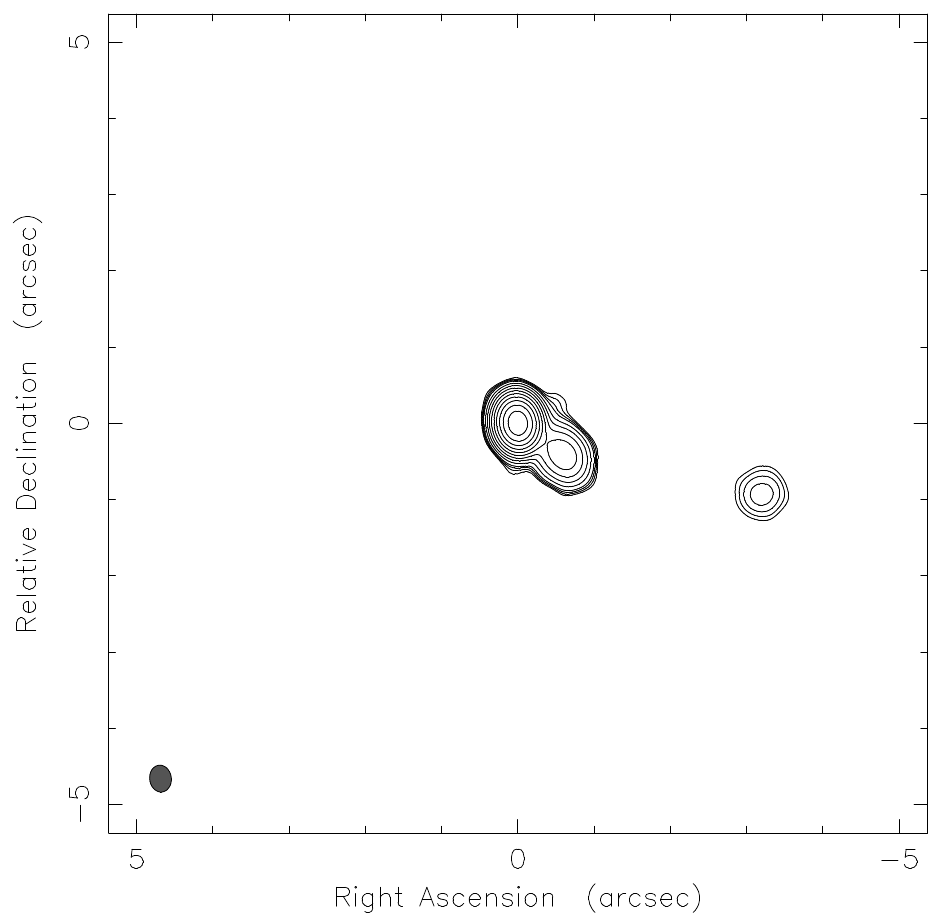} \label{fig:Images4-9}} \\
\caption{}
\label{fig:Images5}
\end{figure}


\begin{figure}
\figurenum{6}
\centering

\subfloat[Part 1][J1429+2607 z = 2.914]{\includegraphics[width=2.2in]{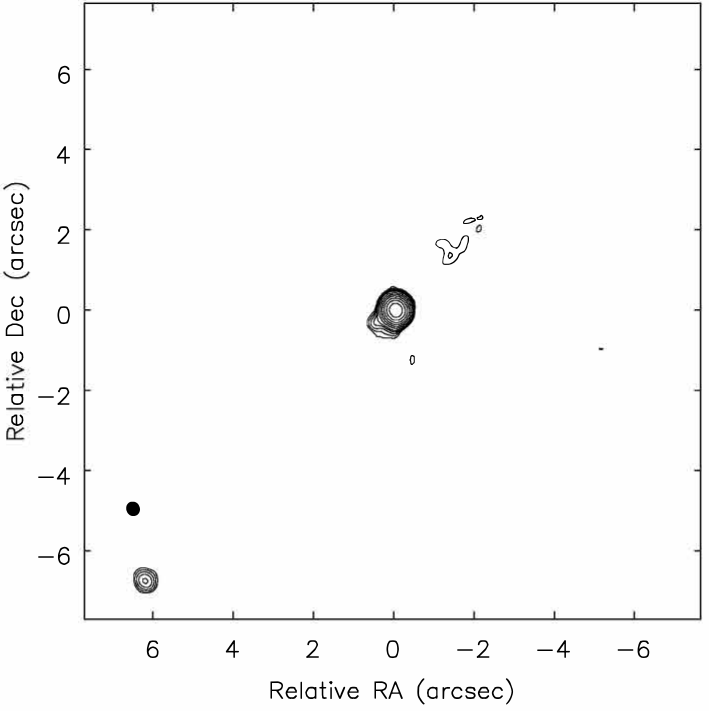} \label{fig:Images5-1}} 
\subfloat[Part 2][J1430+4204 z = 4.715]{\includegraphics[width=2.2in]{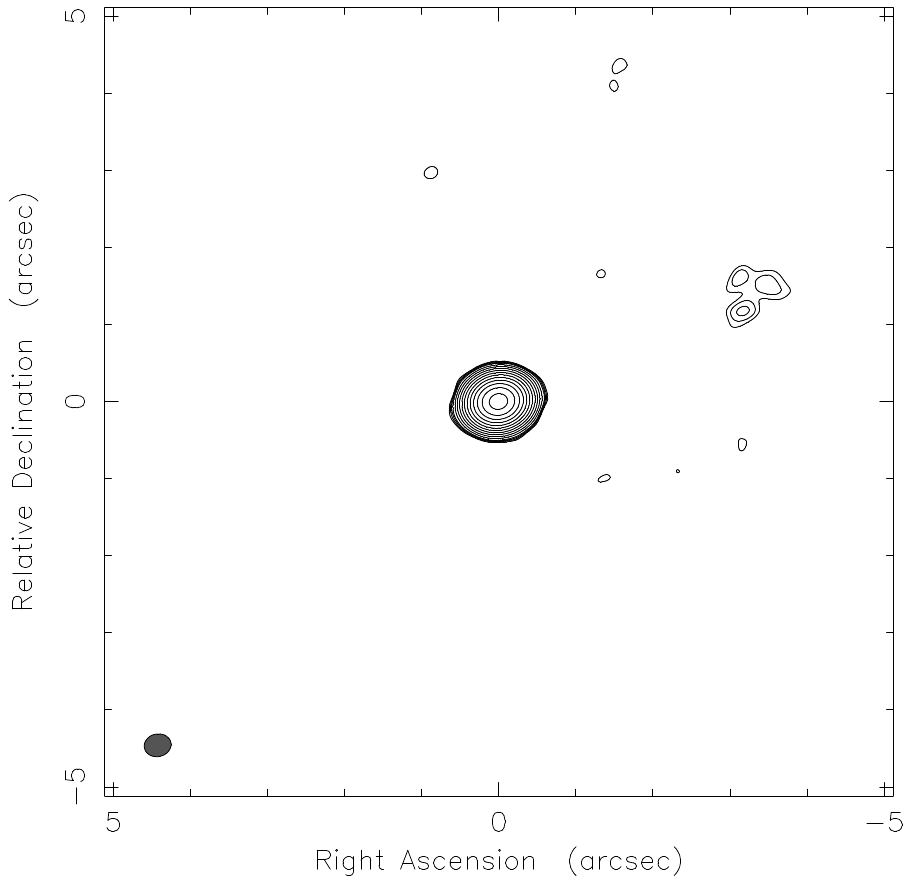} \label{fig:Images5-2}} 
\subfloat[Part 3][J1450+0910 z = 2.611]{\includegraphics[width=2.2in]{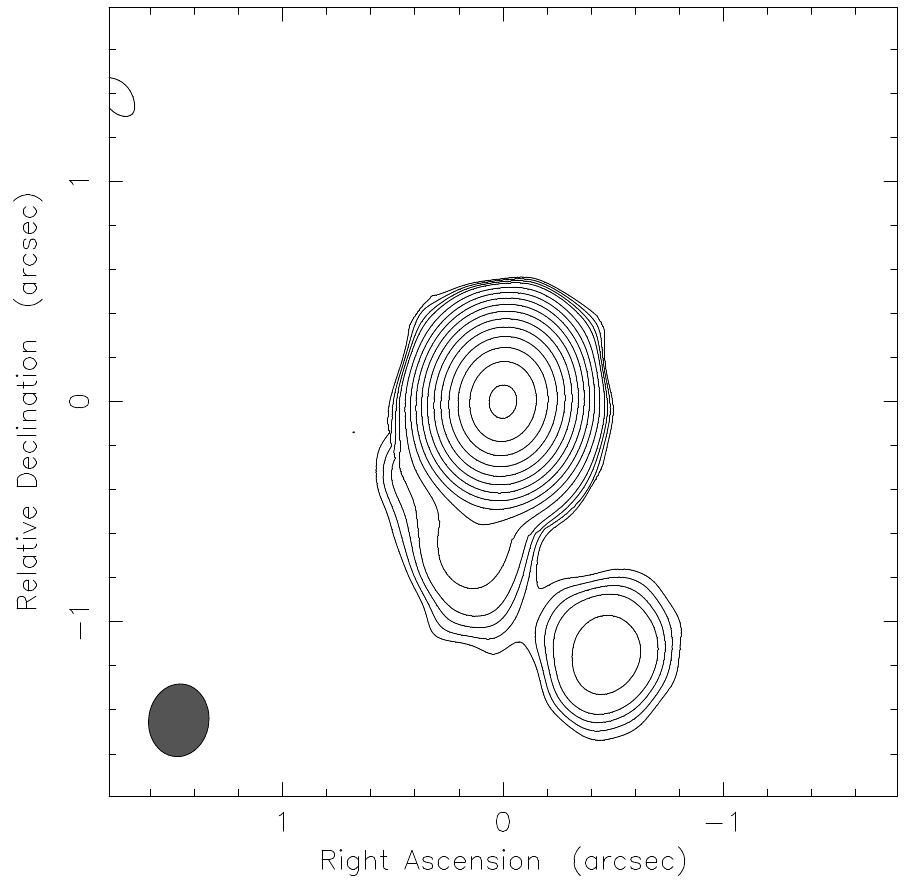} \label{fig:Images5-3}} \\

\subfloat[Part 4][J1457+3439 z= 2.732]{\includegraphics[width=2.2in]{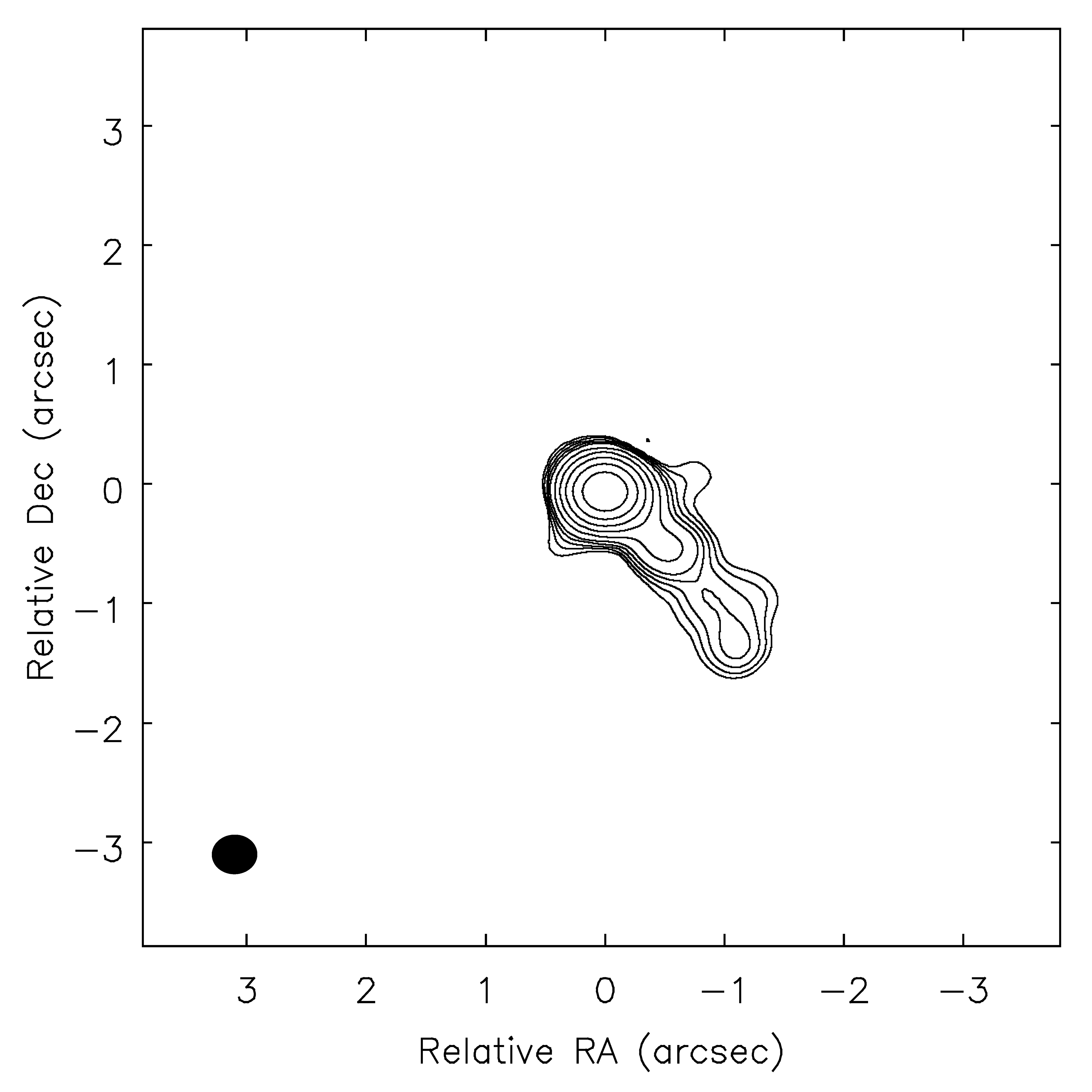} \label{fig:Images5-4}} 
\subfloat[Part 5][J1459+3253 z = 3.324]{\includegraphics[width=2.2in]{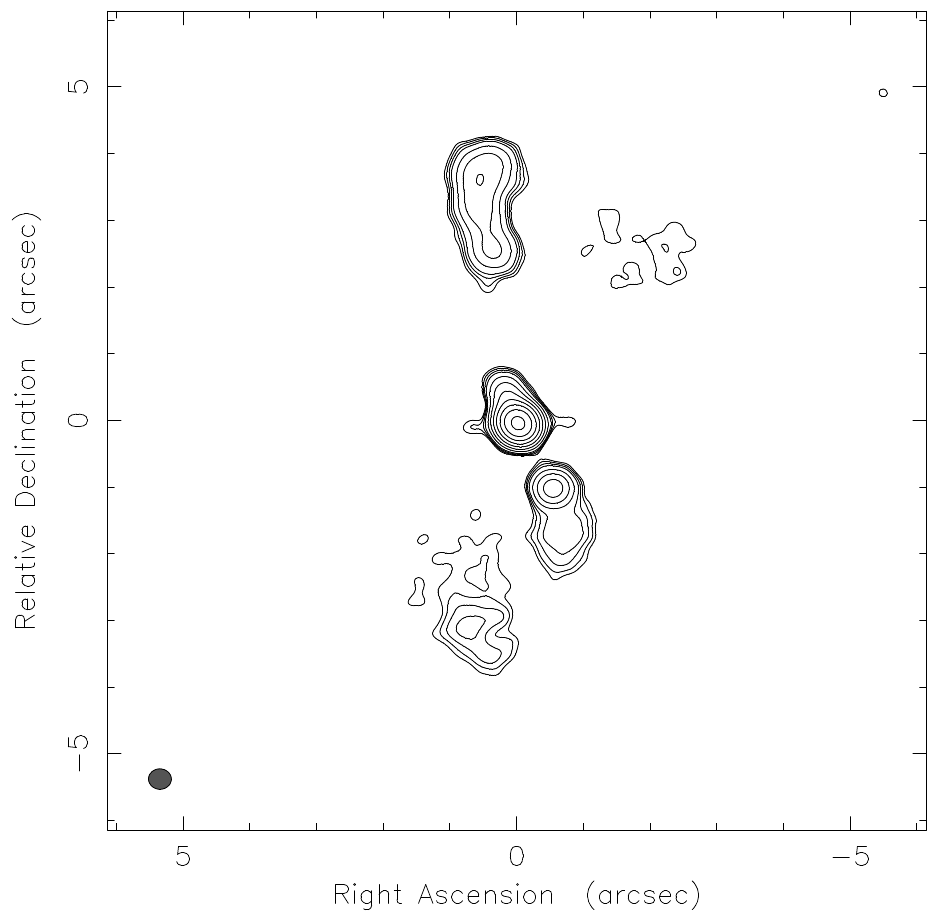} 
\label{fig:Images5-5}} 
\subfloat[Part 6][J1502+5521 z = 3.320]{\includegraphics[width=2.2in]{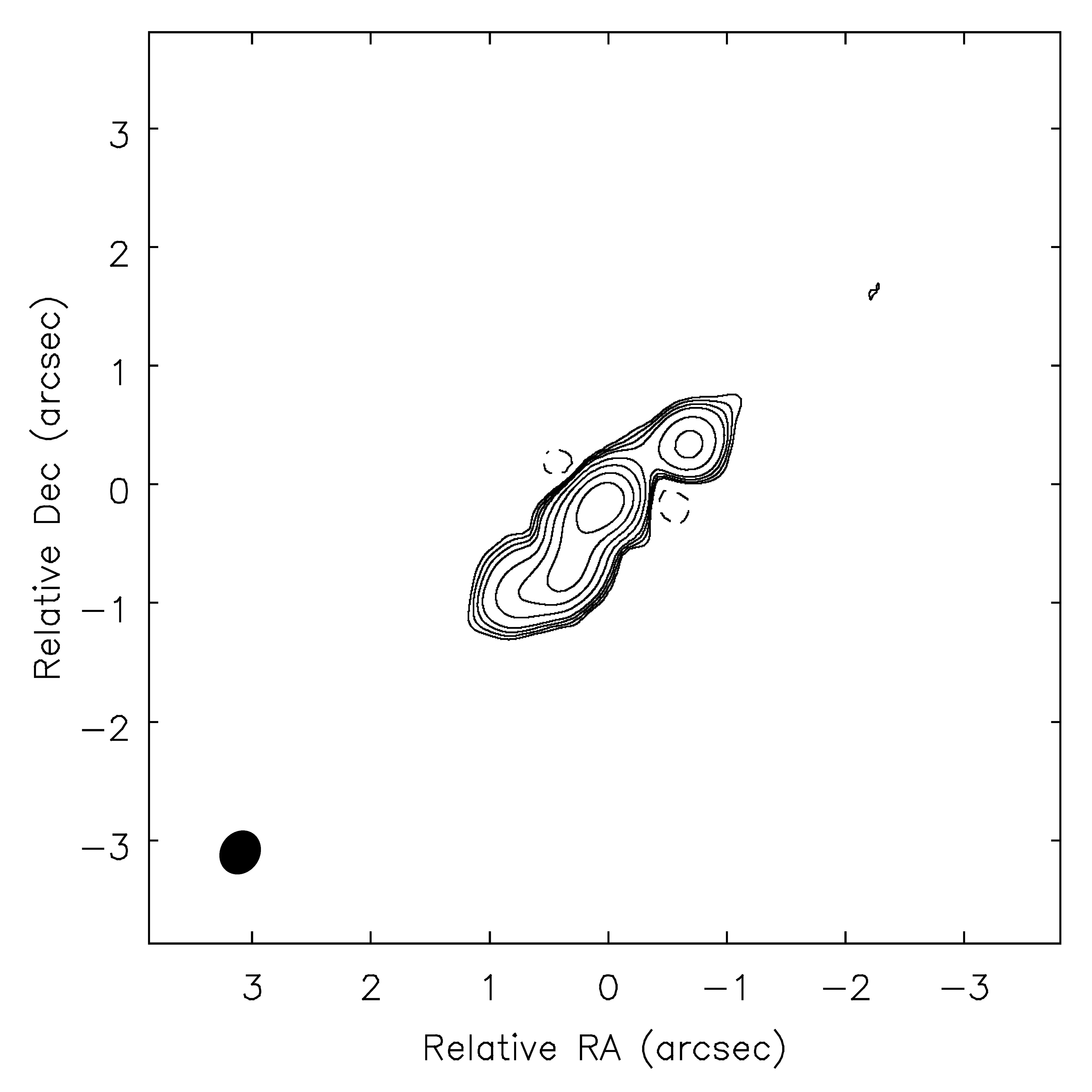} \label{fig:Images5-6}} \\

\subfloat[Part 7][J1528+5310 z =2.822]
{\includegraphics[width=2.2in]{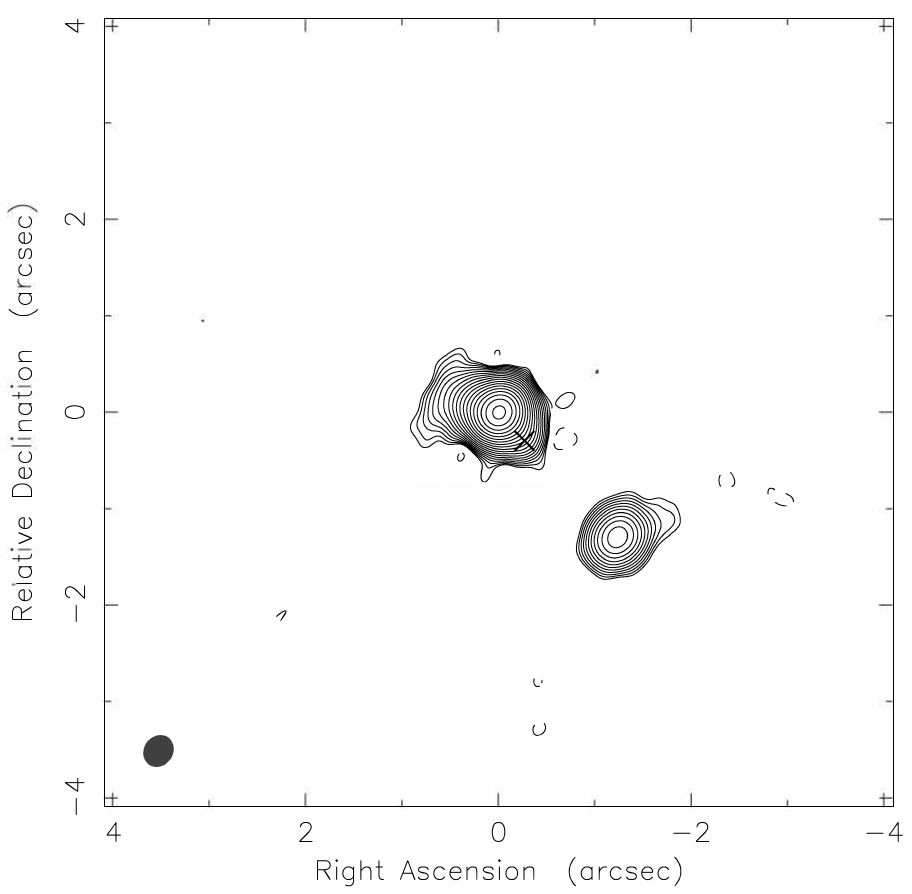} 
\label{fig:Images5-7}} 
\subfloat[Part 8][J1540+4738 z = 2.566]
{\includegraphics[width=2.2in]{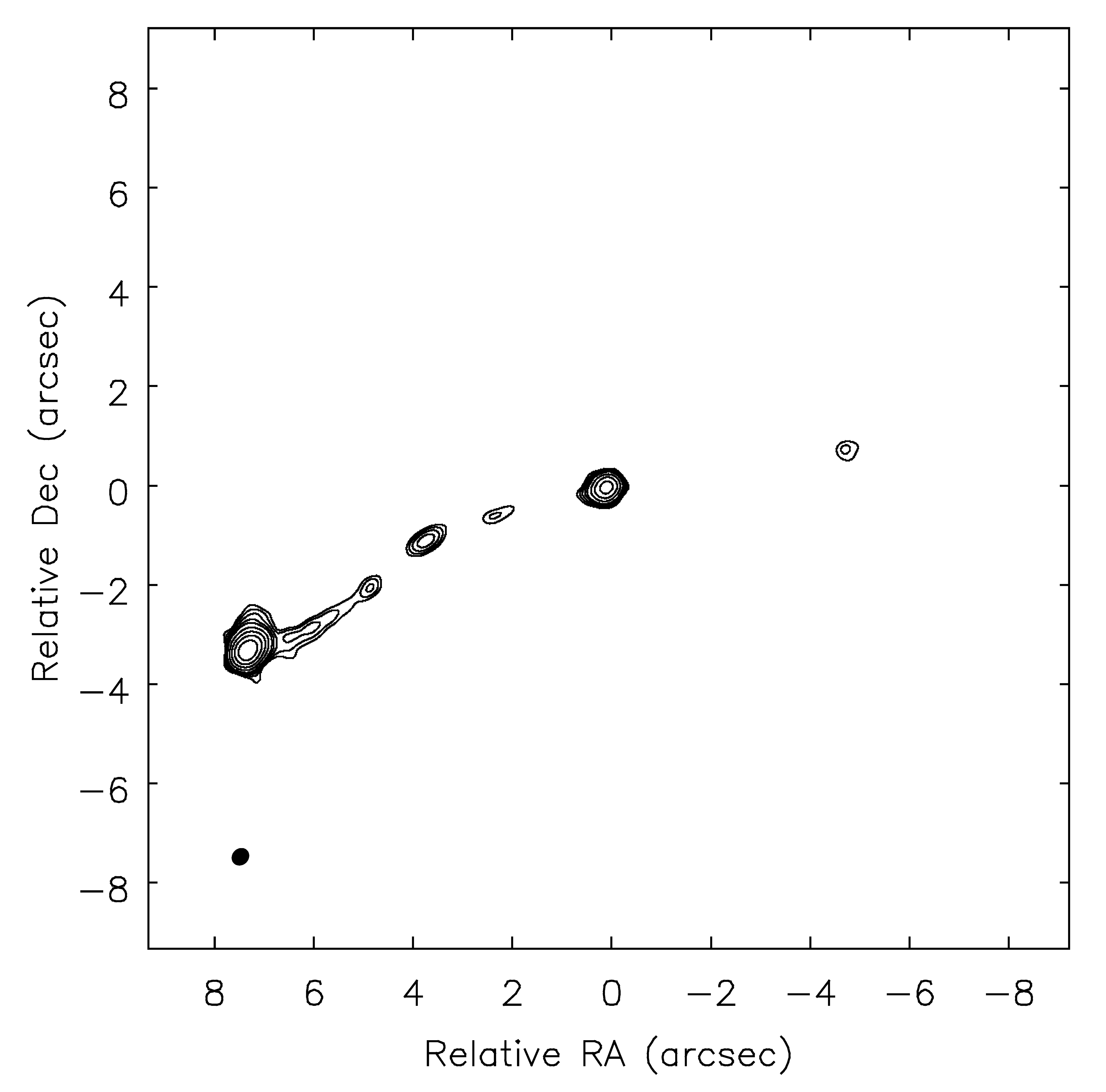} 
\label{fig:Images5-8}} 
\subfloat[Part 9][J1602+2410 z = 2.531]{\includegraphics[width=2.2in]{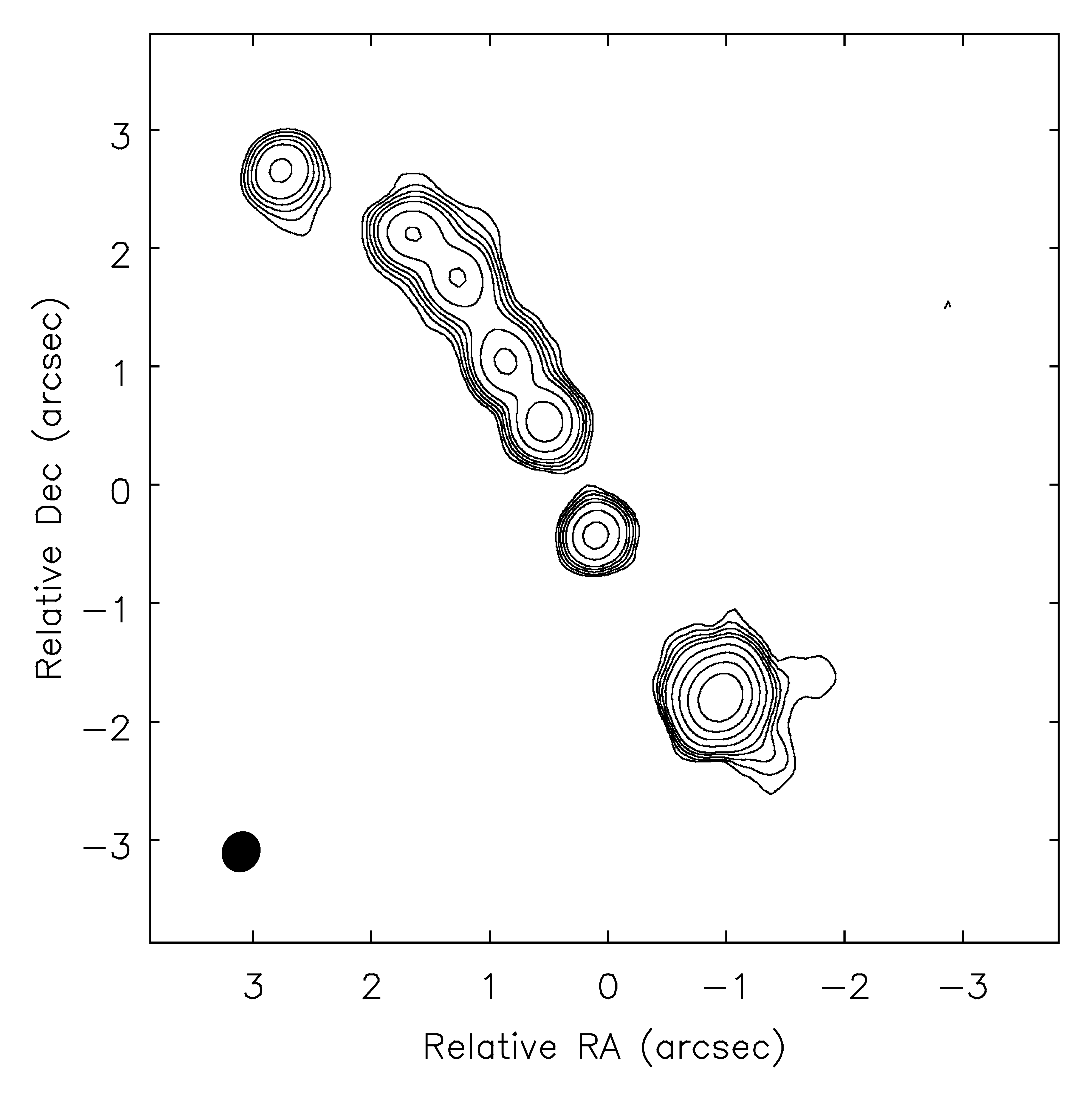} 
\label{fig:Images5-9}} \\
\caption{}
\label{fig:Images6}
\end{figure}


\begin{figure}
\figurenum{7}
\centering

\subfloat[Part 1][J1610+1811 z = 3.118]{\includegraphics[width=2.2in]{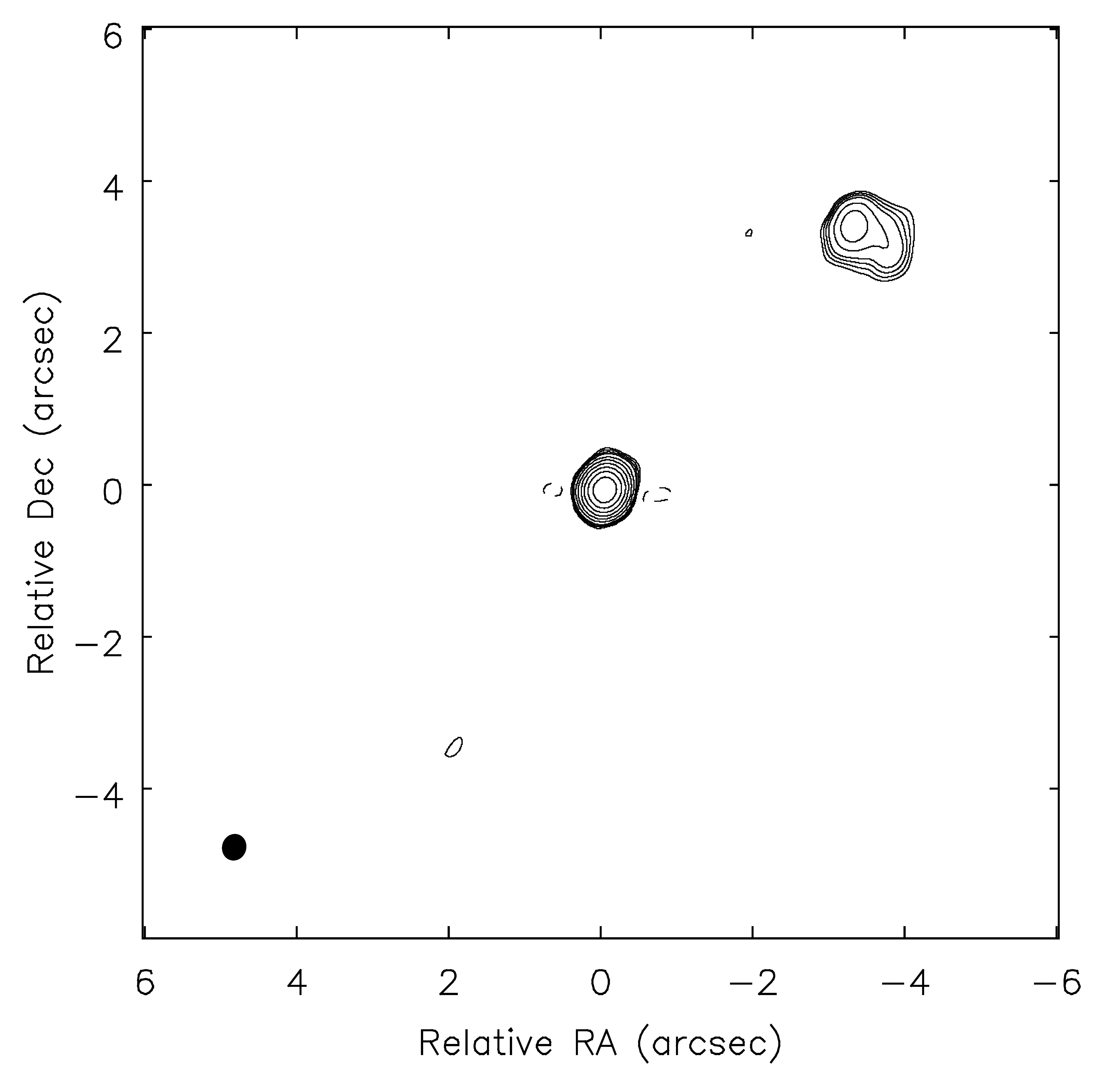} \label{fig:Images6-1}} 
\subfloat[Part 2][J1612+2758 z = 3.535]{\includegraphics[width=2.2in]{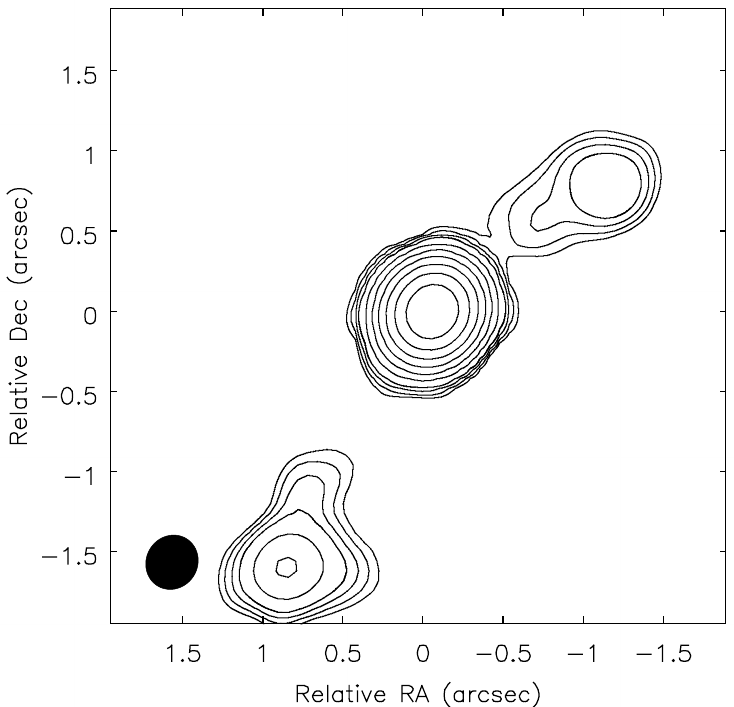} 
\label{fig:Images6-2}} 
\subfloat[Part 3][J1655+3242 z = 3.186]{\includegraphics[width=2.2in]{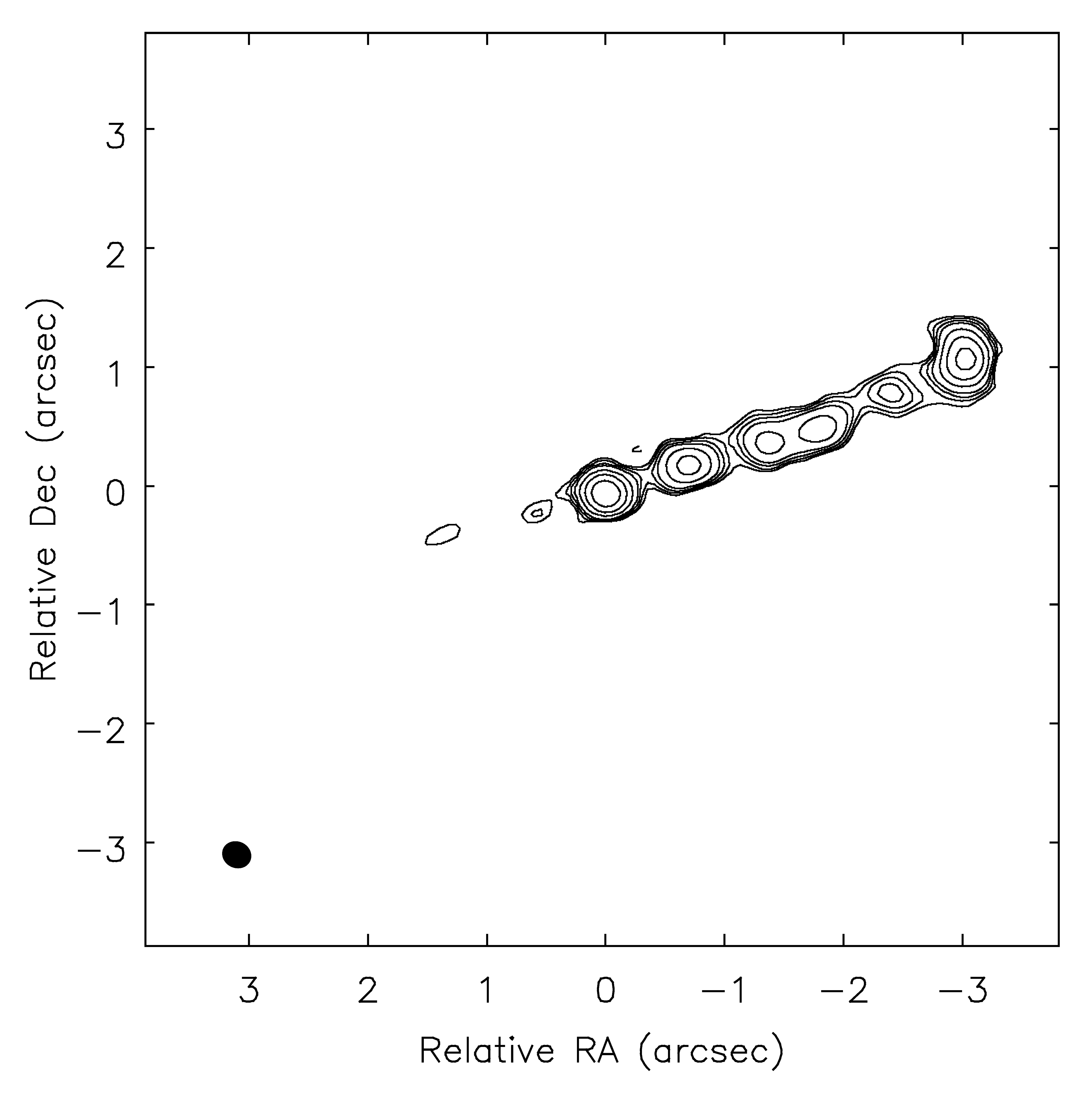}
\label{fig:Images6-3}} 
\caption{}
\label{fig:Images7}
\end{figure}

\begin{figure}
\figurenum{A1}
\plotone{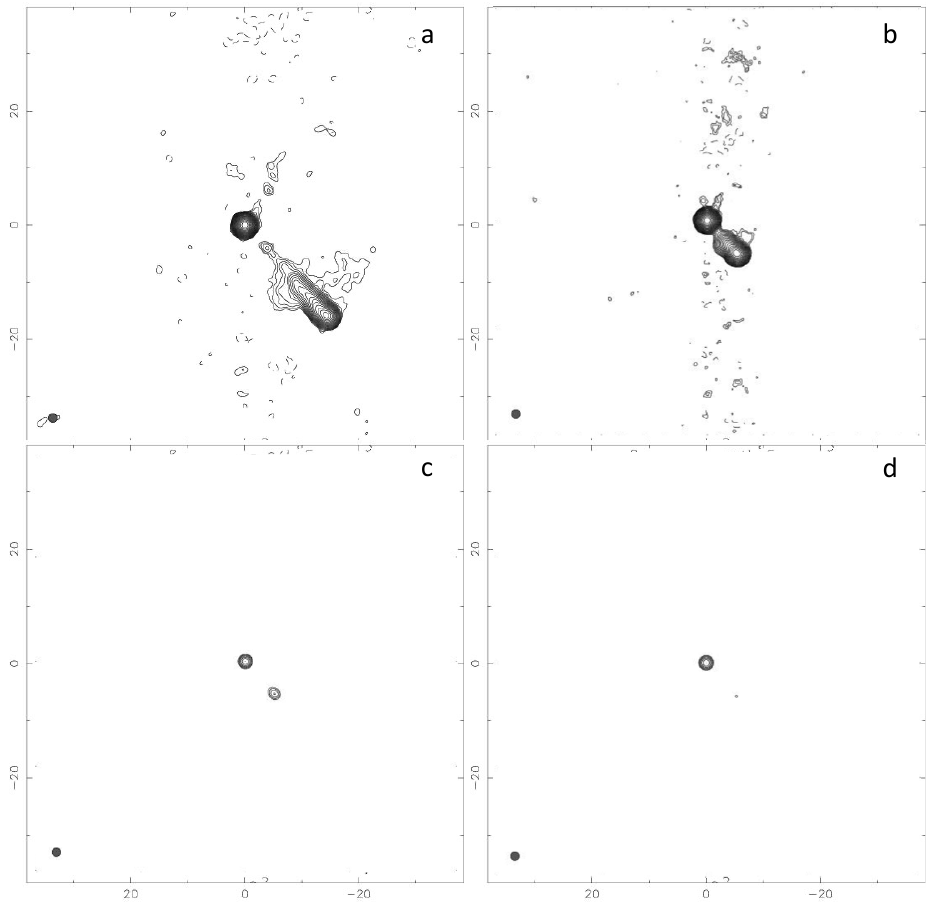}
\caption{Visualization of the $z = 0.158$ quasar 3C~273 jet (panel a) being redshifted to a higher redshift of $z = 3.6$, accounting for image scale (b), cosmological dimming (c), and K-correction (d).We assumed spectral indices of 0.0 for the core and 1.0 for the extended emission.}
\end{figure}

\end{document}